\documentclass{article}

\usepackage[english]{babel}
\usepackage{booktabs}

\usepackage[letterpaper,top=2cm,bottom=2cm,left=3cm,right=3cm,marginparwidth=1.75cm]{geometry}

\usepackage{amsmath}
\usepackage{graphicx}
\usepackage[colorlinks=true, allcolors=blue]{hyperref}

\usepackage{dcolumn}
\usepackage{bm}
\usepackage[numbers,sort&compress]{natbib}

\usepackage[utf8]{inputenc}
\usepackage[T1]{fontenc}
\usepackage{mathptmx}
\usepackage{etoolbox}
\usepackage{makecell}

\usepackage{array}
\newcolumntype{C}[1]{>{\Centering\arraybackslash}m{#1}} 

\usepackage{authblk} 

\newcounter{fifthlevel}[subsubsection]
\renewcommand{\thefifthlevel}{\alph{fifthlevel}.}

\newcommand{\fifthlevel}[1]{%
  \refstepcounter{fifthlevel}%
  \vspace{6pt}
  \noindent\hspace*{1.5em}
  \textbf{\textit{\thefifthlevel\ #1}}
  \par\vspace{3pt}
}

\title{Mechanisms and Opportunities for Tunable High-Purity Single Photon Emitters: A Review of Hybrid Perovskites and Prospects for Bright Squeezed Vacuum}

\author[1]{Galy Yang}
\author[2]{Eric Ashallay}
\author[2]{Abolfazl Bayat}
\author[2]{Zhiming Wang}
\author[1,2]{Arup Neogi\thanks{Corresponding author: arup@uestc.edu.cn}}

\affil[1]{Glasgow College, University of Electronic Science and Technology of China, Chengdu 610054, China}
\affil[2]{Institute of Fundamental and Frontier Sciences, University of Electronic Science and Technology of China, Chengdu 610054, China}

\begin{document}
\maketitle

\begin{abstract}
Single-photon emitters (SPEs) are central to quantum communication, computing, and metrology, yet their development remains constrained by trade-offs in purity, indistinguishability, and tunability. This review presents a mechanism-based classification of SPEs, offering a physics-oriented framework to clarify the performance limitations of conventional sources, including quantum emitters and nonlinear optical processes. Particular attention is given to hybrid organic–inorganic perovskite quantum dots (HOIP QDs), which provide size- and composition-tunable emission with narrow linewidths and room-temperature operation. Through comparative analysis of physical mechanisms and performance metrics, we show how HOIP QDs may address key limitations of established SPE platforms. Recognizing the constraints of current deterministic sources, we introduce a performance framework to guide the development of scalable SPEs, and examine the theoretical potential of bright squeezed vacuum (BSV) states, discussing how BSV mechanisms could serve as a promising avenue for multiplexable, high-purity photon generation beyond conventional heralded schemes. The review concludes by outlining future directions for integrating HOIP- and BSV-based concepts into scalable quantum photonic architectures.
\end{abstract}
\section{Introduction}
\begin{figure}[htbp]
    \centering
    \includegraphics[width=0.7\linewidth]{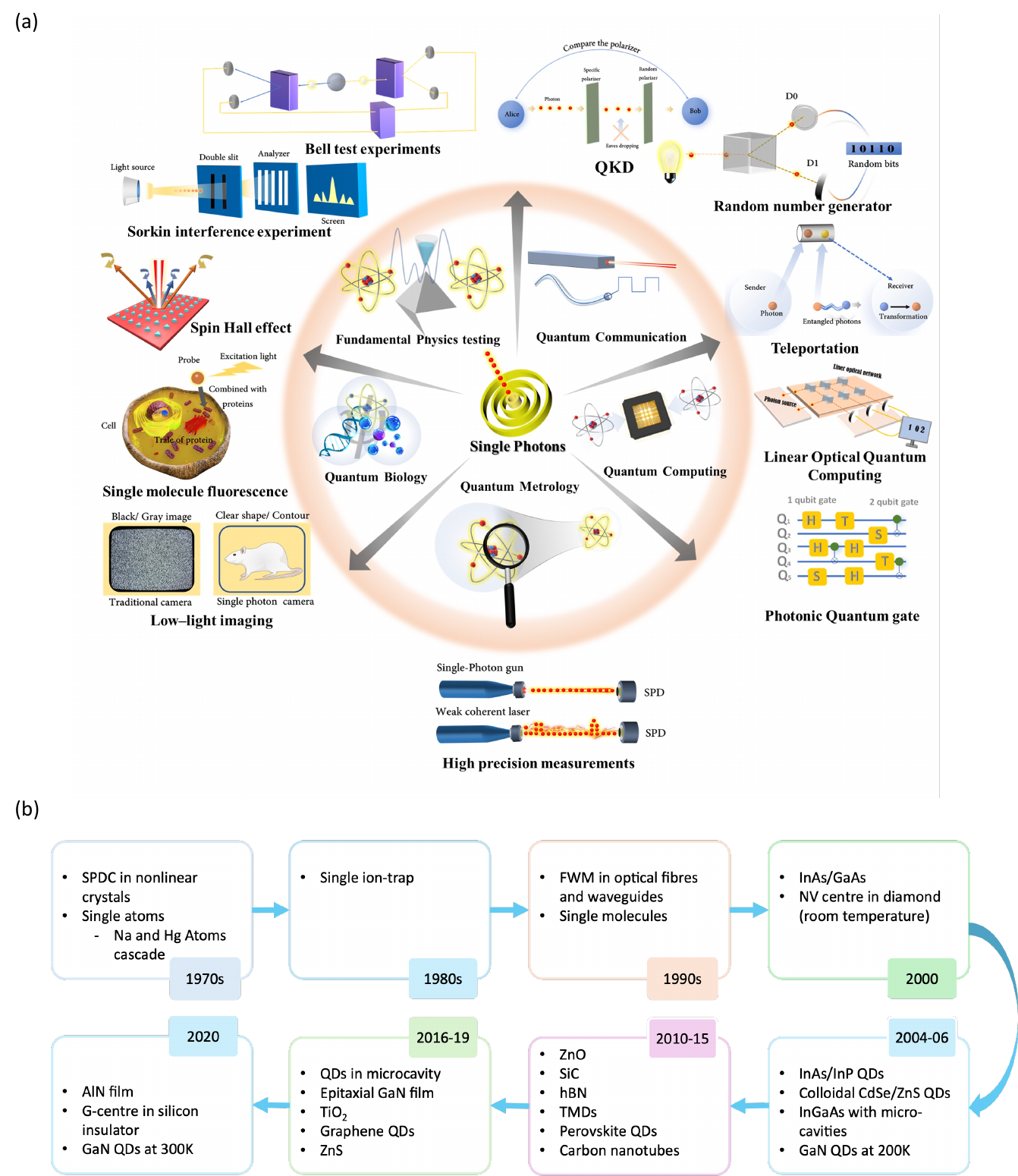}
    \caption{Applications and evolution of single photons sources. (a) Single photons possess indivisibility, low decoherence, fast transmission speed, entanglement capability, and can exist in a superposition of multiple states, which make them highly appealing for advanced applications in quantum communication, computing, metrology, biology, and fundamental physics tests. (b) Timeline highlighting key developments in SPE technologies.}
    \label{fig:applications}
\end{figure}

Quantum systems have driven significant recent advancements in communication, computing, imaging, and spectroscopy, accelerated by photonic and quantum optics \cite{durant2024primer, arakawa2020progress, aslam2023quantum, daley2022practical}. These quantum systems benefit from high purity sources for their efficient operation. The inherent characteristics of single photons (the fundamental quanta of light), such as spin, enable the encoding of information and have become an indispensable component in modern quantum hardware \cite{couteau2023applications1, gupta2023single, hadfield2023single, couteau2023applications2}.  Unlike classical light sources, single photon emitters (SPEs) generate precisely one photon per excitation event, ensuring high-purity. Ideally, these sources also provide high indistinguishability, which is essential for quantum interference. They interact minimally with their environment, making them robust over long distances and easier to manipulate \cite{couteau2023applications1, eisaman2011invited}. These characteristics are fundamental to various quantum applications including quantum key distribution (QKD), quantum networks, and scalable quantum computing applications, as illustrated in Fig. \ref{fig:applications}(a) \cite{cogan2023deterministic, couteau2023applications1, li2023high, liu2023experimental, maring2024versatile, ding2025high}.  Achieving high-purity, on-demand single-photon emission with spectral controllability remains a critical challenge in advancing scalable quantum networks. Among the diverse platforms explored for single photon emission, emerging perovskite systems have garnered significant attention due to their tunable emission properties.

Parallelly, progress in semiconducting and superconducting detectors has also enabled high-resolution detection across various wavelength ranges, particularly in the mid-infrared region, where optimising the balance between spectral response and noise remains crucial \cite{lau2023superconducting, you2020superconducting, esmaeil2021superconducting, liu2023experimental, li2023high}. For an in-depth exploration of detection technologies, readers are directed to existing reviews \cite{lau2023superconducting, hadfield2023single, you2020superconducting, esmaeil2021superconducting}.

\subsection{Evolution of SPEs Techniques}
Over the past decade, diverse approaches for developing SPEs have emerged across optics, nanophysics, chemistry, and material science\cite{gupta2023single, couteau2023applications1, kaplan2023hong, parto2021defect}. Fig. \ref{fig:applications}(b) illustrates the development of solid-state and non solid-state SPEs since the 1970s. Given that the concept of single photons originates from quantum physics, many early-generation methods were inherently probabilistic, offering limited control over photon emission. These methods often featured simpler setups compared to deterministic sources \cite{thomas2022efficient}.

One of the earliest probabilistic techniques, spontaneous parametric down-conversion (SPDC), was demonstrated in 1967 and has since been widely employed in experimental setups \cite{harris1967observation}. SPDC converts an incoming photon beam into pairs of photons, referred to as signal and idler, distributed across two spatially distinct modes. By implementing gating mechanisms, these beams can be filtered to serve as single photon sources. SPDC became the dominant mechanism for single photon generation in the early 2000s and remains prevalent today \cite{kaneda2015time, zhang2021spontaneous, kiessler2025spdc, tseng2022efficient}. Other probabilistic sources, including atomic fluorescence and atomic vapour, predate deterministic techniques and typically involve light interacting with a bulk medium alongside detection systems for heralding photons \cite{dideriksen2021room, thomas2022efficient}.

Nevertheless, the inherent unpredictability of probabilistic SPEs poses scalability challenges, particularly for applications demanding high accuracy and security, such as quantum computing and communication. An ideal SPE should emit single photons on demand, with a 100\% probability of single photon emission and zero probability of multiple-photon emission \cite{couteau2023applications1, senellart2017high, gupta2023single, esmann2024solid}. Probabilistic sources inherently fall short of this criterion, as there is always a non-zero probability of producing multiple photon pairs simultaneously. To overcome these limitations, deterministic SPEs have been developed.

Quantum dots (QDs) represent a prominent class of deterministic sources, with their material properties precisely controlled through molecular beam epitaxy (MBE). MBE-grown QDs can be finely tuned using external electromagnetic (EM) fields or other approaches, and the emitters can be electrically contacted with relative ease \cite{smolka2021optical, shen2024machine, von2022triggered}. Recent studies have demonstrated near-unity quantum purity for QDs integrated into optical cavities, achieving overall efficiencies at least ten times higher than those of heralded sources \cite{thomas2021race, senellart2017high, morozov2023purifying, you2022quantum}. Beyond QDs, P. Thomas \textit{et al.} introduced a deterministic photon generation technique leveraging photonic entanglement with a single memory atom in a cavity, successfully producing large, high-fidelity photonic graph states of the GHz and cluster types \cite{thomas2022efficient}.

In contrast to probabilistic sources, deterministic SPEs require isolating and controlling a single emitter, necessitating techniques such as cryogenic cooling \cite{khramtsov2021bright, martin2023photophysics}, nano-fabrication \cite{hollenbach2022wafer, xu2021creating}, or atomic trapping systems \cite{spencer2023monolithic, kutovyi2022efficient}. Significant advancements in deterministic SPEs have paralleled the maturation of quantum communication and computation, leading to the development of on-demand single photons with enhanced reliability and control.

\subsection{Challenges in SPEs}
Despite considerable advancements, SPEs continue to face critical challenges that impede their practical and scalable applications. A primary obstacle is achieving high purity, ensuring that only one photon is emitted per excitation event. High purity is crucial for boson-sampling applications like QKD, where multi-photon emissions can undermine security by enabling potential eavesdropping. Heralded SPEs derived from SPDC etc. exhibit thermal statistics, resulting in a non-zero probability of producing multiple photon pairs—a notable vulnerability for QKD systems \cite{keller2022cavity, faleo2024entanglement, zahidy2024quantum, meng2024deterministic}.

While techniques such as advanced filtering and cavity quantum electrodynamics (cQED) can mitigate multi-photon emissions, they often do so at the expense of reduced efficiency \cite{pscherer2021single, keller2022cavity, iles2025demand}. This inherent trade-off between purity and efficiency presents a significant barrier to the practical deployment of high-quality photon sources. Overcoming this limitation is essential for the realization of robust and scalable quantum technologies.

Beyond purity, integrating SPEs with existing quantum systems, such as fibre-optic networks, waveguides, or other quantum hardware, poses another critical challenge \cite{gupta2023single, couteau2023applications1, hadfield2023single, nowak2014deterministic}. Many quantum systems operate within the telecom wavelength range (C band and O band), optimised for long-distance communication due to minimal transmission loss. However, not all SPEs naturally align with this range, resulting in inefficiencies in coupling and communication.

Addressing these challenges requires the advancement of tunable SPEs and the development of hybrid integration methods. Tunable emitters can adjust their emission wavelengths to align with specific operational requirements, while hybrid systems enhance compatibility with existing infrastructures. Beyond achieving outstanding performance, it is crucial to develop a single SPE platform that offers broad emission tunability and seamless integration into photonic circuits. Such a platform should also be cost-effective and highly reproducible to enable scalable applications and commercialization. A versatile and standardized SPE system would streamline fabrication, reduce variability, and facilitate mass production, making it practical for widespread deployment in quantum technologies.

\subsection{Scope and Objectives of This Review}
Although numerous reviews have explored various aspects of SPEs, most emphasise material properties or specific applications, often overlooking the fundamental physical mechanisms driving single photon emission \cite{aolita2007quantum, hadfield2023single, lau2023superconducting, gupta2023single, you2020superconducting, esmaeil2021superconducting, liu2025single}. This review addresses this gap by providing an in-depth analysis of the physical principles underlying SPEs, offering insights into their operation and inherent limitations.

We aim to go beyond conventional perspectives by focusing on the challenges of purity and tunability—two key benchmarks for scalable quantum technologies. In particular, we highlight hybrid organic–inorganic perovskite quantum dots (HOIP QDs) as an emerging platform that has already demonstrated high purity and tunability under room-temperature conditions.

To achieve this, we first examine the physical mechanisms of existing SPEs, assessing performance metrics such as purity, indistinguishability, and tunability. We then explore HOIP-based SPEs and their advantages over conventional approaches. Finally, we outline emerging opportunities and future directions, including a framework for evaluating SPE performance and a forward-looking perspective on bright squeezed vacuum (BSV) states. We survey BSV as a high-potential avenue for multiplexable, high-purity photon generation beyond conventional heralded schemes. By connecting physics-driven advances with practical applications, this review highlights potential pathways toward scalable and robust quantum technologies.
\section{Single Photon Emission}
\subsection{Definition of Single Photon}
The concept of a photon, though fundamental, has long been a source of conceptual complexity. The term “photon” was officially introduced in 1926, while the notion of quantized radiation dates back to 1900 \cite{lewis1926conservation, planck1900improvement}. From the perspective of EM radiation, a photon is defined as an elementary excitation of a single mode of the quantized EM field \cite{eisaman2011invited}. More simply, photons are the smallest discrete units of EM radiation—quanta of light.

As quantum objects, photons exhibit wave-particle duality. From the particle perspective, photons are massless bosons and represent indivisible quanta of the EM field. They are characterised by Fock states, which describe a well-defined number of particles in a particular mode of the quantized EM field. A Fock state is an eigenstate of the number operator \(\hat{N}\), with eigenvalue \(N\), representing the number of particles. For a single photon Fock state, \(N=1\). and it is mathematically represented as:
\[\hat{a}^\dagger|0\rangle,\]
where \(\hat{a}^\dagger\) is the photon creation operator and \(|0\rangle\) denotes the vacuum state. From the wave perspective, a photon is represented as a superposition of EM wave modes with quantized amplitudes, localized in both time and space. This description can be expressed as a single-mode or multi-mode wave packet, often written as:
\[\psi (\mathbf{r},t)=\langle\mathbf{r}, t|1\rangle.\]
where \(\psi (\mathbf{r},t)\) denotes the spatial and temporal distribution of the photon’s probability amplitude.

\subsection{Criterion of SPEs}
Before discussing the physical mechanisms of SPEs, it is essential to first define how single photon sources are characterised. Light sources can be classified using the second-order correlation function introduced by Glauber in 1963 \cite{glauber1963coherent, glauber1963quantum}, expressed as:
\begin{equation}
    g^{(2)}(\tau)=\frac{\langle I(t)I(t+\tau)\rangle}{\langle I(t)\rangle^2},
\end{equation}
where \(I\) is the light intensity and \(\tau\) is the time delay between two detectors (output ports). For an ideal SPE, the probability of detecting two photons simultaneously (\(\tau=0\)) is zero, meaning \(g^{(2)}(0)=0\). This distinguishes quantum light, where photon correlations are key, from classical light, where photons are independent and follow a Poisson distribution. 

\begin{figure*}[htbp]
    \centering
    \includegraphics[width=1\linewidth]{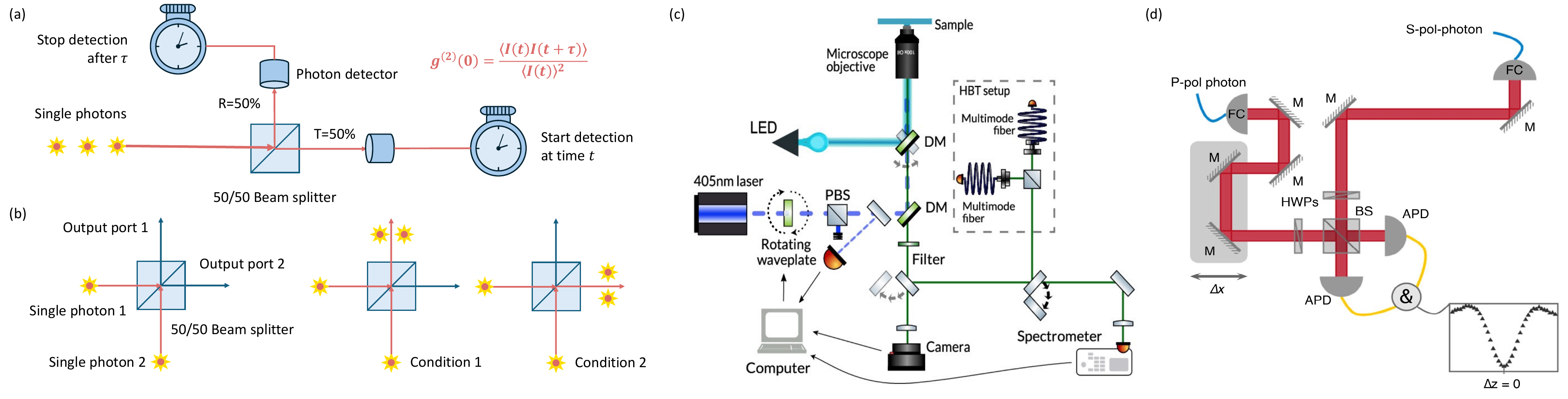}
    \caption{Characterization techniques for SPEs. (a) HBT setup for assessing single-photon purity. A 50/50 beam splitter directs incoming photons to two detectors; the second-order correlation function \(g^{(2)}(0)\) reflects coincidence counts. Lower values indicate higher purity \cite{couteau2023applications1}. Reproduced with permission from Nat. Rev. Phys. \textbf{5}, 326 (2023). Copyright 2023 Springer Nature Limited. (b) HOM setup for testing photon indistinguishability. Two identical photons entering a 50/50 beam splitter from different ports interfere quantum mechanically and exit through the same output, leading to suppressed coincidences (HOM dip) \cite{couteau2023applications1}. Reproduced with permission from Nat. Rev. Phys. \textbf{5}, 326 (2023). Copyright 2023 Springer Nature Limited. (c) Inverted confocal microscope integrating HBT functionality. Emitters are excited via LED or pulsed laser, and emission is analysed via spectrometer or camera. HBT measurements are conducted using multimode fibres and avalanche photodiodes (APDs). DM: dichroic mirror; PBS: polarizing beam splitter \cite{pierini2020highly}. Reproduced with permission from ACS Photonics \textbf{7}, 2265 (2020). Copyright 2020 American Chemical Society. (d) HOM setup for 808 nm photons. Components include FC: fibre coupler; APD: avalanche photodiode; HWP: half-wave plate; BS: beam splitter; M: mirror \cite{euler2021spectral}. S. Euler et al., Eur. Phys. J. Spec. Top. \textbf{230}, 1073 (2021); licensed under a Creative Commons Attribution (CC BY) license.}
    \label{fig:HBT&HOM}
\end{figure*}

In practice, \(g^{(2)}(0)\) is measured using Hanbury-Brown and Twiss (HBT) experiment \cite{brown1956correlation}. This setup, originally designed in the 1950s to study the coherence of starlight,  employs a 50/50 beam splitter, as shown in Fig. \ref{fig:HBT&HOM}(a), (c) \cite{brown1956correlation}. The principle relies on photon detectors, which generate electrons upon receiving photons. Since detectors have a finite recovery time, they struggle to distinguish multiple photons arriving within very short intervals. The HBT experiment overcomes this limitation by splitting incoming photons into two paths. If a second photon arrives shortly after the first, it is directed to the second detector, which is ready to register it. For an ideal SPE, only one photon is emitted at a time, preventing simultaneous detection at both detectors. Experimentally, a source is typically considered an SPE if  \(g^{(2)}(0)<1/2\) \cite{couteau2023applications1}. 

While the HBT experiment confirms whether an emitter produces single photons, the Hong-Ou-Mandel (HOM) interference test assesses photon indistinguishability (See Fig. \ref{fig:HBT&HOM}(b), (d)).  Like the HBT experiment, HOM interference uses the beam splitter, but instead of a single photon, two photons arrive simultaneously \cite{hong1987measurement}. If the photons are indistinguishable, meaning they share the same polarisation, frequency, spatial mode, and temporal mode, they interfere destructively due to the symmetry of the quantum state. This phenomenon is described by the HOM formula:
\begin{equation}
   P (\text{same output port}) = 1 - |\langle \psi_{\text{in}} | \hat{U}_{\text{BS}} | \psi_{\text{in}} \rangle|^2,
\end{equation}
where $\hat{U}_{\text{BS}}$ is the unitary operation of the beam splitter, and $\psi_{\text{in}}$ is the state of the two incoming photons. If the photons are perfectly identical, quantum interference leads to a complete overlap of their wave packets, causing them to always exit through the same output port. This manifests as a dip in the coincidence rate, indicating maximal photon bunching. However, if the photons differ in any way, their wave packets will not fully overlap, reducing the bunching effect and allowing them to exit through different ports with some probability. HOM visibility is the key indicator of indistinguishability, which is given by
\begin{equation}
    V=\frac{C_\text{max}-C_\text{min}}{C_\text{max}},
\end{equation}
where \(C_\text{max}\) is the coincidence rate when photons are distinguishable, and \(C_\text{min}\) is the coincidence rate when they arrive simultaneously. \(V=1\) indicates the perfect indistinguishability. 

The HOM experiment is widely used to verify the indistinguishability of photon pairs generated from SPDC and other quantum light sources. Since photon indistinguishability is crucial for quantum computing and communication, the HOM interference test serves as a fundamental tool in characterizing quantum optical systems. Beyond the widely used HBT and HOM tests, additional characterization methods exist. For example, photon emission spectroscopy analyses spectral purity, coherence, and linewidth, offering insights into the quality of the light source.

In general, three key parameters define SPEs: purity, which ensures only one photon is produced at a time; indistinguishability, which verifies that emitted photons share the same mode; and efficiency/brightness, which accounts for vacuum components that reduce overall emission efficiency \cite{senellart2017high, thomas2021race, couteau2023applications1, gupta2023single, esmann2024solid}. Depending on the application, additional parameters may need to be optimised.

\subsection{Single photons generation}
The most common way to classify SPEs is by their material composition. However, this approach may be less intuitive for those unfamiliar with the field and does not clearly highlight functional differences among SPEs. Instead, we categorize them based on their generation principles: quantum emitter transitions and nonlinear optical processes. The mechanisms behind each method are discussed in the following sections.

\subsubsection{Quantum Emitter Transitions}
The most intuitive and direct way to generate single photons is through quantum emitter transitions. This method relies on transitions between discrete energy levels, a fundamental concept in quantum mechanics.

\begin{figure}[htbp]
    \centering
    \includegraphics[width=0.6\linewidth]{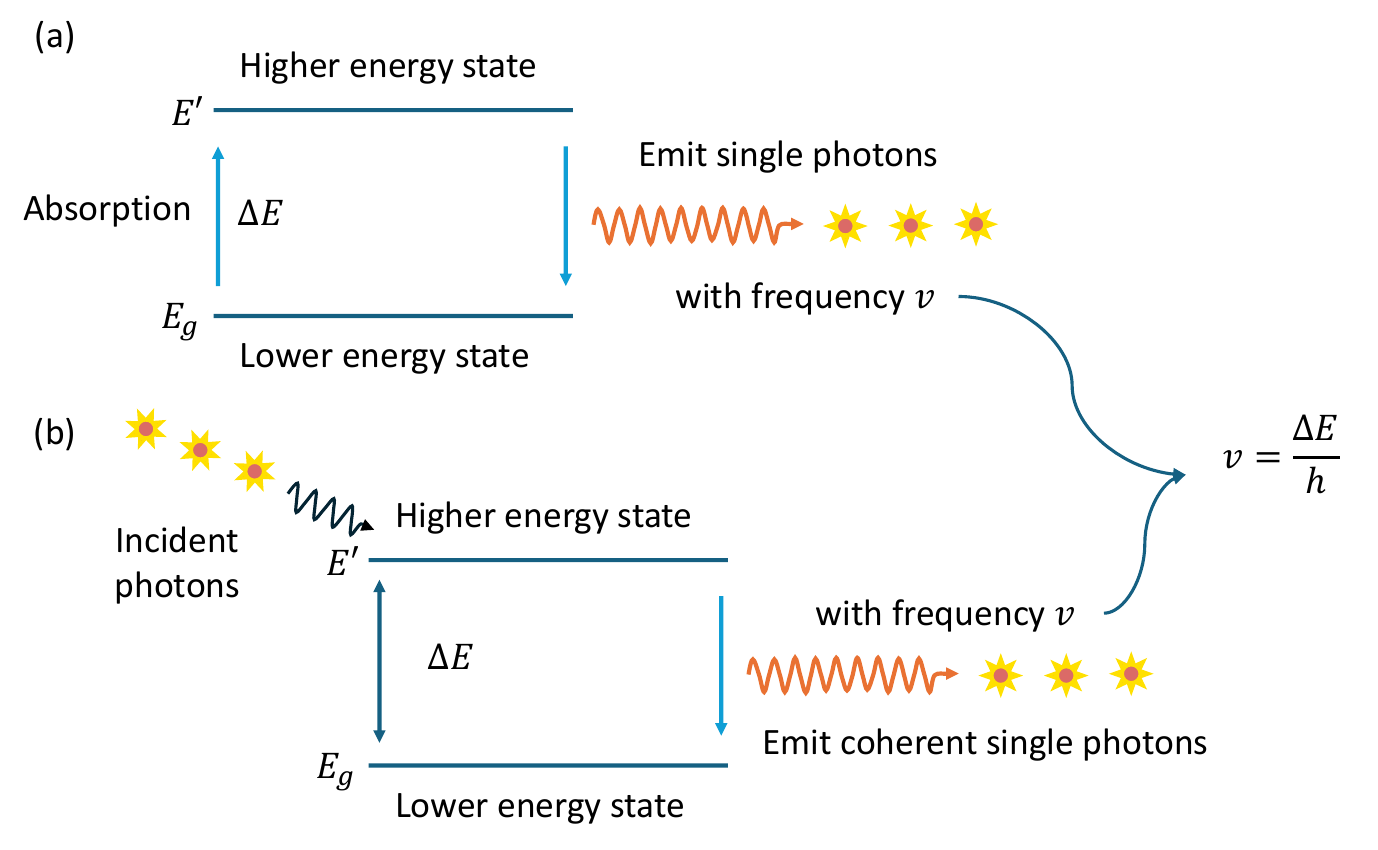}
    \caption{Mechanism of quantum emitters: (a) Spontaneous Emission: It is initially excited to a higher energy state, naturally decays to its ground state without external influence, emitting a single photon in the process. This emission occurs randomly in time and direction. The energy of the emitted photon corresponds to the energy difference between the two states, defined by \(\nu = \frac{\Delta E}{h}\). (b) Stimulated Emission: An incident photon of matching energy interacts with an excited quantum emitter, inducing it to decay to a lower energy state. The result is the emission of a second photon that is coherent with the incident photon, sharing the same phase, frequency, and direction. Although fundamental in laser physics, stimulated emission is less common in SPEs compared to spontaneous emission.}
    \label{fig:quantum emitter}
\end{figure}

A quantum emitter initially resides in its ground state. When it absorbs energy from an external source, it transitions to an excited state (cf. Fig. \ref{fig:quantum emitter}). Since excited states are inherently unstable, the system eventually decays to a lower energy state, releasing energy in the form of a photon. This relaxation occurs through two mechanisms: spontaneous emission and stimulated emission.

\begin{itemize}
    \item Spontaneous emission occurs naturally when an emitter releases excess energy as a photon without external influence.
    \item Stimulated emission is an induced process in which an incoming photon of the right energy prompts the excited emitter to release a second photon that is coherent with the incident one.
\end{itemize}

For SPEs, spontaneous emission is typically the dominant process \cite{senellart2017high, muller2018quantum, esmann2024solid, gupta2023single, arakawa2020progress}. The emitted photon’s frequency is determined by the energy difference between states, given by \(\nu = \frac{\Delta E}{h}\), where \(h\) is the Plank constant. The key feature of this method is that each decay event results in the emission of a single photon, ensuring true single photon operation.

Many widely used SPEs follow this principle, including single atoms, single ions, single molecules, QDs, and colour centres. The specific mechanisms and design considerations for these emitters are discussed in the following sections.

\subparagraph{Single Atoms.}
In trapped atom-based SPEs, neutral atoms with a $\Lambda$-type level scheme, such as Rb, Cs, and Na, are preferred due to their suitability for strong coupling regimes. The $\Lambda$-type configuration consists of three energy levels: an excited state $|e\rangle$, a metastable ground state $|u\rangle$, and a lower ground state $|g\rangle$. Two-photon Raman transitions occur between the ground states via the excited state, enabling efficient photon generation and coherent control of atomic states \cite{dideriksen2021room, higginbottom2016pure, doherty2000trapping, mckeever2004deterministic, vitanov2017stimulated}.

The process begins with atom trapping and cooling using a magneto-optical trap (MOT), which combines laser cooling and magnetic fields. The laser cooling helps to reduce the thermal motion of the atoms, lowering their kinetic energy to trap them, and the magnetic fields can spatially confine the atoms \cite{hijlkema2007single}.  Once the atoms are sufficiently cooled and localized, the MOT is turned off, allowing them to fall freely under gravity. As they pass through a high-finesse optical cavity, an optical trap is activated to capture a single atom inside the cavity. This cavity, designed with highly reflective mirrors, allows light to bounce back and forth multiple times, significantly enhancing the interaction between the trapped atom and the cavity field.

\begin{figure}[htbp]
    \centering
    \includegraphics[width=0.8\linewidth]{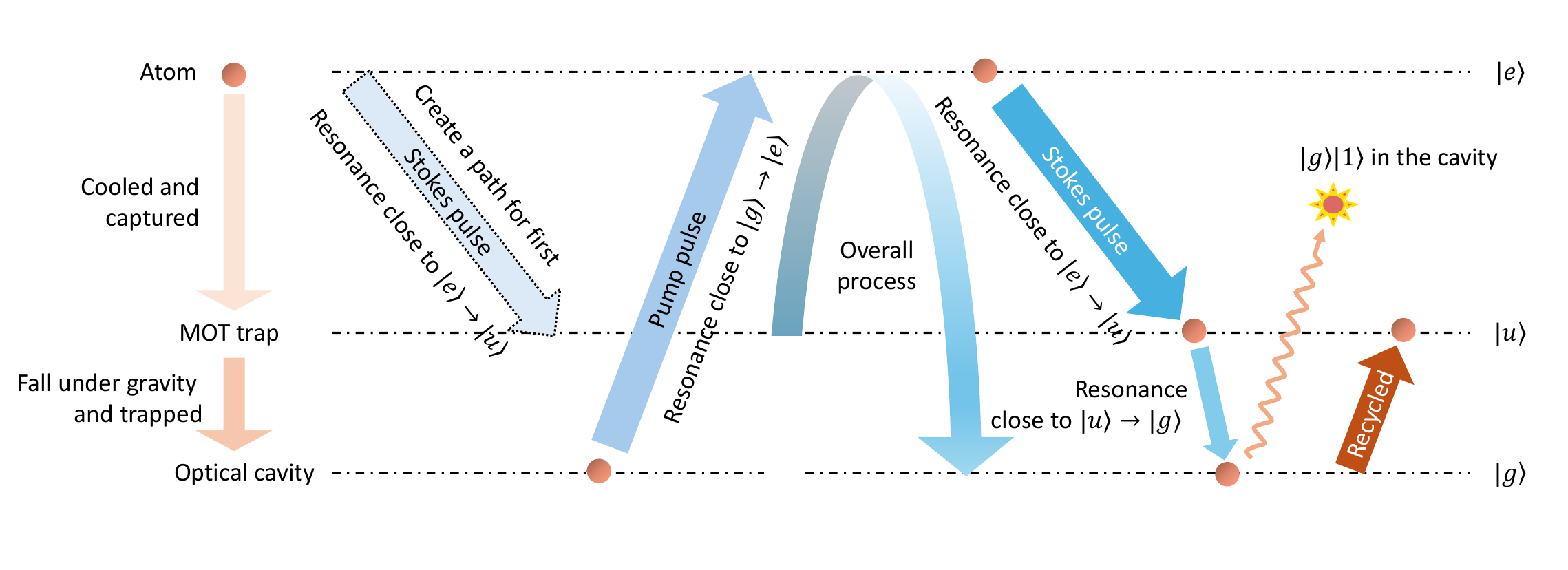}
    \caption{Single photon Generation Using STIRAP Process. Atoms are first captured and cooled using an MOT, which utilizes laser cooling and magnetic fields to reduce thermal motion and spatially confine the atoms. Once cooled, the MOT is turned off, and the atoms fall under gravity into a high-finesse optical cavity. Here, an optical dipole trap holds a single atom in place. The energy transition is driven by STIRAP, employing two laser pulses: the Stokes pulse (coupling \(|e\rangle \longrightarrow |u\rangle\)) and the pump pulse (coupling \(|g\rangle \longrightarrow |e\rangle\)). The Stokes pulse is applied first, creating a pathway, followed by the pump pulse that adiabatically transfers the population to \(|u\rangle\)while avoiding  \(|e\rangle\). The atom, now in \(|u\rangle\), undergoes a \(|u\rangle \longrightarrow |g\rangle\) transition, emitting a photon into the cavity mode enhanced by cQED, resulting in the final state \(|g\rangle |1\rangle\).}
    \label{fig:STIRAP}
\end{figure}

The energy transition during the above process is called the stimulated Raman adiabatic passage (STIRAP), as shown in Fig. \ref{fig:STIRAP}. The key idea behind STIRAP is to achieve coherent population transfer between two states \(|g\rangle \) and \(|u\rangle \) while avoiding significant population in the intermediate state \(|e\rangle \). This is done through a counterintuitive pulse sequence, meaning the Stokes pulse (coupling \(|e\rangle \longrightarrow |u\rangle\)) is applied before the pump pulse (coupling \(|g\rangle \longrightarrow |u\rangle\)). Before the pump pulse turns on, the Stokes laser creates a "path" between \(|e\rangle \) and \(|u\rangle \). However, since the pump pulse is still weak or off, there is no direct excitation from \(|g\rangle\). As the pump pulse turns on gradually, the system starts to transition from \(|g\rangle \) to \(|e\rangle \). Since the Stokes field is already present, the system immediately follows a dark state, which is a superposition of  \(|g\rangle \)and  \(|u\rangle \), bypassing \(|e\rangle \). As the pump pulse reaches its peak, and then the Stokes pulse gradually turns off, the population is fully transferred to \(|u\rangle\). The atom is now inside a high-finesse optical cavity, which enhances the interaction between the atom and the cavity mode. The \(|u\rangle \longrightarrow |g\rangle\) transition is coupled to the cavity mode rather than free-space radiation. Because the cavity is designed to be resonant with the \(|u\rangle \longrightarrow |g\rangle\) transition, the atom prefers to emit its photon into the cavity mode rather than in a random direction. This process is known as cQED, where the interaction between the atom and cavity field dictates the emission properties. Once the atom decays to \(|g\rangle\), it emits exactly one photon into the cavity mode. The final system state is \(|g\rangle |1\rangle\) \cite{mckeever2004deterministic, eisaman2011invited, hijlkema2007single}. This means the atom is in the ground state, and one photon is stored in the cavity mode. The photon can then escape the cavity through a partially reflective mirror, where it can be detected or used for quantum applications. The escape rate is controlled to ensure that only one well-defined photon is emitted at a time.

Obviously, this process avoids the spontaneous emission from \(|e\rangle \). If the atom was to transition directly from \(|e\rangle \longrightarrow |g\rangle\) via spontaneous emission, it could reduce the efficiency and indistinguishability of the single photon source. The laser fields are carefully controlled such that the atom follows a non-adiabatic evolution, avoiding spontaneous decay to lower states. This method allows for coherent transfer between the ground states, leading to the emission of a photon in a highly controlled manner. J. McKeever \textit{et al.} have proved that the efficiency of this method can be even close to unity \cite{mckeever2004deterministic}. 

However, this requires advanced experimental setups and meticulous control. The maintenance of coherence during the STIRAP process is sensitive to noise, thermal fluctuations, and imperfections in the alignment of the laser and the cavity \cite{mckeever2004deterministic, thomas2022efficient, higginbottom2016pure, dideriksen2021room}. In addition, single atom-based emitters face issues about low efficiency and decoherence. Their performance is significantly compromised by the trapping time, fluctuating coupling, and intended emission caused by thermal population. For scalable applications, these issues should first be resolved.

\subparagraph{Quantum Dots (QDs)}
\begin{figure*}[htbp]
    \centering
    \includegraphics[width=1.0\linewidth]{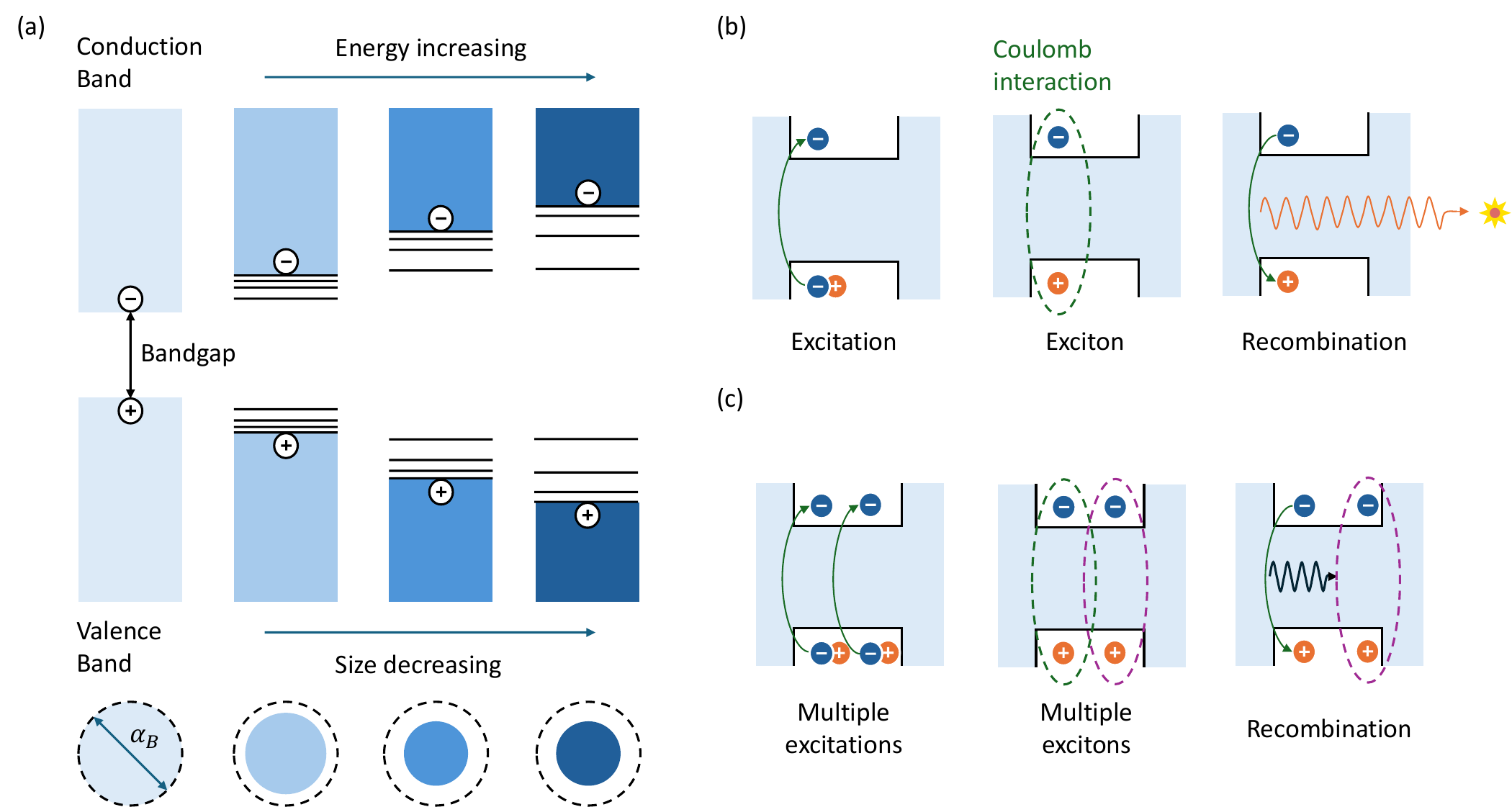}
    \caption{QDs confinement and emission mechanisms (a) Electron-hole pairs in QDs and their size-dependent optical and electrical properties. The strong Coulomb attraction between the hole and electron will make them bonded as a quasi-particle. As the size of a QD decreases, its bandgap inversely increases. When the size of the semiconductor crystal is smaller the exciton Bohr radius, it is considered as a quantum dots \cite{garcia2021semiconductor}. Reproduced with permission from Science \textbf{373}, eaaz8541 (2021). Copyright 2021 The American Association for the Advancement of Science. (b) The radiative decay of QDs. The electron is first excited to the conduction band, and a hole is formed at the valence band. The exciton they formed will then lose energy gradually in the form of phonons. Finally, the electron and hole will recombine and emit a photon. (c) The non-radiative (Auger) decay of QDs. In the recombination state, instead being in the form of photons, the energy is transferred to another charge carrier, hindering single photon emission.}
    \label{fig:QDs}
\end{figure*}

In addition to natural atoms, QDs, often referred to as “artificial atoms”, play a crucial role in single photon emission. They are nanoscale semiconductor particles or nanocrystals (NCs) that exhibit remarkable optical properties due to quantum confinement from all three spatial directions. When a semiconductor’s dimensions are reduced to a size where electronic states become discrete rather than continuous, electrons and holes are confined within a small volume, akin to particles in a box. This results in quantised energy levels instead of the continuous energy bands seen in bulk semiconductors \cite{alivisatos1988electronic}. A key parameter in this behaviour is the exciton Bohr radius—the characteristic size of an electron-hole pair in a bulk material. When a QD’s size is smaller than this radius, strong quantum confinement occurs, leading to discrete energy levels and modified optical properties (cf. Fig. \ref{fig:QDs}(a)).

In a QD, electrons excited across the bandgap experience strong Coulomb attraction and spin-exchange coupling with the holes left in the valence band. This interaction forms excitons, electron-hole pairs that give rise to excitonic many-body states, as illustrated in Fig. \ref{fig:QDs}. QD sizes typically range from 2 to 10 nanometres in diameter, and at this scale, excitons are confined in all three spatial dimensions. This confinement results in discrete, atom-like electronic structures, where the energy levels depend on the QD’s size \cite{alivisatos1988electronic}.

\begin{figure*}[htbp]
    \centering
    \includegraphics[width=1\linewidth]{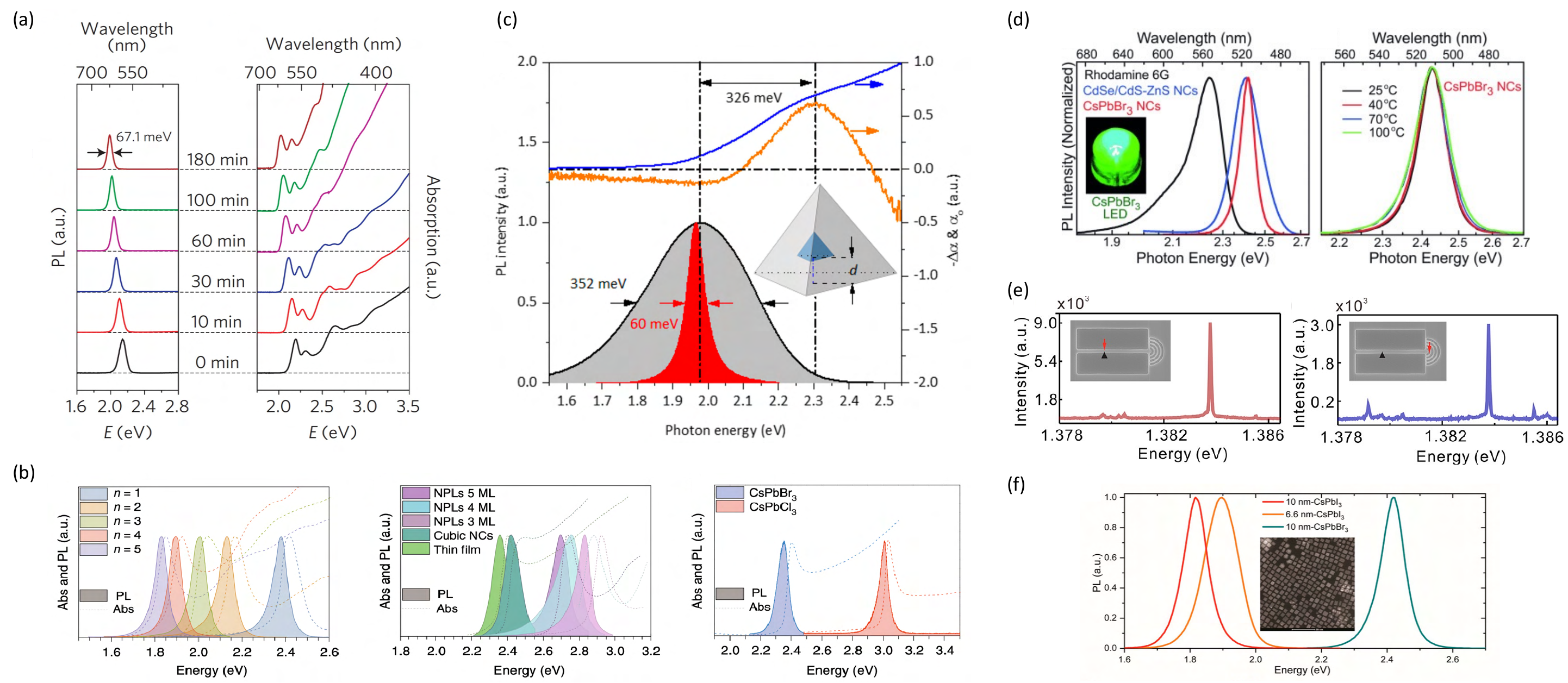}
    \caption{Narrow and tunable emission spectra of various types of QDs. (a) PL spectra of ensemble CdSe-CdS core-shell QDs, showing different temporal evolution. The ensemble PL peaks exhibit full-width at half-maximum (FWHM) values as narrow as 67.1 meV ($\sim$20 nm) \cite{chen2013compact}. Reproduced with permission from Nat. Mater. \textbf{12}, 445 (2013). Copyright 2013 Springer Nature Limited. (b) Optical absorption and PL spectra of halide perovskite materials with varying composition: 2D exfoliated perovskite crystals of the form $(\text{CH}_3{(\text{CH}_2)}_3\text{NH}_3)_2(\text{CH}_3\text{NH}_3)_{n–1}\text{Pb}_n\text{I}_{3n+1}$, as a function of n. The emission remains narrow across different compositions and structures. NPLs: nanoplatelets; NCs: nanocrystals; ML: monolayer \cite{su2021perovskite}. Reproduced with permission from Nat. Mater. \textbf{20}, 1315 (2021). Copyright 2021 Springer Nature Limited. (c) PL spectra of single $\text{CuInS}_2$ QDs with thick ZnS shells. The red curve represents single-dot emission, while the gray shading denotes the ensemble. By adding a thick ZnS shell, Zang \textit{et al.} achieved single-dot PL linewidths as narrow as 60 meV, significantly reduced compared to ensemble values \cite{zang2017thick}. Reproduced with permission from Nano Lett. \textbf{17}, 1787 (2017). Copyright 2017 American Chemical Society. (d) PL linewidth comparison among Rhodamine 6G, CdSe/CdS-ZnS core/hybrid-shell QDs, and $\text{CsPbBr}_3$ NCs. All demonstrate narrow emission bandwidths. The temperature-dependent PL spectra of colloidal $\text{CsPbBr}_3$ QDs are also shown \cite{swarnkar2015colloidal}.  Reproduced with permission from Angew. Chem. Int. Ed. \textbf{54}, 15424 (2015). Copyright 2015 WILEY-VCH. (e) \(\mu\)-PL spectra of QDs embedded in the GaAs waveguide. Upper panel: on top of the source position. Lower panel: on top of the grating coupler. The red markers indicate the light collection position, and the black ones represent the laser excitation position \cite{tao2020chip}. Reproduced with permission from ACS Photonics \textbf{7}, 2682 (2020). Copyright 2020 American Chemical Society. (f) PL spectra at RT from ensembles of 10 nm $\text{CsPbBr}_3$ QDs in toluene (centred at 2.419 eV), 6.6 nm $\text{CsPbBr}_3$ in hexane (1.895 eV), and 10 nm $\text{CsPbI}_3$ in toluene (1.816 eV) \cite{zhu2022room}. C. Zhu et al., Nano Lett. \textbf{22}, 3751 (2022); licensed under a Creative Commons Attribution-NonCommercial-NoDerivatives 4.0 International (CC BY-NC-ND) license.}
    \label{fig:QD emission}
\end{figure*}

More specifically, the optical and electronic properties of QDs, including their absorption and emission wavelengths, are highly dependent on their size. Smaller QDs have a larger effective bandgap and emit light at shorter wavelengths, leading to a blue shift in their absorption and emission spectra. Conversely, larger QDs have a smaller bandgap, resulting in a red shift and emission at longer wavelengths \cite{klimov2014multicarrier, alivisatos1988electronic}. In addition to this tunability, QDs exhibit narrow emission spectra and high colour purity due to their discrete energy levels (cf. Fig. \ref{fig:QD emission}).

Unlike traditional bulk semiconductors, where optical properties are primarily determined by material composition, crystal structure, and impurities, QDs offer an extra degree of tunability through size control. This makes them highly attractive for scalable applications, including SPEs.

\fifthlevel{General Physical Mechanisms of QDs}

Single photon emission in QDs relies on radiative decay (recombination) (cf. Fig. \ref{fig:QDs}(b)). QDs can be excited either optically or electrically, with the optical process typically occurring in three stages: excitation, relaxation, and recombination. Upon absorbing a photon, an electron is excited to the conduction band, leaving a hole in the valence band. The electron-hole pair, known as an exciton, remains bound and undergoes relaxation, where it loses excess energy, primarily through phonon emission, until it settles into a lower energy state. Finally, the electron recombines with its corresponding hole, releasing energy. For SPE applications, this released energy takes the form of a single photon (radiative decay), whereas non-radiative decay processes dissipate the energy without photon emission.

Just like other SPEs, the fundamental requirement for QD-based SPEs is that each excitation event must result in the emission of exactly one photon. Ideally, only one exciton should be present in the QD at any given time. Once the exciton recombines, it should emit a single photon before the QD is re-excited. The exciton lifetime, defined by the radiative decay rate, is crucial in this process. Typically, the radiative lifetime is on the order of 1 nanosecond or less \cite{huang2024scalable, almeida2023size, you2025developing}. If the recombination rate is too slow, the exciton persists longer, increasing the likelihood that the QD will absorb another photon before the first exciton recombines. This can lead to biexciton formation, where multiple excitons exist simultaneously, resulting in the emission of more than one photon and compromising the single photon purity.

Excitons in QDs can also decay through the non-radiative process of Auger recombination (AR) (cf. Fig. \ref{fig:QDs}(c)). In AR, when an electron-hole pair recombines, its energy is transferred to a third carrier (either an electron or a hole), exciting it to a higher energy state instead of emitting a photon \cite{klimov2000optical, klimov2000quantization}. This process is particularly significant in QDs due to quantum confinement, which enhances Coulombic interactions between carriers and results in extremely short lifetimes (on the order of picoseconds) for multi-exciton states such as biexcitons \cite{klimov2000quantization, vaxenburg2016biexciton}. Consequently, AR leads to non-radiative energy dissipation, reducing the photoluminescence quantum yields (PLQY), defined as:
\begin{equation}
    \text{PLQY}=\frac{\text{Number of emitted photons}}{\text{Number of absorbed photons}}=\frac{k_{\text{rad}}}{k_{\text{rad}}+k_{\text{nrad}}},
\end{equation}
where \(k_{\text{rad}}\) and \(k_{\text{nrad}}\) are the radiative and non-radiative decay rates, respectively \cite{vaxenburg2016biexciton, leblanc2013enhancement, galland2011two}. 

\begin{figure}[htbp]
    \centering
    \includegraphics[width=0.7\linewidth]{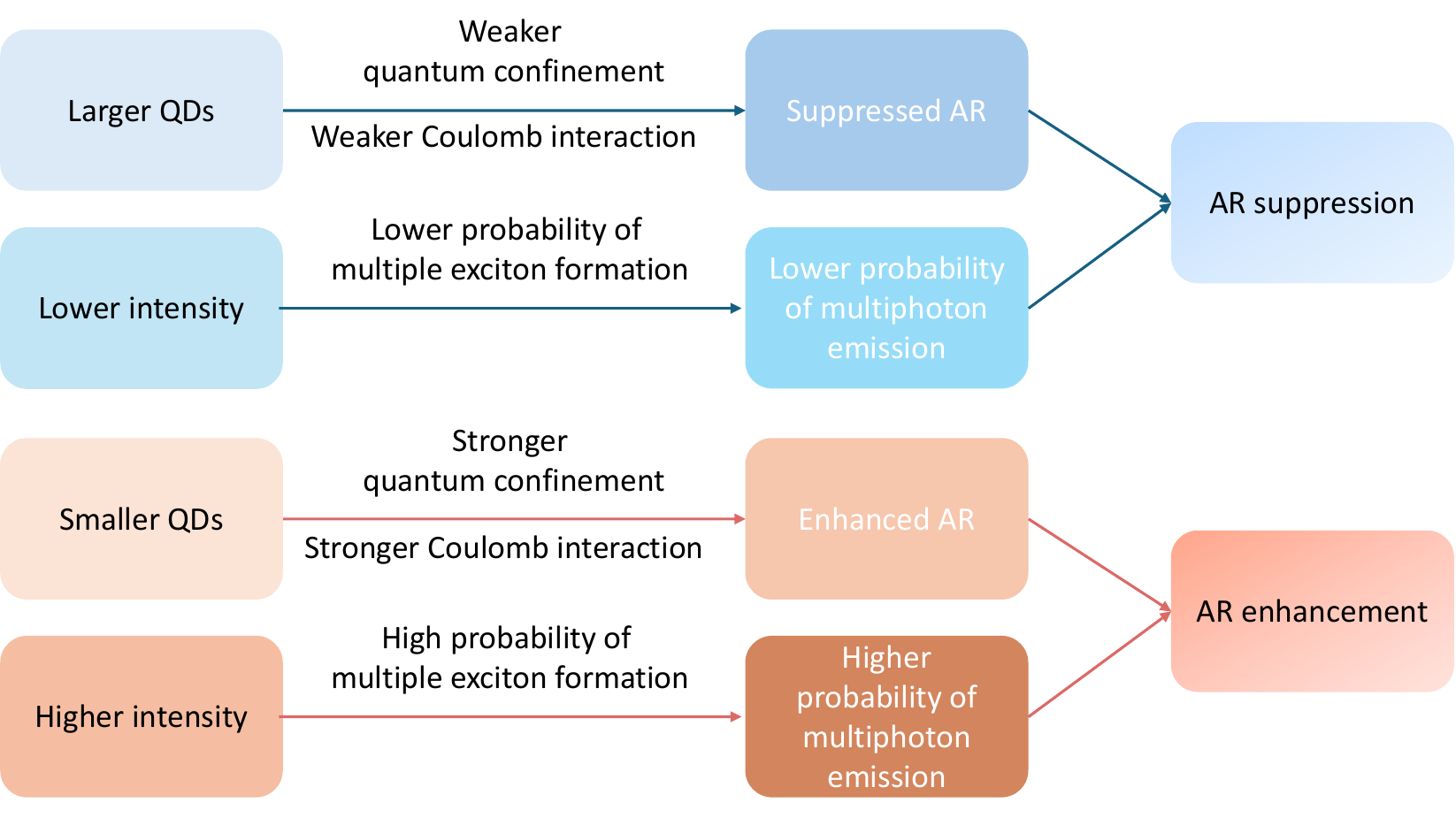}
    \caption{Schematic representation of simplified preference of AR in different scenarios: (a) In low-intensity and bigger QDs. First, for bigger QDs, the quantum confinement is weaker, leading to weaker AR. Second, in the low intensity scheme, only one electron is excited normally. The probability of multiple-exciton formation is relatively low. Aimed at sufficient efficiency, AR should be suppressed. (b) In high-intensity and smaller QDs. First, for smaller QDs, the the quantum confinement is stronger, leading to stronger AR. Second, due to high intensity, the probability of the existence of multiple excitons is higher. AR can effectively transferred the energy to another charge carrier, effectively suppressing multiple photon emission.}
    \label{fig:AR preference}
\end{figure}

However, AR does not always need to be suppressed; its impact hugely depends on both the excitation intensity and the size of the QDs. The Auger rate increases as the QD size decreases, following an inverse square relationship with the NCs radius ($1/\tau_{\text{Auger}} \propto R^2$). Smaller QDs typically have a larger bandgap and a higher likelihood of biexciton formation, which enhances AR. Interestingly, enhanced AR in smaller QDs can actually suppress biexciton emission. Experimental studies have shown that AR can occur at rates several orders of magnitude faster than radiative recombination, particularly in smaller QDs \cite{klimov2000quantization, cragg2010suppression, efros2016origin, neogi2005size}. The accelerated AR rate significantly shortens the biexciton lifetime, often reducing it to just a few picoseconds\cite{zhu2016screening, trinh2018organic, park2015room, zhu2022room}. For instance, in Cs-Pb-halide quantum dots, biexcitons decay so quickly through AR that radiative recombination cannot occur in time \cite{zhu2022room}. This process, known as Auger blockade, prevents the emission of multiple photons by ensuring that biexcitons (XX) rapidly annihilate into single excitons (X), which then emit a single photon even under high-intensity excitation \cite{makarov2016spectral, klimov2000quantization, chandrasekaran2017nearly, neogi2005exciton, neogi2005near}. This creates a delicate balance, where the AR rate must be carefully optimised depending on the operational context (See Fig. \ref{fig:AR preference}).

By contrast, in low-intensity regimes and larger QDs, AR should be suppressed to preserve PLQY, as discussed earlier. Ideally, biexciton AR should be accelerated while single-exciton AR is minimized. In practice, however, strong AR in small single-dot QDs, particularly in colloidal systems, can compromise photostability and lead to blinking.

Blinking is a phenomenon where individual QDs randomly switch between bright neutral (“ON”) and dark charged (“OFF”) states under continuous excitation. It arises from stochastic PL intensity fluctuations due to nonradiative recombination pathways activated by defects or environmental factors. Galland \textit{et al.} classified blinking mechanisms into two types: : type A, driven by conventional charging/discharging of the NC core, and type B, associated with hot-electron trapping \cite{galland2011two, nirmal1996fluorescence, efros2016origin, ahmed2019mechanistic}. 

\begin{figure*}[htbp]
    \centering
    \includegraphics[width=0.9\linewidth]{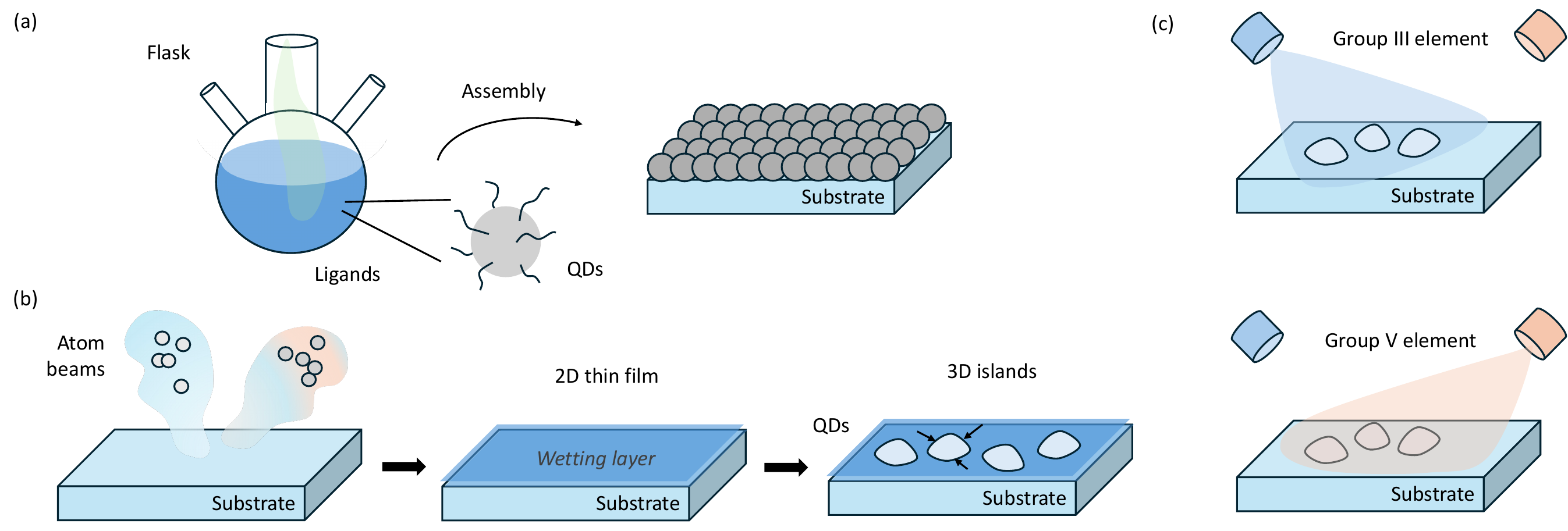}
    \caption{Three different fabrication mechanisms for QDs. (a) Solution based colloidal synthesis method for cQDs. Adapted from \cite{garcia2021semiconductor}. Reproduced with permission from Science \textbf{373}, eaaz8541 (2021). Copyright 2021 The American Association for the Advancement of Science. (b), (c) Epitaxial growth methods for eQDs. (b) SK growth mode. (c) DE growth mode.}
    \label{fig:Fab of QDs}
\end{figure*}

Importantly, blinking is not universal; its presence or absence is intrinsically tied to how the QDs are fabricated. Based on fabrication, QDs are generally grouped into epitaxial QDs (eQDs) and colloidal QDs (cQDs). eQDs are highly stable and essentially blinking-free, whereas cQDs are notorious for strong blinking. To understand the physical origins of blinking, and how it affects emission mechanisms, photostability, operating temperature, and scalability, the fabrication processes must be examined in detail (cf. Fig. \ref{fig:Fab of QDs}) \cite{garcia2021semiconductor, skiba2017universal, keizer2010atomic, strobel2023unipolar}.

\fifthlevel{Colloidal QDs (cQDs)}
cQDs are tiny NCs synthesized in a chemical solution, referring to Fig. \ref{fig:Fab of QDs}(a). The most common and precise fabrication route is a bottom-up chemical process named colloidal synthesis, typically via the hot-injection method. A high-boiling-point solvent and special stabilizing molecules called ligands are added to a reaction flask, and one chemical precursor is dissolved in this mixture. The mixture is heated to a very high temperature, 150 to 300°C, under an inert atmosphere (like nitrogen) to prevent unwanted side reactions. A second precursor is then rapidly injected into the hot flask, which is essentially the hot injection step. This sudden injection causes the chemical concentration to become supersaturated, triggering a "burst" of nucleation. In seconds, millions of tiny seeds of the crystal form all at once. The temperature is then lowered slightly to a growth regime. At this stage, no new seeds form, but the existing ones slowly grow by adding more precursor atoms from the solution. The reaction is allowed to continue until the QDs reach the desired size. The reaction is finally quenched, yielding a colloidal suspension of ligand-stabilized NCs with narrow size dispersion.

It is obvious that this rapid process is excellent for controlling the size of the QD core, but it's terrible for creating a perfect crystal. The NC is formed with an incomplete atomic surface, due to the sudden quenching. Atoms on the surface lack the bonds they would have in a perfect, bulk crystal and are exposed to vacuum or solutions. These atoms are still electronically unsatisfied and highly reactive to form bond, which are known as dangling bonds. To reduce dangling bonds, ligands added before hot injection help to passivate them by bonding to surface atoms, as well as preventing the formation of clumps. However, the bulky ligands can't physically cover every single defect site on the QD's complex, faceted surface. What's worse is that they are not permanently glued on. They are in a constant state of "popping off" (desorbing) and "sticking back on" (adsorbing). Therefore, a charge carrier can be easily caught in a surface trap that is failed to be passivated by ligands. This is the reason that induces both type A and type B blinking. The key difference between these two blinking mechanisms is time that a charge is trapped in defects. For type A, the charge is captured relatively long (milliseconds or even seconds), leaving the core electrically charged. While for type B, the charge will be immediately released instead of being held, so the QD remains neutral throughout the process. 

Type B blinking now can be well mitigated by core-shell structure. However, there is limited success in suppressing A-type blinking \cite{galland2011two, mi2025towards}. If the QD core is charged, it will lead to fully OFF state and type A blinking begins. This process is described as the defect-mediated AR, which is discussed in detail in the third section. As the size of QDs get smaller, the surface to volume ratio and AR rate increase, which exacerbates the blinking behaviour \cite{hines2014recent, giansante2018surface}. 

In short, cQDs are essentially microscopic crystal fragments floating in a liquid or embedded in a polymer. Because they are so small, a huge fraction of their atoms are on the surface. This imperfect, dangling-bond-filled surface is the root cause of blinking. 

\fifthlevel{Epitaxial QDs (eQDs)}
In comparison, eQDs are instead grown inside of a larger, highly-ordered semiconductor crystal. eQDs are typically III-V semiconductor structures and some of oxide perovskite. The atoms are delivered via vapour deposition process. The two main physical growth mechanisms are Stranski-Krastanov (SK) and Droplet Epitaxy (DE). 

Between them, SK growth is the more dominant and commercially established one for forming self-assembled eQDs. It is completely driven by strain energy, which requires two materials with a significant lattice mismatch (different atomic spacing or lattice constant). The growth begins in a 2D, layer-by-layer fashion, as illustrated in Fig. \ref{fig:Fab of QDs}(b). The deposited material is forced to adopt the lattice spacing of the substrate, causing it to become compressively strained. For instance, when depositing InAs onto GaAs, the InAs atoms are forced to stretch and match the GaAs lattice. As more material is deposited, strain energy accumulates in this thin, strained layer, known as the wetting layer (WL). After a critical thickness is reached, it becomes high-enregy and unstable, like a rubber stretched too tight. It becomes energetically favourable for the system to relieve this strain by transitioning from 2D growth to the formation of 3D islands, which are the QDs. These islands are coherent (not dislocated) and are still strained, but the 3D geometry allows for more effective strain relaxation \cite{baskaran2012mechanisms, eaglesham1990dislocation}.

DE growth mode, however, is a more flexible mode that is not independent on lattice strain, applicable for lattice-matched materials. It is a two-step process, driven by surface energy and chemical reaction kinetics (cf. Fig. \ref{fig:Fab of QDs}(c)). First, a beam of only the group-III element, the metal atoms, are directed at the substrate. With no group-V element present to react with, the group-III atoms migrate on the surface and coalesce into nanoscale liquid metal droplets. Their density is primarily controlled by the substrate temperature. In the second step, the substrate is exposed to a flux of the group-V element, which controls the size of QDs. They react with the liquid metal droplets, crystallizing them into semiconductor nanostructures—the QDs \cite{wu2014droplet, keizer2010atomic, gurioli2019droplet}.  Obviously, both of them require a precise control over atom beams. To realise this atom-level control, Metal-Organic Chemical Vapour Deposition (MOCVD) and MBE are introduced, one chemical and one physical respectively. For detailed explanation of MBE and MOCVD, we direct readers to \cite{hidayat2024applications,sogabe2023high,petroff1994mbe}. Comparing these two growth modes, the biggest difference lies in the presence of the WL and the interlinked density \& size properties. WL hugely affects excitation and relaxation dynamics of QDs. For SK grown QDs, carriers are generated first in the WL, then they can move along in it until they are captured by QDs. Once inside a QD, carriers become localized again. The WL dramatically accelerates the carrier capture rate, the process of getting electrons and holes into the dot in the first place. It serves like a large reservoir that stores excessive charge carriers and provides them for the upper dots. However, this strain based approach limits the variability of QDs—size-density dependence. Higher temperature increases the mobility of atoms. They move so fast that they cannot stick to each other and are more likely to join other already-formed islands. This results in larger and less dense QDs on the WL. Whereas, when temperature is low, the atoms can't travel far. They quickly find their nearest neighbours and form many new, tiny islands. The nucleation rate is very high, resulting in high-density of small dots. 

In terms of blinking, specifically, eQDs do not present intrinsic blinking behaviour. Both of SK grown and DE grown QDs are buried into a semiconductor matrix. They always have a capping/barrier layer on top of the islands (QD assemble). This overgrowth passivates everything, by reconstructing the crystal bonds and eliminating surface states. Although the WL does provide an escape route, making QDs susceptible to these charging/discharging events, it is has been effectively resolved \cite{hu2015defect, liu2018single, wang2009non, schimpf2021entanglement}. Furthermore, this semiconductor matrix structure enables QDs to emit at longer wavelengths without losing confinement, not only relying on small physical size. It serves as a 3D potential well around the QD. For example, in terms of InAs/GaAs QDs, the valence band of GaAs is lower than that of InAs, while the conduction band of GaAs is higher than InAs. This confines excitons within InAs, also known as the core. Therefore, eQDs can be fabricated in much larger sizes than cQDs.

To enhance radiative decay and spatial confinement, QDs are often coupled to high-quality optical cavities at the microscale or nanoscale, including micropillars, bullseye, and open cavities \cite{peinke2021tailoring, xia2021enhanced, ding2025high}. Proper cavity design can be tailored to match the wavelength and polarisation of the QD, enhancing the Purcell effect, which significantly increases the radiative decay rate while suppressing non-radiative processes. By confining light to a small region, the cavity strengthens the local EM field at the QD’s position, improving photon collection efficiency. Additionally, it helps direct emitted photons into a specific spatial mode, improving their coherence and usability in quantum applications.

Compared to other SPEs such as SPDC, QD-based emitters can produce brighter and more efficient photons. Their discrete energy levels result in narrow emission spectra, leading to sharp, well-defined emission lines. Fig. \ref{fig:QD emission} illustrates different emission spectra of both ensemble ad single-dot QDs, which are all under 70 meV. Furthermore, QDs can now operate across a wide temperature range, from cryogenic to room temperature, by tuning their bandgap \cite{raino2022ultra, gosain2022quantitative, fedin2021enhanced, park2015room}.

Despite these advantages, several challenges hinder the scalability of QD-based SPEs. First, while QDs can achieve near-unity PLQY, their overall emission efficiency remains limited due to optical losses and imperfect cavity coupling. Second, inhomogeneous surface patterns introduce variability among QDs, leading to decoherence and reduced indistinguishability. Because it is nearly impossible to fabricate identical QDs, achieving uniform performance is a challenge. Although post-processing techniques can improve indistinguishability, P. Senellart \textit{et al.} demonstrated that increasing indistinguishability often compromises efficiency \cite{senellart2017high}. Additionally, they are not stable, seriously affected by blinking and bleaching. Although some of QDs can work at room temperature (RT), most of them have to work under cryogenic conditions. Lastly, many colloidal QDs contain toxic heavy metals such as Cd, Pb, and Hg, posing environmental and health risks \cite{zhu2022room, fedin2021enhanced, mi2025towards}. Developing non-toxic alternatives is essential for broader adoption. 

\subsubsection{Colour Centres}
Unlike single atoms and QDs, colour centres are not discrete particles but atomic-scale defects within a crystal lattice. These defects, found primarily in wide-bandgap materials like diamond, can act as isolated quantum light emitters. They arise from vacancies, interstitial atoms, or impurity atoms that alter the electronic structure of the host material  \cite{kurtsiefer2000stable, gruber1997scanning}. When excited, colour centres absorb specific wavelengths and re-emit light, often producing visible colouration, which gives them their name.

In a perfect crystal, the periodic potential creates delocalized electronic states, forming continuous energy bands. However, the presence of a defect disrupts this periodicity, introducing a localised perturbation in the potential energy landscape. The defect site behaves like a spatially confined potential well, trapping electrons in discrete energy levels, much like electrons in an atom \cite{brus1983simple}. According to quantum mechanics, an electron confined within such a finite potential well can only occupy specific quantised energy states, with energy gaps determined by the shape and depth of the defect’s potential well. 

When a colour centre is optically or electrically excited, an electron is promoted from its ground state to an excited state associated with the defect. This excitation is typically induced by a laser tuned to the energy difference between these states. Once excited, the electron remains confined to the defect site due to the localised potential well created by the crystal defect \cite{sorman2000silicon, smith2019colour}. Unlike free electrons in a conduction band, its wavefunction is spatially restricted and decays exponentially away from the defect.

The electronic states introduced by the defect lie within the bandgap of the host material. This isolation prevents the electron from merging with the delocalized states of the conduction or valence bands unless sufficient external energy is provided. Consequently, the electron remains trapped at the defect site until it relaxes back to its ground state, emitting a photon in the process. Still, the energy difference between these states determines the wavelength of the emitted photon, producing spectrally sharp emission due to the discrete nature of these transitions.

While the fundamental emission process is similar across different materials, crystal field effects and phonon interactions introduce variations. For example, negatively charged nitrogen-vacancy ($\text{NV}^-$) centres in diamond can be engineered to form a $\Lambda$-type system, similar to single atoms discussed earlier, whereas neutral NV centres ($\text{NV}^0$) cannot \cite{gali2019ab}. The host material’s lattice structure and local strain modify the defect’s electronic states, shifting emission wavelengths, while interactions with lattice vibrations (phonons) influence spectral linewidth and overall emission properties. Some defects also exhibit spin-related optical transitions, which are useful for quantum applications, while others experience Jahn-Teller distortions, leading to energy level splitting \cite{stern2024quantum, stern2022room, bathen2021resolving, bathen2021manipulating}. This paper does not cover the effects of different host materials on colour centres. For a detailed discussion, we refer readers to \cite{bradac2019quantum, redjem2023all, smith2019colour, ou2024silicon}.

Among their primary strengths, defect-based SPEs feature a robust solid-state architecture that enhances durability and simplifies integration into photonic circuits compared to single atoms and QDs \cite{peyskens2019integration, li2021integration}. Defect-based SPEs like hexagonal boron nitride (hBN) operate efficiently at RT, eliminating the need for costly cryogenic setups \cite{su2022tuning, shaik2021optical, li2019size}. In addition, these sources exhibit exceptional photostability, resisting degradation under prolonged optical excitation. Certain defects, such as NV centres, also offer spin-photon interfaces that enable entanglement between photon emission and the defect’s quantum spin state, expanding their utility beyond photon generation to include applications in quantum memory, sensing, and distributed quantum networks \cite{michaelis2022single, samaner2022free, li2023arbitrarily}. Mature nanofabrication techniques further benefit defect-based systems, allowing for precise positioning within nanostructured environments to enhance photon extraction efficiency \cite{rosenberger2019quantum, li2021integration}.

However, these systems face notable limitations. Achieving high brightness and collection efficiency remains challenging, as defect emission is often isotropic or hindered by the host material’s refractive index, which necessitates complex photonic engineering \cite{li2021integration, swarnkar2015colloidal, esmann2024solid, körber2024fluorescence}. Spectral instability, driven by interactions with lattice vibrations or charge fluctuations, can degrade photon indistinguishability. Inhomogeneities in defect properties such as emission wavelength or decay rates complicate large-scale device integration \cite{michaelis2022single}. While RT operation is advantageous, it frequently exacerbates decoherence effects, limiting spin-photon entanglement fidelity compared to cryogenically cooled alternatives \cite{ji2022proposal}. Moreover, scalable manufacturing is hindered by the stochastic nature of defect creation and the difficulty of achieving deterministic placement with nanoscale precision \cite{abidi2019selective, su2022tuning}. These made defect-based SPEs promising but imperfect for scalable quantum technologies. Further research on advanced materials and hybrid photonic designs should be conducted.

\subsubsection{Nonlinear Optical Process}
Unlike quantum emitters, which are deterministic sources, SPEs based on nonlinear optical processes are probabilistic and rely on multi-photon interactions in nonlinear media. Instead of emitting one photon per excitation cycle, these sources generate correlated photon pairs through nonlinear interactions in a crystal. A high-energy photon, typically from a laser, is converted into two lower-energy photons within a nonlinear medium.

In this process, the detection of one photon in the pair serves as an indicator of the other’s existence. These are referred to as the heralding photon and heralded photon, respectively, and such sources are known as heralded SPEs. While probabilistic in nature and seemingly not ideal for generating pure single photon states, heralded photons play a crucial role in information-related applications, particularly in quantum communication and cryptography \cite{zhang2021spontaneous, faleo2024entanglement, zhan2023measurement, murakami2023quantum}.

\subsubsection{Spontaneous Parametric Down Conversion (SPDC)}
As early as in 1961, the pair production by SPDC was predicted and then experimentally demonstrated in 1970 \cite{louisell1961quantum, burnham1970observation}. Since then, SPDC has been at the core of quantum optics. Although early efforts have been made on the atomic-cascade scheme, generated photon pairs carry away part of the momentum, making their directions not well-correlated \cite{zhang2015advances, kwiat1999ultrabright}. SPDC is now a predominant heralded way to produce correlated photons, entangled photons, and single photons. \cite{couteau2018spontaneous, kiessler2025spdc, santiago2021entangled, faleo2024entanglement, zhan2023measurement}.

\begin{figure*}[htbp]
    \centering
    \includegraphics[width=1\linewidth]{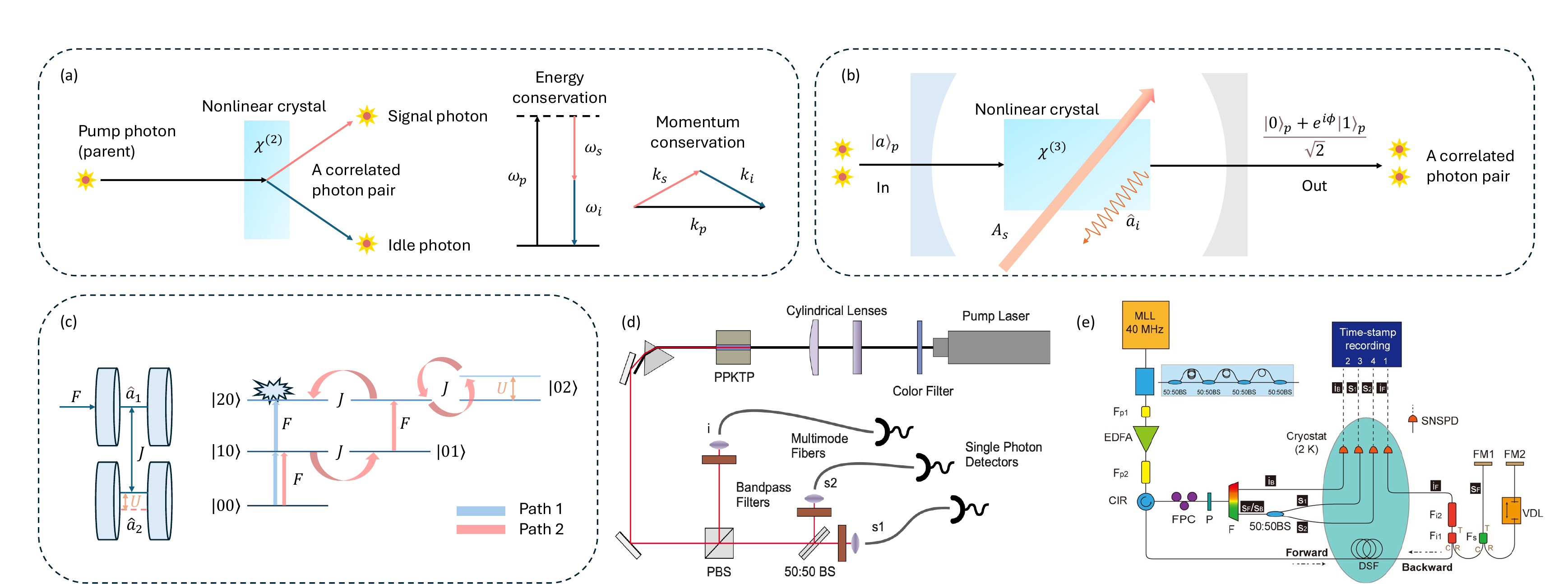}
    \caption{Sources based on nonlinear optical processes. (a) Schematic of the SPDC process. (b) Schematic of the FWM process. (c) Schematic representation of an asymmetrically driven photonic molecule, where each cavity supports a single resonant mode in the spectral region of interest. Only the first cavity is driven by an external coherent field. The inset shows the corresponding energy level ladder, illustrating the photon number states of the two cavities. Adapted from \cite{flayac2015all}. H. Flayac et al., Sci. Rep. \textbf{5}, 11224 (2015); licensed under a Creative Commons Attribution 4.0 International (CC BY 4.0) license. (d) Experimental setup for SPDC-based single-photon generation. A periodically poled potassium titanyl phosphate (PPKTP) crystal generates collinear photon pairs with orthogonal polarisations. A polarising beamsplitter (PBS) separates the idler and signal photons; the idler acts as a herald for the signal photon, which is then directed to a non-polarising beamsplitter (BS) for second-order correlation measurement $g^{(2)}(0)$ \cite{couteau2018spontaneous}. Reproduced with permission from Contemp. Phys. \textbf{59}, 291 (2018). Copyright 2018 Taylor \& Francis. (e) Experimental setup for stimulated FWM-based single-photon generation. Key components: MLL – mode-locked laser; 50:50 BS – 50/50 optical beamsplitter; Fp1, Fp2, Fi1, Fi2, Fs, F – filters implemented via (cascaded) 3-port 100 GHz DWDMs; CIR – optical circulator; FPC – fiber polarization controller; FM – Faraday mirror; VDL – variable optical delay line; P – polarizer; SNSPD – superconducting nanowire single-photon detector. C (T, R) refer to the common, transmission, and reflection ports of the DWDM devices. For detailed system description, refer to \cite{dong2017true}. Reproduced with permission from ACS Photonics \textbf{4}, 746 (2017). Copyright 2017 American Chemical Society.}
    \label{fig:herald}
\end{figure*}
SPDC is a nonlinear optical process in which an incident pump photon splits into two lower-energy photons, referred to as the signal and idler photons. This is essentially the reverse process of second harmonic generation, as illustrated in Fig. \ref{fig:herald}(a) . The process is governed by the interaction Hamiltonian:
\begin{equation}
H_{\text{int}} = \chi^{(2)} \int \int \int d^3 \mathbf{r} \, E_{\text{pump}} (\mathbf{r}, t) a_s^\dagger (\mathbf{r}, t) a_i^\dagger (\mathbf{r}, t) + \text{h.c.}
\end{equation}
where \(\chi^{(2)}\) is the second-order nonlinear susceptibility, \(E_\text{pump}\) is the electric field of the pump photon, and \(a_s^\dagger (\mathbf{r}, t)\), \( a_i^\dagger (\mathbf{r}, t)\) are the creation operators for the signal and idler photons at position \(\mathbf{r}\) and time \(\mathbf{t}\). The term \(\text{h.c.}\) denotes the Hermitian conjugate. 

The second susceptibility \(\chi^{(2)}\) characterises the material’s polarization response to the pump field and determines the probability amplitude for photon-pair generation. A simplified representation of the interaction Hamiltonian is:
\begin{equation}
H_{\text{int}} = \hbar g (a_p a_s^\dagger a_i^\dagger+ a_p^\dagger a_s a_i). 
\label{H_{int}}
\end{equation}
where \(g\) is a coupling constant dependent on \(\chi^{(2)}\), and \(a_p\), \(a_p^\dagger\) are the creation and annihilation operators of the pump photon, respectively \cite{edamatsu2007entangled}. Eq. (\ref{H_{int}}) shows that there is a non-zero possibility for the annihilation of a pump photon, leading to the simultaneous creation of a signal and idler photon. The reverse process, where two photons combine to form a single pump photon, is also theoretically possible but does not occur in a classical SPDC model. 

However, SPDC does obey momentum and energy conservation as in the classical case for coherent generation, leading to the phase-matching constraint:
\begin{equation}
    \begin{split}
           \omega_p=\omega_s+\omega_i,\\
    \mathbf{k_p}=\mathbf{k_s}+\mathbf{k_i},
    \end{split}
\end{equation}
where $\omega_p$, $\omega_s$, and $\omega_i$ are the angular frequencies of the pump, signal, and idler photons, respectively, and $\mathbf{k_p}$, $\mathbf{k_s}$, and $\mathbf{k_i}$ are their respective wave vectors.

This constraint has both advantages and challenges. On the one hand, it ensures that the generated photon pairs are highly directional, which is a significant advantage over many other single photon sources. On the other hand, due to recoil effects, not all wavelength combinations satisfy this condition, making phase matching a critical aspect of SPDC. Achieving phase matching for specific target wavelengths is a key challenge and will be discussed in detail later \cite{zhang2021spontaneous, kwiat1999ultrabright, edamatsu2007entangled}.

Phase matching in SPDC can be classified into different polarization settings. In type 0, all three photons (pump, signal, and idler) share the same polarization. In type I, the generated signal and idler photons have the same polarization but are orthogonal to the pump photon. In type II, the signal and idler photons have perpendicular polarizations. Among these, type I SPDC is most commonly used for single photon emission, as it provides a more predictable emission pattern. A possible quantum state of type I is given by
\begin{equation}
    |\psi\rangle =\frac{1}{\sqrt{2}}(|HH\rangle +e^{i\theta}|VV\rangle)
    \label{type 1}
\end{equation}
The total number of photon pairs generated in SPDC follows Poisson statistics, whereas individual modes (signal or idler) exhibit sub-Poissonian statistics. The entanglement between the signal and idler photons reduces fluctuations in the photon number detected in one mode when conditioned on the detection of the other. This effect is known as a squeezed vacuum, which will be explored further in BSV SPEs section.

Since the wavelengths of the pump (parent) photon and the signal/idler (daughter) photons are fixed by energy conservation, phase matching can only be controlled through the incident angle and the properties of the nonlinear medium. Various techniques exist for material selection and tuning, but they all rely on the same fundamental principle: using materials with direction-dependent refractive indices to satisfy momentum and energy conservation conditions. One of the most effective approaches is the use of birefringent nonlinear crystals \cite{zhang2021spontaneous, couteau2018spontaneous, lopez2022exact}. In these materials, the refractive index \(n\) varies with polarization direction relative to the crystal’s optical axis. The refractive index for light polarized along the extraordinary axis \(n_e\) differs from that along the ordinary axis \(n_o\) \cite{edamatsu2007entangled, kwiat1999ultrabright}.

For type I SPDC, the pump photon typically propagates as an extraordinary wave, while the signal and idler photons propagate as ordinary waves. Phase matching is achieved when the effective refractive index of the pump photon matches the ordinary refractive index of the signal and idler photons:
\begin{equation}
    n_e^\text{eff}(\text{pump})=n_o(\text{signal})=n_o(\text{idler}).
\end{equation}
By carefully selecting the crystal orientation and propagation angle, the refractive indices can be tuned to meet the phase-matching condition. Additionally, temperature adjustments can further optimise phase matching for a given wavelength range.

Since birefringent adjustment depends on the orientation of the crystal’s optical axis, it also influences the effective \(\chi^{(2)}\) value. The \(\chi^{(2)}\) tensor is anisotropic in birefringent materials, meaning that for a given crystal, the interaction between the polarizations of the pump, signal, and idler photons is dictated by the specific orientation of \(\chi^{(2)}\). Even if birefringence enables phase matching by adjusting the refractive indices, the efficiency of SPDC depends on both the magnitude and orientation of \(\chi^{(2)}\). The birefringence must be strong enough to allow phase matching at the desired wavelengths and polarizations, while \(\chi^{(2)}\) must be sufficiently large to ensure efficient photon conversion. Therefore, suitable birefringent materials are chosen based on both their nonlinear strength and their ability to satisfy phase-matching conditions \cite{zhang2021spontaneous}. Two commonly used birefringent materials are beta barium borate (BBO) and potassium titanyl phosphate (KTP). More detailed theoretical derivations can be found in \cite{couteau2018spontaneous, hong1985theory, edamatsu2007entangled, kwiat1999ultrabright}.

Once the SPDC process occurs, the two daughter photons are separated and directed to single photon detectors. The idler photon can either be detected separately or used in entanglement-based experiments. A coincidence counter is employed to identify correlated photon pairs. The system registers a “success” only when both detectors detect a photon within a specific time window, confirming the creation of a photon pair \cite{couteau2018spontaneous}.

Besides birefringent materials, periodically poled crystals, such as periodically poled lithium niobate (PPLN), are also commonly used for phase matching. These crystals contain periodically alternating regions of high and low  \(\chi^{(2)}\), following the same fundamental principle as birefringent materials. Since this paper focuses on the physical mechanisms of SPDC, we refer readers to \cite{armstrong1962interactions, bock2016highly, murakami2023quantum} for an in-depth discussion of PPLN-based phase matching. 

It is important to note that both birefringent and periodically poled phase-matching techniques involve a trade-off between photon indistinguishability and control over wavelength. Birefringence in the nonlinear crystal makes it difficult to maintain identical polarization states for both photons, which can degrade polarization entanglement or introduce a mixed state rather than a pure one. Additionally, as mentioned earlier, multiple photon emissions are always present in heralded sources. Consequently, the average pair production rate is deliberately kept much lower than one, significantly limiting efficiency. Since both pump power and \(\chi^{(2)}\) values are relatively low, the nonlinear crystal must be thin to minimise the interaction length and reduce unwanted effects \cite{armstrong1962interactions, kwiat1999ultrabright, couteau2018spontaneous, edamatsu2007entangled}.

\subparagraph{Four-Wave Mixing (FWM)}
Since higher-order nonlinear interactions cannot be entirely eliminated, they can instead be harnessed for single photon emission. Four-wave mixing (FWM) is a \(\chi^{(3)}\) nonlinear process in which two pump photons are converted into a correlated photon pair, as illustrated in Fig. \ref{fig:herald}(b). Like SPDC, FWM satisfies both energy and momentum conservation and is widely used in optical fibres and waveguides. By tuning the pump wavelength and leveraging the dispersion properties of the medium, the wavelengths of the generated signal and idler photons can be precisely controlled \cite{wang2001generation, fang2013state}.

Traditionally, FWM occurs spontaneously. However, stimulated FWM can be used to enhance single photon emission, as shown in Fig. \ref{fig:herald}(e). In this process, two pump photons are annihilated, and a seeded single photon stimulates the emission of signal and idler photons. The generated photons remain correlated, with one photon occupying the same mode as the seeded photon \cite{dong2017true}.

Despite its potential, FWM has faced increasing challenges as a viable source of single photons in resent years. Fabrication imperfections in waveguides introduce spectral mismatches between independent FWM sources, reducing photon indistinguishability, a critical factor for scalable quantum computing. While methods such as tapered waveguides and pump-delay tuning can mitigate this issue, achieving indistinguishability levels exceeding 99.5\% requires extremely precise fabrication tolerances \cite{borghi2022mitigating}. Additionally, alternative nonlinear processes, such as SPDC, often provide simpler engineering or higher performance, further limiting the widespread adoption of FWM for single photon sources.

\subparagraph{Unconventional Photon Blockade (UPB)}
Besides the dominant SPEs discussed earlier, unconventional photon blockade (UPB) has been explored as an alternative method for generating single photons. However, it is not yet widely adopted in practical applications. Unlike conventional photon blockade, which relies on strong anharmonicity in a nonlinear system, UPB arises from quantum interference effects in weakly nonlinear systems \cite{flayac2017unconventional}. Instead of preventing multi-photon occupation through strong nonlinear interactions, UPB suppresses two-photon excitation by destructive interference between multiple excitation pathways \cite{flayac2015all}. In UPB, a weak coherent field drives a system where two or more pathways exist for excitation from the ground state to multi-photon states, a mechanism known as pathway competition. This effect can be achieved through indirect excitation, where photons tunnel between coupled resonators or interact with \(\chi^{(3)}\) nonlinear elements such as Kerr media \cite{wang2021giant, flayac2017unconventional}.

A common implementation of UPB utilises two coupled optical cavities, often referred to as a “photonic molecule”, where the EM fields of the cavities overlap (cf. Fig. \ref{fig:herald}(c)). This enables photon tunnelling or coupling between the cavities, described by a tunnel coupling rate J in the system Hamiltonian:
\begin{equation}
    \hat{H}_{\text{coupling}}=J(\hat{a}_1^\dagger \hat{a}_2+\hat{a}_2^\dagger \hat{a}_1).
\end{equation}
The parameter \(J\) determines the strength of energy exchange between cavities. These pathway acquire phase  differences due to detuning parameter \(\Delta\), which is influenced by \(J\) and is defined as:
\begin{equation}
    \Delta = \omega_c-\omega_L,
\end{equation}
where \(\omega_c\) is the resonance frequency of the cavity, and \(\omega_L\) is the driving laser frequency. The optimal detuning condition varies for different systems. When interference between pathways becomes destructive, two-photon excitation is suppressed, leading to photon antibunching. Notably, this effect persists even with weak nonlinearity, as nonlinearity primarily modulates the relative phase between pathways \cite{zou2020enhancement, shen2015exact}. By carefully tuning \(J\) and \(\Delta\), the system can be engineered to minimize two-photon occupation, enhancing the purity of single photon emission.

After the interference process, pulsed driving combined with temporal filtering enables controlled single photon emission. The emitted photon pulse has a finite duration, determined by the cavity lifetime. However, antibunching is only preserved within a short time window (e.g., <100 ps) \cite{flayac2015all}. Without filtering, photons emitted outside this window degrade the single photon purity of the output. Temporal filtering removes multi-photon events occurring at later times within the same pulse, ensuring a higher-purity single photon stream.

\subsection{Current Technological Advancements and Performance Metrics}
In this section, we introduce the state-of-the-art of the SPEs above and analyse their performance statistics. Also, we specifically analyse the tunability and purity for each SPE. Overall, Table \ref{tab:performance metrics} summarises the key performance metrics of each SPE. Below the atom systems and probabilistic sources are investigated in detail.

\begin{table*}[t]
\caption{\label{tab:performance_metrics}
Summary of reported performance parameters of SPEs over the past five years.}
\centering
\scriptsize
\setlength{\tabcolsep}{3pt}        
\renewcommand{\arraystretch}{1.15} 

\begin{tabular}{ccccccccc}
\\
\toprule
\makecell{\textbf{Source}} & 
\makecell{\textbf{Wavelength}} & 
\makecell{\(\boldsymbol{g^{(2)}(0)}\)} & 
\makecell{\textbf{Tuning}\\\textbf{Range}\\(meV)} & 
\makecell{\textbf{Indistin-}\\\textbf{guishability}} & 
\makecell{\textbf{Clock}\\\textbf{Rate}} & 
\makecell{\textbf{Transmission}\\\textbf{Distance}\\(km)} & 
\makecell{\textbf{Efficiency}} & 
\makecell{\textbf{Type}} \\ 
\midrule
\midrule

Atoms & 
\makecell{Vis to\\C band} & 
0.002 \cite{higginbottom2016pure} & 
– & 
95\% \cite{comandar2016near} & 
8.1 MHz \cite{maunz2007quantum} & 
\makecell{0.7 \cite{hensen2015loophole}\\(fibre)} & 
– & 
QEs \\

\makecell{Single-dot QD\\coupled with a \\tunable open microcavity} & 890 nm \cite{ding2025high} & 
0.0205 \cite{ding2025high} & 
- & 
98.56\% \cite{ding2025high} & 
25.38 MHz \cite{ding2025high} & 
- & 
71.2\% \cite{ding2025high} & 
QEs \\

\makecell{InGaAs nanowire} & 
\makecell{890–1550 nm,\\optimised at\\O/C band} & 
<0.02 \cite{haffouz2018bright} & 
1.1 \cite{munnelly2017electrically} & 
97\% \cite{tomm2021bright} & 
1 GHz \cite{tomm2021bright} & 
\makecell{<50\cite{haffouz2018bright}\\(fibre) } & 
40\%–90\% & 
QEs \\

\makecell{InAs/GaAs QDs} & 
C band & 
\makecell{0.184 \cite{portalupi2019inas};\\0.07 \cite{moczala2019strain};\\0.012 \cite{he2019coherently}} & 
\makecell{0.75\\(via strain) \cite{moczala2019strain};\\1.1} & 
98.5\% \cite{ding2016demand} & 
100 MHz \cite{moczala2019strain} & 
\makecell{50–80\cite{motohisa2024characterization}\\(fibre) } & 
>80\% (QE) \cite{liu2018single} & 
QEs \\

InAs/InP & 
C band & 
0.0051 \cite{takemoto2015quantum} & 
– & 
– & 
– & 
\makecell{120\cite{takemoto2015quantum}\\(fibre) } & 
>20\% (CE) \cite{musial2020high} & 
QEs \\

\makecell{GaAs/AlGaAs\\via local droplet etching} & 
750–850 nm & 
\makecell{0.00075 \cite{schweickert2018demand};\\0.024 \cite{da2022pure}} & 
– & 
94.8\% \cite{da2022pure} & 
– & 
– & 
– & 
QEs \\

\makecell{CdSe-CdS\\core-shell NC} & 
Vis & 
\makecell{0.03 \cite{morozov2023purifying};\\<0.05 \cite{lin2017electrically}} & 
0.1–0.5 \cite{eich2022single} & 
– & 
MHz \cite{morozov2023purifying} & 
– & 
60\% \cite{morozov2023purifying} & 
QEs \\

\makecell{Perovskite \(\text{CsPbX}_3\)\\(X = Cl, Br, I)} & 
400–730 nm (Vis) & 
\makecell{0.02 \cite{zhu2022room};\\0.019 \cite{kaplan2023hong}} & 
0.3–0.8 \cite{farrow2023ultranarrow} & 
99.2\% \cite{kaplan2023hong} & 
\makecell{5 MHz \cite{farrow2023ultranarrow};\\1 GHz \cite{zhu2022room}} & 
\makecell{$\sim$0.5\\(free space)} & 
60\% \cite{zhu2022room} & 
QEs \\

NV centre & 
637 nm (ZPL) & 
0.08 \cite{andersen2017ultrabright} & 
<1 & 
95\% \cite{grange2017reducing} & 
168 kHz \cite{radulaski2017nonclassical} & 
– & 
82\% (CE) \cite{lubotzky2025approaching} & 
QEs \\

hBN & 
\makecell{300–800 nm\\(UV to near IR)} & 
\makecell{0.45 \cite{xia2019room};\\0.01 \cite{shaik2021optical}} & 
\makecell{31 \cite{xia2019room};\\80 \cite{haussler2021tunable}} &
98\% & 
10 MHz \cite{li2019near} & 
– & 
\makecell{10\% (CE) \cite{schell2017coupling};\\87\% (QE) \cite{nikolay2019direct}} & 
QEs \\

SPDC & 
Vis to C band & 
\makecell{0.0028 \cite{somaschi2016near};\\<0.003 \cite{kiessler2025spdc}} & 
– & 
99.56\% \cite{somaschi2016near} & 
2.1 MHz \cite{ngah2015ultra} & 
\makecell{1700\cite{yin2017satellite}\\(free space) } & 
\makecell{45\% \cite{kiessler2025spdc};\\65\% (CE) \cite{somaschi2016near}} & 
NOP \\

FWM & 
Vis to C band & 
0.09 \cite{sun2016quantum} & 
\makecell{132.5\\(near IR) \cite{kowligy2018tunable}} & 
0.74\% \cite{xiong2008polarized} & 
100 MHz \cite{sun2016quantum} & 
\makecell{15.7\cite{sun2016quantum}\\(fibre) } & 
20\%–60\% & 
NOP \\
\bottomrule
\\
\multicolumn{9}{p{\textwidth}}{\footnotesize\textit{Note.}Values listed in each row may originate from different experiments or from a single experiment conducted under varying conditions. As such, the table serves as an indicative comparison of the potential performance achievable by different physical systems rather than a direct benchmark. The overall efficiency is defined as the product of quantum efficiency (QE) and collection efficiency (CE). Unless otherwise specified, the efficiency values refer to this overall efficiency. Vis: visible spectrum. QEs: quantum emitters. NOP: non-linear optical process.}\\

\end{tabular}
\vspace{1mm}
\label{tab:performance metrics}
\end{table*}

\subsubsection{Natural Atoms and Artificial Atoms} 
Over the past decade, research interest in single photon sources based on natural atoms has significantly declined, while artificial atoms, such as quantum dots, diamond NV centres, and hBN defects, have become mainstream \cite{wilk2007single, piro2011heralded, mckeever2004deterministic}. This shift is driven by three main factors: the intrinsic technological challenges of working with single atoms, breakthroughs in the performance of solid-state quantum systems, and a theoretical shift in research paradigms.

As J. McKeever \textit{et al.} discovered in the early 2000s, single atom experiments require ultra-high vacuum environments combined with laser cooling, magneto-optical traps, or optical tweezers, especially for trapping Cs and Rb atoms, to maintain stable atomic positioning within a cavity \cite{mckeever2004deterministic, darquie2005controlled}. These systems are highly sensitive to vibrations, temperature fluctuations, and laser phase noise, necessitating complex feedback control. In contrast, QDs and defect centres are fixed to substrates using nanofabrication techniques, which inherently provides mechanical stability and scalability \cite{tawfik2017first, senellart2017high, garcia2021semiconductor}. Additionally, photons emitted by single atoms are typically distributed isotropically, leading to low collection efficiencies. The photon collection efficiency in Rb experiments is often below 10\% \cite{darquie2005controlled}. Although the Purcell effect in optical cavities can enhance the radiation rate, such as coupling Cs with a cavity to achieve an output efficiency of 69\% \cite{mckeever2004deterministic}, challenges such as cavity mode matching and long-term atom-cavity alignment remain significant technical bottlenecks.

In comparison, QDs have gained recognition for their high PLQY, suggesting potential for higher efficiency in the future (cf. Table \ref{tab:performance metrics}). Additionally, in recent years, novel materials such as defects in hBN have become a research focus due to their ability to emit single photons at RT. First-principles calculations indicate that the CBVN defect in hBN exhibits phonon coupling characteristics consistent with experimental observations, providing a new platform for controllable SPEs \cite{tawfik2017first, su2022tuning, shaik2021optical}. In contrast, material selection for single-atom systems is limited, and optimising performance through chemical modifications is challenging. Furthermore, solid-state systems are inherently suited for on-chip integration, such as the combination of silicon-based photonic crystals with quantum dots or the direct coupling of diamond photonic cavities with optical fibres \cite{kim2017hybrid, kim2020hybrid}. This is essential for developing quantum networks and photonic integrated circuits. By contrast, integrating multi-component single-atom systems, such as optical tweezer arrays with fibre networks, remains at an experimental stage.

Early single photon sources were primarily used for fundamental quantum optics demonstrations \cite{bergquist1986observation}. However, research has since shifted toward practical performance metrics, including photon generation rate, compatibility, and robustness against interference \cite{arcari2014near, tawfik2017first}. QDs and defect centres offer advantages in terms of higher repetition rates (see Table \ref{tab:performance metrics}) and wavelength stability, which can be further optimised through strain engineering.

Although interest in single-atom systems has declined, their core concepts, such as cQED, continue to shape the design of solid-state systems. The theory of selective radiative states originates from subradiant studies in atomic arrays, while the trigger control mechanisms of QDs based SPEs, such as electrical pulse modulation, draw inspiration from optical pulse excitation techniques in single-atom experiments \cite{asenjo2017exponential, darquie2005controlled}. In the future, cold atomic arrays may find renewed application in specific areas like quantum simulation or repeater nodes, but single atoms as standalone photon sources are likely to be entirely replaced by solid-state systems.

 \subparagraph{QDs}
As for QDs, significant progress has been made in their development as SPEs, particularly in improving photon indistinguishability and brightness, positioning them at the forefront of quantum communication and computing \cite{couteau2023applications1, arakawa2020progress, ding2025high, chen2006quantum}. III-V semiconductors such as InAs and InGaAs, along with III-nitrides like InGaN, are the leading materials for high-performance QD-based SPEs, while cQDs and hybrid systems are gaining attention for scalable integration.

Research on QDs-based SPEs follows two main directions: material innovation and structural optimisation. Before 2019, the single photon emission properties of traditional materials such as GaAs, CdSe/ZnS, InP/ZnSe, and InAs/InP were extensively studied, revealing both advantages and limitations. GaAs QDs, with their high exciton binding energy, are well suited for low-temperature single photon emission. InAs/InP QDs have become a key focus for quantum communication due to their compatibility with the telecom wavelength (1550 nm). CdSe/ZnS core-shell QDs have suppressed blinking effect, but are less suitable for high-purity single photon applications \cite{smolka2021optical, holtkemper2020selection}.

To improve performance, researchers have employed shell passivation, such as ZnS coatings to suppress surface defects, and doping, such as \(\text{Mn}^{2+}\) to regulate spin states, pushing quantum yield and single photon purity toward theoretical limits \cite{beaulac2008mn2+, holtkemper2020selection}. However, intrinsic efficiency remains constrained by low light extraction (typically below 10\%) and temperature sensitivity, where phonon scattering at high temperatures leads to performance degradation \cite{aghaeimeibodi2018integration, haffouz2024photoluminescence}. Thus, over the past five years, the focus has shifted toward optimising photonic structures to overcome these material limitations and enhance device-level performance.

Microcavities and nanowire waveguides have significantly improved efficiency through the Purcell effect. Common cavity designs, including micropillars, bullseye cavities, open cavities, and adiabatic pillars, have led to substantial performance gains. As early as 2016, Jian-Wei Pan’s group increased efficiency from 10\% to 66\% using micropillar cavities, and by 2025, they achieved single-dot QD coupling with a tunable open microcavity, reaching the threshold for scalable photonic quantum computing for the first time \cite{ding2025high, ding2016demand}. Plasmonic nanoantennas have further enhanced brightness by a factor of seven, halved emission lifetimes, and significantly suppressed blinking \cite{jiang2021single}. Additionally,  Li \textit{et al.} demonstrated high directionality using line-array plasmonic antenna coupling, with predicted collection efficiencies approaching 85\% \cite{li2022directional}.

Beyond efficiency breakthroughs, compatibility with optical communication has improved, particularly in telecom band optimisation. InAs/InP QDs now achieve low-noise single photon emission at 1550 nm through advancements like in MBE growth, making them fully compatible with optical fibre networks \cite{smolka2021optical, joos2024coherently}. Additionally, as previously mentioned, silicon-based photonic crystals integrated with quantum dots offer a path toward scalable quantum light source arrays. 

Despite progress in photonic structures, continued engineering improvements in traditional materials remain essential. O ptimising core-shell structures and advancing fabrication techniques such as droplet epitaxy and site-controlled growth have led to enhanced performance \cite{da2021gaas, haffouz2024photoluminescence}.

While research on conventional materials has slowed, novel materials have gained attention, linking back to the first research direction. QDs in monolayers of transition metal dichalcogenides (TMDs) have attracted significant interest due to their flexibility. Their weak van der Waals binding forces enable rich possibilities for heterointegration. For a detailed discussion, we refer readers to \cite{esmann2024solid}. Perovskites represent another promising material for quantum dots. Unlike traditional quantum dots and TMDs, perovskite quantum dots (PQDs) maintain the high purity characteristic of other quantum dots while offering exceptional tunability at RT. Their ability to easily couple with cavities makes them particularly attractive for scalable quantum photonics applications.

 \subparagraph{Defect Centres}
As for defects-based SPEs, diamond remains a cornerstone, with NV centres offering long spin coherence times and stable optical transitions. However, their broad emission and moderate brightness limit quantum communication efficiency. This has spurred interest in diamond’s group-IV vacancies, such as silicon- (SiV), germanium- (GeV), and tin-vacancy (SnV) centres. SiV centres exhibit narrow zero-phonon lines (ZPLs) with high Debye-Waller factors, enabling efficient photon extraction \cite{redjem2023all}. SnV centers, with even narrower ZPLs and spin-photon interfaces, show promise for quantum networks but require cryogenic temperatures. Beyond diamond, silicon carbide (SiC) defects, such as divacancies, provide telecom-wavelength emission and CMOS compatibility, while hBN hosts ultra-bright, room-temperature defects with high spatial resolution, ideal for on-chip integration \cite{castelletto2020silicon, redjem2023all, shaik2021optical, su2022tuning}. 

Especially, hBN is a very promising material. Its planar structure makes it very suitable for integration with other photonic circuits. It is a layered material with strong in-plane bonding and weak interlayer interactions like TMDs, making it inherently compatible with planar integration techniques. Defects in hBN, such as those emitting sharp ZPLs, can be embedded in multilayer flakes while maintaining their optical properties \cite{jungwirth2016temperature, shaik2021optical, su2022tuning}. Significantly, hBN exhibits broad emission range from UV to near IR region and the first signature of Rabi oscillations and resonant fluorescence emission was observed from a resonantly drive hBN quantum emitter (cf. Table \ref{tab:performance metrics}) \cite{shaik2021optical, konthasinghe2019rabi}. Contemporarily, an attractive phenomenon of second harmonic generation (SHG) was discovered in hBN multilayer coupled to circular Bragg grating (CBG) photonic microstructures which, encourages the delightful applications \cite{bernhardt2021large, cunha2020second}. Although its performance has not reached the mark, it is still a very appealing alternative and worth researching. 

\subsubsection{Probabilistic Sources}
Probabilistic sources are no longer the most popular SPEs while still widely used. Techniques like SPDC, FWM, and UPB rely on nonlinear optical processes to generate photon pairs, as previously discussed. These sources offer high scalability and tunability. For instance, SPDC remains widely used in quantum optics experiments for single photon generation. However, because photon pair generation in SPDC and FWM follows a Poisson distribution, achieving deterministic single photon output is impossible. At high pump power, the likelihood of multi-pair generation increases significantly, requiring complex post-selection techniques such as photon-number-resolving detectors (PNRDs). Also, improving efficiency through multiplexing or active modulation increases device size and costs \cite{zhang2021spontaneous, rohde2015multiplexed, stasi2022enhanced, stasi2023high}. In contrast, quantum emitters now provide sufficient efficiency without significantly sacrificing purity. This shift has made QDs and defect-based SPEs the research focus. 

Although SPDC and FWM are limited as SPEs, they remain irreplaceable for multiphoton state generation and quantum information processing. In quantum entanglement and multiphoton experiments, SPDC is the standard method for generating polarization-entangled photon pairs, such as in Bell state preparation. Notable demonstrations include 12-photon entanglement experiments and Scattershot Boson Sampling, showcasing their role in complex quantum protocols \cite{zhong201812, bock2016highly}. For quantum communication and QKD, SPDC’s broad wavelength tunability, such as in the telecom C-band, makes it well-suited for fibre-optic communication (cf. Table \ref{tab:performance metrics}) \cite{faleo2024entanglement, zhan2023measurement, euler2021spectral}. In quantum networks and nonlinear interference, FWM enables the generation of multimode quantum-correlated beams via cascaded processes, making it valuable for continuous-variable quantum networks \cite{pscherer2021single, cheng2021quantum}.

As for UPB, research progress has been relatively limited due to theoretical complexity, competition with other sources, and experimental challenges. Table \ref{tab:upb} compares UPB with other photon sources in terms of performance. Clearly, UPB lags behind in purity and brightness, and its experimental difficulty makes research costly in both time and resources.

\begin{table*}[t]
\caption{
Performance comparison among UPB, SPDC/FWM, and QDs.}
\centering
\scriptsize
\setlength{\tabcolsep}{3pt}        
\renewcommand{\arraystretch}{1.15} 
\begin{tabular}{ccccccccc}
\\
\toprule
\makecell{\textbf{Source}} & 
\makecell{\(\boldsymbol{g^{(2)}(0)}\)} &  
\makecell{\textbf{Brightness}\\\textbf{(photons per second)}} & 
\makecell{\textbf{Integrability}} & 
\makecell{\textbf{Experimental Complexity}}
\\ 
\midrule
\midrule

UPB & 
<0.5 \cite{flayac2015all} $\sim$ 0.2 & 
\(1\times 10^3\) & 
\makecell{Medium \\(requires complex cavity design)} & 
Very high \\

SPDC/FWM & 
\makecell{0.01 $\sim$ 0.1\\(with post-selection)} & 
\makecell{\(1\times 10^4\) \footnote{From Ref. \cite{kaneda2019high}};\\ \(10^3\sim10^8\)} & 
\makecell{Low\\(large volume)} & 
\makecell{High\\(requires multiplexing techniques)} \\

QDs & 
\makecell{<0.01 \cite{abudayyeh2019purification,ihara2019superior};\\ 0.2 \cite{zhu2022room};\\0.0075\%\\at cryogenic temperature \cite{schweickert2018demand}}& 
\makecell{\(1\times 10^6\) \footnote{From Ref.\cite{jiang2021easy}};\\ \(10^6\sim10^9\)} & 
\makecell{High\\(on-chip integration)} & 
\makecell{Medium \\(usually requires low temperature)} \\
\bottomrule
\end{tabular}
\vspace{1mm}
\label{tab:upb}
\end{table*}

Furthermore, nonlinear optical processes offer outstanding tunability in scenarios requiring wavelength matching, bandwidth compression, and multiphoton entanglement. As shown in Table \ref{tab:performance metrics}, the tuning range of FWM reaches an impressive 132.5 meV, whereas QDs typically have a tuning range below 1 meV. This broad wavelength tunability makes FWM well-suited for cross-band applications, such as visible-to-telecom interfaces \cite{bock2016highly, anwar2021entangled}. Additionally, SPDC enables flexible bandwidth compression through dynamic techniques like cavity enhancement and group velocity matching, whereas QDs inherently have a narrow bandwidth \cite{kaneda2016heralded}. Nonlinear optical processes also allow multi-parameter optimisation and high-dimensional encoding, making them essential in quantum technologies \cite{chang2020observation, zhang2021spontaneous, couteau2018spontaneous}. Thus, quantum emitters and nonlinear optical processes serve complementary roles—emitters excel in scalable quantum networks, while nonlinear processes are better suited for multiphoton protocols and wavelength-sensitive applications.

Specifically, for tunability and purity, Table \ref{tab:performance metrics} indicates that purity has reached theoretical limits for some sources, marking a major breakthrough. However, these sources, particularly QDs, exhibit limited tunability. Conversely, sources with broad tuning ranges often suffer from low efficiency and poor on-chip integrability. This highlights the need for a versatile SPE that integrates high purity, tunability, and efficiency. Thus, we explored HOIP QDs, which maintain the high purity of conventional QDs while offering enhanced tunability and reduced blinking. The following section will explore HOIP QDs in detail.
\section{Hybrid Perovskite SPEs}
\begin{figure}[htbp]
    \centering
    \includegraphics[width=0.8\linewidth]{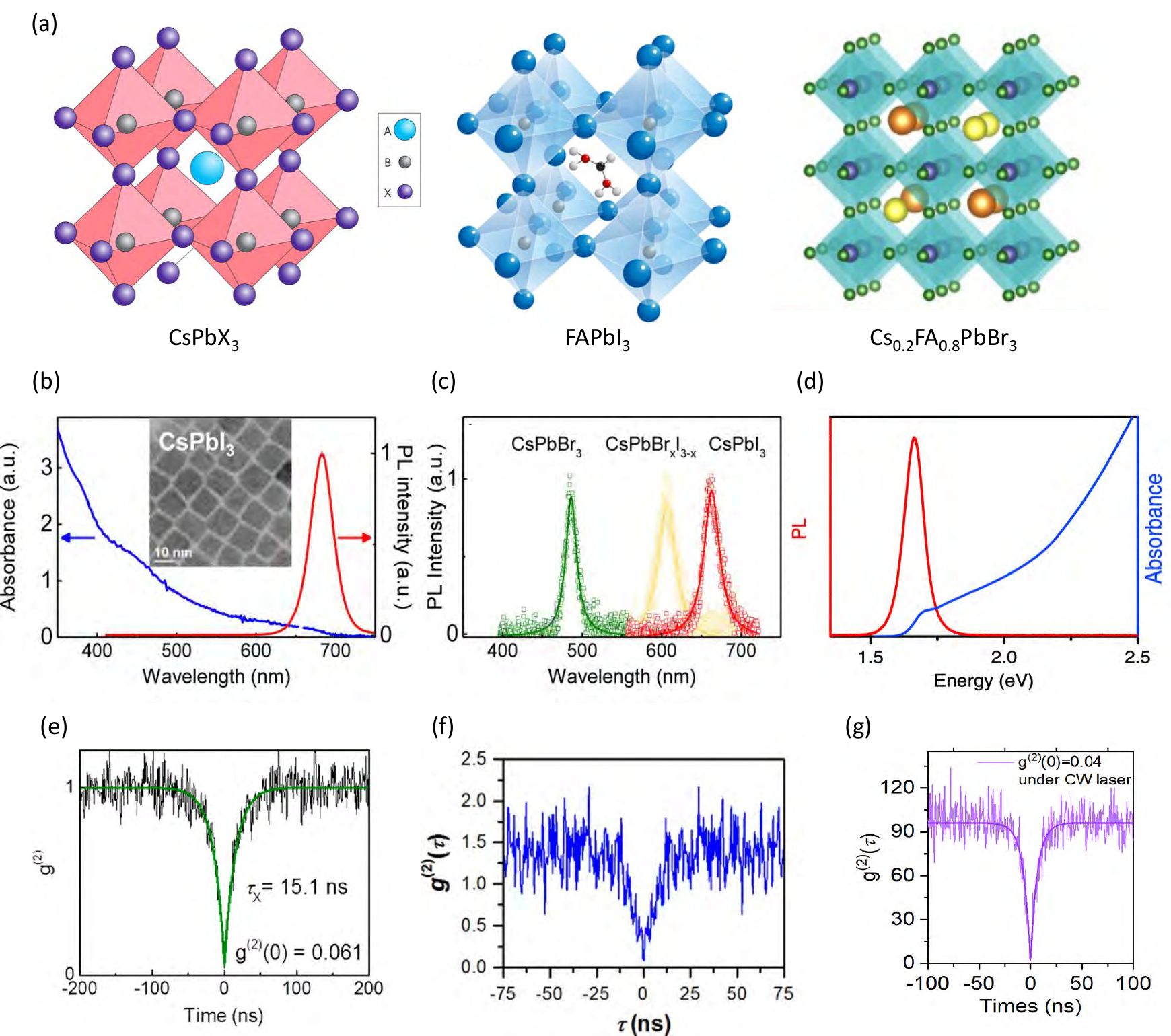}
    \caption{Perovskites used for single photon emission: (a) Perovskite crystal structure. Left: generic structure for perovskite. All of the halide inorganic \(\text{CsPbX}_3\) share the same structure \cite{green2014emergence}. Reproduced with permission from Nat. Photon. \textbf{8}, 506 (2014). Copyright 2014 Springer Nature Limited. Middle: crystal structure for organic perovskite \(\text{FAPbI}_3\) \cite{fu2018unraveling}. M. Fu et al., Nat. Commun. \textbf{9}, 3348 (2018); licensed under a Creative Commons Attribution (CC BY) license. Right: crystal structure for HOIP \(\text{Cs}_{0.2} \text{FA}_{0.8}\text{PbBr}_3\) \cite{d2023color}. Reproduced with permission from ACS Photonics \textbf{10}, 197 (2023). Copyright 2023 American Chemical Society. (b), (c), (d) Absorption and PL spectra of PQDs: (b) Absorption and PL spectra of \(\text{CsPbX}_3\) \cite{park2015room}. Reproduced with permission from ACS Nano \textbf{9}, 10386 (2015). Copyright 2015 American Chemical Society. (c) PL intensity of \(\text{CsPbBr}_{x} \text{I}_{3-x}\). With \(x\) decreasing, the spectrum do the red shift\cite{park2015room}. Reproduced with permission from ACS Nano \textbf{9}, 10386 (2015). Copyright 2015 American Chemical Society. (d) Absorption and PL spectra of \(\text{FAPbI}_3\) in a toluene solution \cite{fu2018unraveling}. M. Fu et al., Nat. Commun. \textbf{9}, 3348 (2018); licensed under a Creative Commons Attribution (CC BY) license. (e), (f), (g) Second-order PL intensity correlation functions measured for PQDs under continuous wave (c.w.) excitation: (e) A single \(\text{CsPbI}_3\) QD, \(g^{(2)}(0)=0.061\) \cite{park2015room}. Reproduced with permission from ACS Nano \textbf{9}, 10386 (2015). Copyright 2015 American Chemical Society. (f) A single \(\text{CsPbBr}_3\) NC, \(g^{(2)}(0)=0.06\) \cite{hu2015superior}. Reproduced with permission from ACS Nano \textbf{9}, 12410 (2015). Copyright 2015 American Chemical Society. (g) A single \(\text{FAPbBr}_3\) QD, \(g^{(2)}(0)=0.04\) \cite{wang2024weakly}. Reproduced with permission from ACS Nano \textbf{18}, 10807 (2024). Copyright 2024 American Chemical Society.}
    \label{fig:perovskite}
\end{figure}

Hybrid organic-inorganic perovskites are promising materials for single photon emission and belong to the broader perovskite family. This material has sparked wide interest for its use in solar cells, lasing, and LEDs, especially in PV, due to its high conversion efficiency and emission properties \cite{jena2019halide, tu2021perovskite, huang2021perovskite, zhao2018opportunities, xu2020halide, makarov2019halide, raino2016single}. As single photon sources, HOIP QDs have very good composition and size tuning as other PQDs, which is a huge advantage over other QDs. More importantly, compared to other PQDs, it offers better tuning range and keeps high purity without inducing serious blinking (cf. Fig. \ref{fig:perovskite}). This section introduces HOIP materials, examines three key mechanisms underlying HOIP SPEs, and rigorously analyses different implementation schemes.

Perovskites refer to a class of materials with a distinctive crystal structure, described by the general formula  \(\text{ABX}_3\). The A-site is occupied by a monovalent cation, either a metal or an organic molecule, such as \(\text{Cs}^+\) or \(\text{CH}_3\text{NH}_3^+\). The B-site contains a smaller divalent metal cation, typically \(\text{Pb}^{2+}\) or \(\text{Sn}^{2+}\), while the X-site consists of an anion, often a halide like \(\text{Cl}^-\), \(\text{Br}^-\) or \(\text{I}^-\) as illustrated in Fig. \ref{fig:perovskite}(a). Depending on the choice of A, B, and X, perovskites are categorized into three types: organic, all-inorganic, and hybrid organic-inorganic perovskites. 

HOIPs are distinguished by their organic A-site cations, such as methylammonium \(\text{CH}_3\text{NH}_3\) or formamidinium \(\text{CH}_3\text{NH}_2\), combined with inorganic metal B-site cations like lead or tin. More recently, perovskites have attracted significant interest as single photon sources, particularly in the form of colloidal perovskite quantum dots (cPQDs) \cite{zhu2022room, park2015room}. The preference for HOIP materials stems from both fabrication and photostability. The choice of colloidal perovskites is guided by their ease of fabrication, while the selection of HOIPs within the perovskite family is motivated by their superior optical and electronic characteristics.

\subsubsection{Ease of Fabrication}
Over the past decade, III-V semiconductor QDs have led SPEs in terms of indistinguishability and efficiency \cite{senellart2017high, park2015room, zhu2022room, garcia2021semiconductor}. However, this outstanding performance comes with significant complexity in fabrication. As mentioned in the previous section, III-V QDs, as a kind of eQDs, rely on MBE and MOCVD, which require precise control, operate at cryogenic temperatures, and are costly \cite{senellart2017high, park2015room, zhu2022room, garcia2021semiconductor}. Furthermore, cryogenic cooling is necessary to suppress lattice vibrations and maintain coherence \cite{garcia2021semiconductor}.

Colloidal PQDs, by contrast, offer a low-cost alternative that can be fabricated under mild conditions at RT \cite{park2015room, zhu2022room, castelletto2022prospects}. They also benefit from ultrafast AR, which effectively suppresses multiphoton emission, making them temperature-independent and robust. Another significant advantage of cPQDs is their exceptionally high emission quantum yield, reaching up to 90\%, which is considerably higher than that of traditional QDs like CdSe \cite{protesescu2015nanocrystals, song2018organic}. This high efficiency, combined with their simple fabrication process, makes cPQDs well-suited for mass production. Here we do not discuss detailed synthesis methods of PQDs. For a detailed review, we direct readers to \cite{castelletto2022prospects, garcia2021semiconductor}.

Most importantly, cPQDs are highly tunable. By adjusting their halide composition (X = Cl, Br, I) or leveraging quantum size effects, their emission wavelength can be tuned across the entire visible spectrum (410–700 nm), with particularly strong performance in the blue-green region (410–530 nm). In contrast, traditional metal chalcogenide QDs degrade rapidly in this range \cite{protesescu2015nanocrystals} and often require mechanical strain or external fields for tuning \cite{d2023color}. Additionally, conventional eQDs depend on material selection (e.g., InAs, GaN) or quantum-confined Stark effects for wavelength tuning, but their range is limited and requires external electric fields \cite{he2019quantum, tamariz2020toward, xia2019room}.

To date, cPQDs have demonstrated pure, bright, and narrow-linewidth single photon emission at RT \cite{zhu2022room}. Their accessibility and commercial availability contribute to their rising popularity since 2015. Unlike III-V QDs, which require complex lattice matching, perovskites integrate easily with photonic structures such as waveguides, making them a promising platform for scalable quantum photonics.

\subsubsection{Enhanced Photostability}
\begin{figure*}[htbp]
    \centering
    \includegraphics[width=0.8\linewidth]{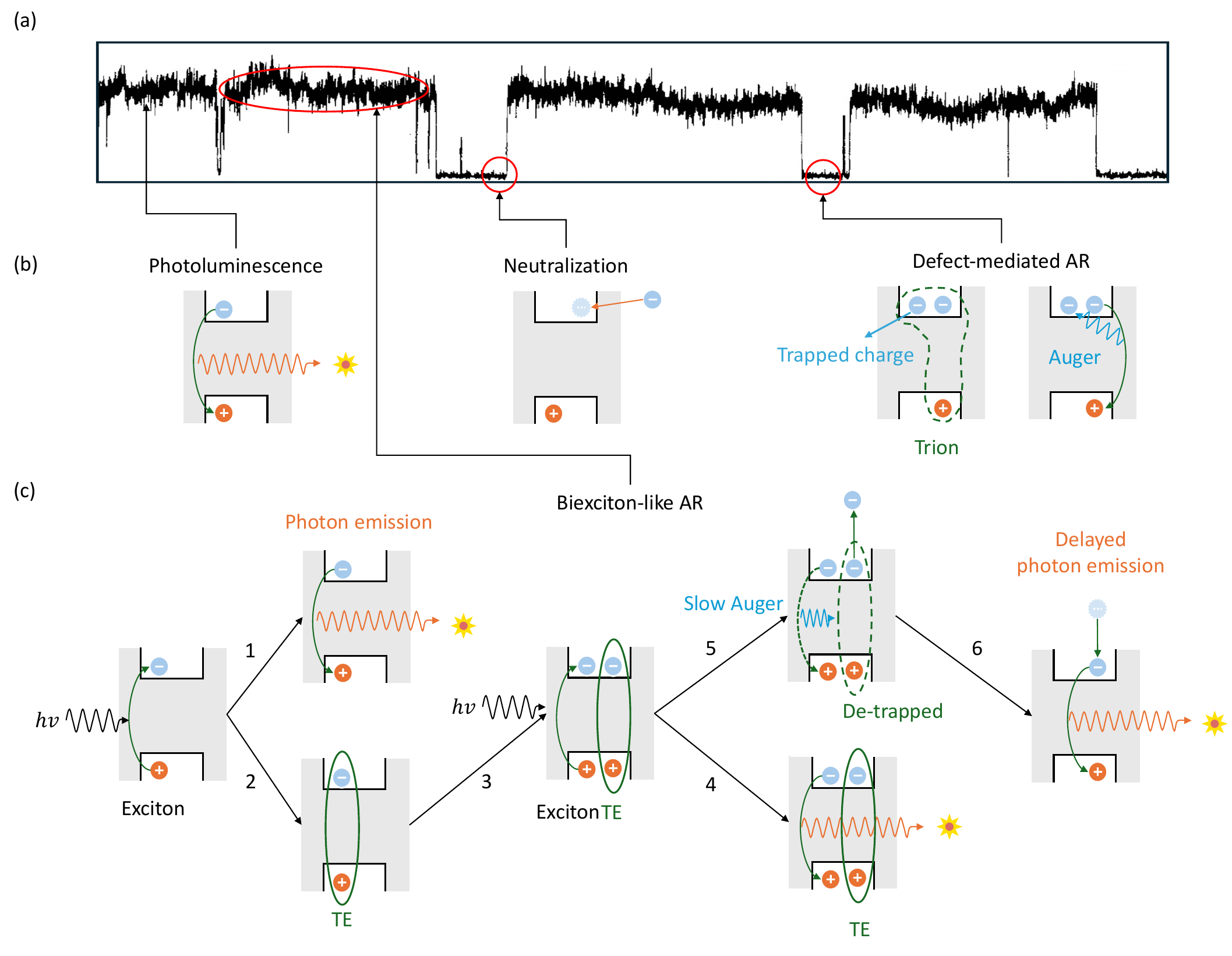}
    \caption{AR induced blinking: (a) The time dependence of the PL intensity of a single CdSe NC coated with a seven-monolayer-thick ZnS shell with a radius of 2.1 nm under a c.w. excitation intensity of 0.7 kW \(\text{cm}^2\) \cite{nirmal1996fluorescence}. Reproduced with permission from Nature \textbf{383}, 802 (1996). Copyright 1996 Springer Nature Limited. (b) Schematic representation of type A blinking. It goes through three stages: photoluminescence, defect-mediated AR, and neutralization. One of the charge in an exciton is trapped and form a trion with another exciton. When the exciton recombines, the energy transfer to the trapped charge, instead of emitting light. This form a long OFF state. The trapped charge then is ejected and eventually returns to the QD, ending the blinking cycle. Adapted from \cite{efros2016origin, galland2011two}. Reproduced with permission from Nat. Nanotechnol. \textbf{11}, 661 (2016). Copyright 2016 Springer Nature Limited. Reproduced with permission from Nature \textbf{479}, 203 (2011). Copyright 2011 Springer Nature Limited. (c) Scheme of proposed biexciton-like Auger interaction between an exciton and a trapped exciton (TE): When an exciton is generated in a QD, it will emit a photon (route 1) or be trapped (route 2). The QD that contains a TE can be excited again (route 3), forming an exciton, TE pair. The exciton can emit a photon under the influence of the TE (route 4). Alternatively, biexciton-like Auger recombination can happen in the exciton TE pair (route 5). Then the TE will gain energy to be detrapped, resulting in a recovered normal exciton (route 6) accompanied by delayed PL. Adapted from \cite{mi2023biexciton}. Reproduced with permission from J. Phys. Chem. Lett. \textbf{14}, 5466 (2023). Copyright 2023 American Chemical Society.}
    \label{fig:blinking}
\end{figure*}

However, as suggested before, cQDs are inherently prone to blinking behaviours, including PQDs. Since the first demonstration of $\text{CsPbI}_3$ PQDs in 2015, strongly confined all-inorganic lead halide perovskites have been the dominant choice for QDs due to their excellent optoelectronic properties \cite{zhu2022room, park2015room}. These PQDs range in size from 5 to 15 nm, with emission tunability across approximately 400–700 nm \cite{esmann2024solid}. Those strongly confined PQDs typically rely on fast Auger rate to significantly annihilate multiphoton emission \cite{boehme2023strongly, zhu2022room}. Their second-order correlation function can be approximated by:
\begin{equation}
    g^{(2)}(0)\approx \eta_{\text{XX}}/\eta_{\text{X}}\approx 4\tau_{\text{XX}}/\tau_{\text{X}}.
\end{equation}
This enables PQDs to achieve high single photon purity with $g^{(2)}(0) = 2\%$ \cite{nair2011biexciton, wang2024weakly}. Ensemble-level PL behaviour, including intensity scaling and spectral modulation under varying dielectric environments or external perturbations, has been systematically explored in earlier studies \cite{garner2008electric, garner2009light, neogi2009control}. However, the strong confinement also makes small single-dot PQDs suffer from blinking (cf. Fig. \ref{fig:blinking}(a)) \cite{nirmal1996fluorescence, trinh2018organic, chuang2024highly, efros2016origin, cragg2010suppression}. 

The defect-mediated AR is a major contributor to type A blinking, particularly in PQDs, where the charge carrier dynamics is highly sensitive to defects, external fields, and excitonic interactions. After exciton formation, surface defects, such as halide vacancies and incomplete ligand passivation, create mid-gap energy states. One charge carrier from the exciton become trapped at these defect sites. This trapping prevents radiative recombination and leaves the QD charged. It is charged positively if a hole is trapped or negatively if an electron is trapped, creating a net charge on the QD. A subsequent excitation generates another electron-hole pair in the charged QD. The new exciton combines with the trapped charge to form a trion, which undergoes non-radiative AR. The energy from the electron-hole recombination is transferred to the trapped charge, which is excited out of the QD or into a high-energy state. During this whole process, no photon is emitted, causing the QD to enter a long OFF state. 

The trapped charge now with excess energy will be ejected from the QD into surrounding matrix like ligands or substrate. Note that the 'charged QD' refers to the charge separation in a neutral QD, so the ejection switches on the electric field that further exacerbate AR. But the blinking process does not end here. Due to charge neutrality, the ejected charge needs to return to the QD later, which completes the blinking cycle \cite{efros2016origin, galland2011two}. Fig. \ref{fig:blinking}(a) present a typical blinking spectrum and illustrate the process of type A blinking. It is observable that fluctuations exist during the ON state, which cannot be explained by the trion model. Mi \textit{et al.} (2024) proposed a microscopic model to explain the fluctuations during photoluminescence (cf. Fig. \ref{fig:blinking}(c)) \cite{mi2023biexciton}.

To suppress blinking, the most extensively studied approach is the addition of shells to QDs, forming core-shell structures \cite{chen2013compact, zang2017thick, he2019suppression, efros2016origin, cragg2010suppression}. However, this method often comes at the expense of reduced single photon purity \cite{wang2024weakly, nair2011biexciton}. For example, Tang \textit{et al.} have used CdS shell to cap \(\text{CsPbBr}_3\) NCs, whose PL intermittency has been reduced, while \(g^{(2)}(0)\) was 0.43. Moreover, in 2016, Rainò \textit{et al.} has demonstrated a single \(\text{CsPbX}_3\) (X=Cl, Br) QDs of 9.5 nm diameter which presented ultrafast radiative decay, thus suppressing blinking free emission. Though this is realised under cryogenic temperature (6K) instead of RT, where phonon interactions are suppressed, so the coherence volume of the exciton increases \cite{raino2016single}. This enhances the oscillator strength. As a result, the radiative lifetime becomes much shorter. Since the radiative decay is ultrafast, the excitons recombine before nonradiative process, suppressing Auger decay. Similarly, its \(g^{(2)}(0)\) is approximately 0.3, which is a solid value for a completely unpassivated, bare NC without dielectric engineering, but it is not as optimal as low values like 0.06 \cite{park2015room, raino2016single}. 

HOIP QDs, on the other hand, can achieve blinking-free emission with minimal trade-offs under RT. Instead of relying on size-dependent confinement, HOIP QDs mitigate blinking through the effects of organic cations. Additionally, the A-site cation in perovskite QDs does not have to be strictly organic. By incorporating organic cations into all-inorganic PQDs, such as $\text{CsPbBr}_3$, their spectral tunability and structural stability can be enhanced \cite{d2023color}. Table \ref{tab:perovskite_comparison} compares the performance of these different types of QDs.

\begin{table*}[t]
\caption{
Comparison among perovskite QDs, hybrid perovskite QDs, and III-V QDs}
\centering
\scriptsize
\setlength{\tabcolsep}{3pt}        
\renewcommand{\arraystretch}{1.15} 

\begin{tabular}{ccccccccc}
\\
\toprule
\makecell{\textbf{Property}} &  
\makecell{\textbf{Perovskite QDs}} & 
\makecell{\textbf{Hybrid Perovskite QDs}} & 
\makecell{\textbf{III-V QDs}}
\\ \hline
\hline

\makecell{Synthesis \\Method \cite{esmann2024solid, garcia2021semiconductor, vighnesh2022hot, shamsi2019metal}} & 
\makecell{Colloidal \\(hot-injection,\\ LARP, post-synthesis\\ modifications)}  & 
\makecell{Colloidal \\(hot-injection method,\\LARP, post-synthesis\\treatments and modifications)} & 
\makecell{Epitaxial \\(MBE,\\MOCVD)}
\\

\makecell{Synthesis \\Complexity} & Low & Low & High \\

\makecell{Operating \\Temperature \cite{zhu2022room, park2015room, raino2022ultra}}& Room temperature & Room temperature & Cryogenic \\
 
\makecell{Composition \\Tunability \cite{d2023color, su2022tuning, xia2019room}}& High & High & Low \\

\makecell{Photostability \cite{park2015room, chouhan2021real, wang2024weakly}} & \makecell{Low} & High & High\\

\makecell{Integration\\Potential} & High (solution-processable) & High (on-chip compatible) & Medium (requires cooling setup) \\

\bottomrule
\\

\end{tabular}

\footnotesize\textit{Note.} This comparison is indicative, applying to general, representative cases of each QD class. Specific formulations, advanced synthesis techniques, and device architectures can significantly alter the properties listed. For example, some highly stabilized perovskite QDs may exhibit photostability comparable to III-V QDs.\\
\vspace{1mm}
\label{tab:perovskite_comparison}
\end{table*}

\subsection{Mechanisms of Photon Emission in HOIP SPEs}
Generally, HOIP SPEs can operate through three distinct emission mechanisms, depending on their size. HOIPs are commonly engineered as NCs (1–100 nm) to enhance their optical properties. When their size is reduced below the excitonic Bohr radius (~2–20 nm for perovskites), quantum confinement effects emerge, making them QDs. In contrast, larger NCs may still emit single photons but lack quantum confinement, instead relying on defect states, such as vacancies, interstitials, or surface states to trap excitons.

The most promising approach for achieving high-performance HOIP SPEs is the formation of HOIP QDs, leveraging extensive research on inorganic PQDs. Two primary configurations exist: strongly confined and weakly confined QDs.
    \begin{itemize}
        \item  Strongly confined QDs exploit strong Coulomb interactions and fast AR to enhance tunability. These QDs benefit from highly efficient charge carrier dynamics, enabling precise spectral control.
        \item Weakly confined QDs rely on dielectric screening or exciton-exciton repulsion to suppress AR, minimising PL blinking without sacrificing single photon purity.
    \end{itemize}
This distinction between AR-driven and reduced-AR QDs allows for tailored engineering of HOIP SPEs to optimise performance for different quantum photonics applications.

\subsubsection{Strongly confined HOIP QDs}
The strongly confined HOIP QDs operate through mechanisms similar to those of inorganic PQD-based SPEs, with one key distinction: the incorporation of organic $\text{FA}^+$ cations into inorganic $\text{CsPbBr}_3$ PQDs. This doping strategy enables broader spectral tunability, making these QDs comparable to conventional colour-tunable SPEs \cite{d2023color}. Importantly, the addition of $\text{FA}^+$ does not significantly alter the size or shape of the PQD, preserving the same quantum confinement and dimensionality as before.

\begin{figure*}
    \centering
    \includegraphics[width=1\linewidth]{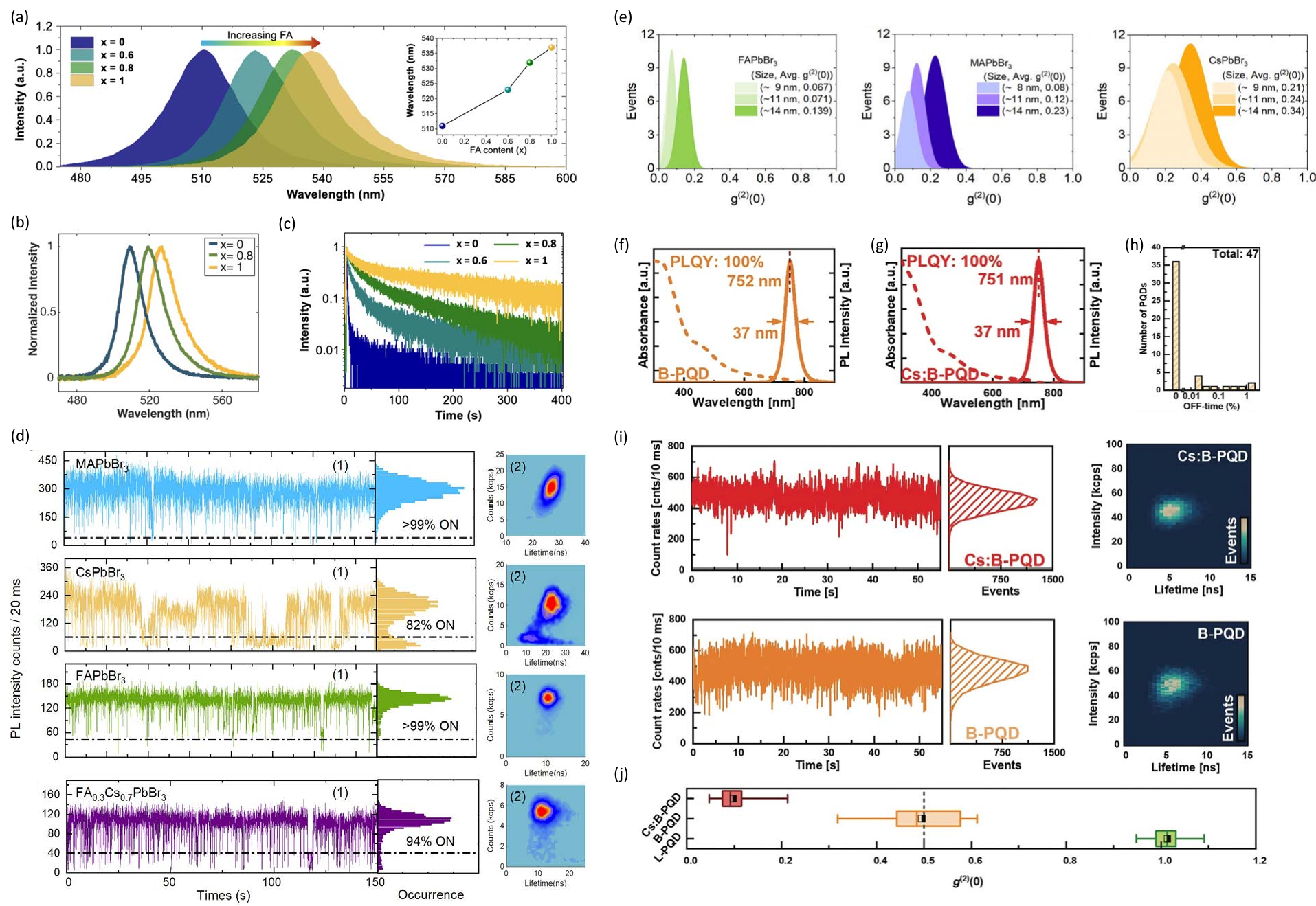}
    \caption{Performance of different schemes of HOIP QDs. \textbf{(a)-(c)} Performance of strongly confined HOIP QDs: (a) Normalized PL spectra of the \(\text{CsPbBr}_3\) (blue), \(\text{Cs}_{0.4} \text{FA}_{0.6}\text{PbBr}_3\) (cyan), \(\text{Cs}_{0.2} \text{FA}_{0.8}\text{PbBr}_3\) (green), and  \(\text{FAPbBr}_3\) (yellow) QD ensembles. (b) Normalized PL spectra of  an individual \(\text{CsPbBr}_3\) (blue), \(\text{Cs}_{0.4} \text{FA}_{0.6}\text{PbBr}_3\) (cyan), \(\text{Cs}_{0.2} \text{FA}_{0.8}\text{PbBr}_3\) (green), and  \(\text{FAPbBr}_3\) (yellow) QD. (c) Time-resolved PL lifetime measurements of the \(\text{Cs}_{1-x} \text{FA}_{x}\text{PbBr}_3\) QD \cite{d2023color}. Reproduced with permission from ACS Photonics \textbf{10}, 197 (2023). Copyright 2023 American Chemical Society. \textbf{(d)-(e)} Performance of the first scheme of weakly confined HOIP QDs: (d) ON-OFF blinking behaviour observed in fluorescence time traces for 11 nm  \(\text{MAPbBr}_3\), \(\text{CsPbBr}_3\), \(\text{FAPbBr}_3\), and \(\text{FA}_{0.3} \text{Cs}_{0.7}\text{PbBr}_3\) single PQDs. (e) Histogram distributions of \(g^{(2)}(0)\) histogram distributions for \(\text{APbBr}_3\) (A = FA, MA, Cs) PQDs with varying compositions and sizes \cite{wang2024weakly}. Reproduced with permission from ACS Nano \textbf{18}, 10807 (2024). Copyright 2024 American Chemical Society. \textbf{(f)-(j)} Performance of the second scheme of weakly confined HOIP QDs: (f) Absorption and PL spectra of \(\text{MAPbI}_3\). (g) Absorption and PL spectra of Cs-doped \(\text{MAPbI}_3\). (h) OFF time histogram distributions from 47 individual Cs-doped \(\text{MAPbI}_3\) PQDs in total. (i) Time traces of single photon emission over 60 seconds and corresponding fluorescence lifetime intensity distribution (FLID) plots. Upper panel: Cs-doped \(\text{MAPbI}_3\); lower panel: \(\text{MAPbI}_3\). (j)  \(g^{(2)}(0)\) of Cs-doped \(\text{MAPbI}_3\) (red), \(\text{MAPbI}_3\) (orange), and \(\text{MAPbI}_3\) PQDs synthesised via the LARP method (green) \cite{chuang2024highly}. Reproduced with permission from Small \textbf{20}, 2308676 (2024). Copyright 2023 Wiley-VCH GmbH.}
    \label{fig:schemes}
\end{figure*}

PL measurements by D’Amato \textit{et al.} (2024) reveal that increasing $\text{FA}^+$ content leads to broadening of the full width at half maximum (FWHM) and an increase in the average lifetime of QDs \cite{d2023color}. This effect arises because the substitution of $\text{Cs}^+$ with the larger $\text{FA}^+$ cation distorts the $\text{PbBr}_6$ octahedral framework, increasing Pb-Br bond lengths and altering bond angles. Since the valence band (Br 4p orbitals) and conduction band (Pb 6p orbitals) are highly sensitive to structural changes, the bandgap narrows as Pb-Br bond lengths increase, shifting emission toward longer wavelengths \cite{chen2017full}. D’Amato \textit{et al.} demonstrated that the bandgap decreases progressively with increasing $\text{FA}^+$ concentration ($x$ in $\text{FA}x \text{Cs}_{1-x}\text{PbBr}_3$), enabling precise spectral tuning from approximately 505 nm (pure $\text{CsPbBr}_3$) to 537 nm (pure $\text{FAPbBr}_3$), as shown in Fig. \ref{fig:schemes}(a), (b). This 30 nm (150 meV) tuning range surpasses conventional tuning methods, such as electric field or strain tuning, which typically offer shifts of less than 1 meV (cf. Table \ref{tab:performance metrics}) \cite{xia2019room, d2023color, moczala2019strain}.

While they are not entirely blinking-free, their PL intensity surpasses that of $\text{CsPbBr}_3$, and they maintain strong anti-bunching characteristics ($g^{(2)}(0)<0.5$ for 95\% of emitters) (cf. Fig. \ref{fig:schemes}(c)) \cite{d2023color}. These findings establish strongly confined HOIP QDs as a promising platform for scalable SPEs, combining spectral tunability, stability, and high-purity quantum emission.

\subsubsection{Weakly confined HOIP QDs}
Instead of relying on extremely fast AR to suppress multiphoton emission, weakly confined HOIP QDs can realise blinking free performance through two main schemes, either by reducing rapid AR or by avoiding it altogether. They are the key focus of this section.

\subparagraph{First Scheme.}
The first approach is to use organic cations $\text{FA}^+$ and $\text{MA}^+$ to modulate the effective dielectric constant $\varepsilon_{\text{eff}}$, thus to achieve high ON-state fractions and high-purity $\text{FAPbBr}_3$PQDs at RT \cite{wang2024weakly}.

The presence of organic cations increases $\varepsilon_{\text{eff}}$, enhancing dielectric screening, which reduces the impact of Coulomb interactions. Dielectric screening refers to a material’s ability to weaken Coulomb interactions between charged particles due to the material’s polarizability that "shields" the charges \cite{zhu2016screening, miyata2017large, franceschetti2000addition}. Coulomb interactions significantly influence key processes such as exciton and biexciton formation, which, in turn, affect AR and blinking behaviour in PL emission. A strong Coulomb interaction promotes biexciton and trion formation, leading to AR and fluctuations in emission intensity.

Fermi’s golden rule describes the dependence of the Auger rate on Coulomb interactions:
\begin{equation}
    \frac{1}{\tau_\text{A}}=\frac{2\pi}{\hbar}\int |G^{\text{Aug}}|^2\rho(\varepsilon_\text{f})=\frac{2\pi}{\hbar}\int | \langle\Psi_{\text{i}}|\frac{e^2}{\kappa_\text{Aug}|r_1-r_2|}|\Psi_{\text{j}}\rangle|^2\rho(\varepsilon_\text{f}),
    \label{fermi}
\end{equation}
where \(\rho(\varepsilon_\text{f})\) is the density of carrier final states, \(G^{\text{Aug}}\) is the Auger coulomb matrix element, \(r_1\) and \(r_2\) are coordinates of electrons, and \(\kappa_\text{Aug}\) is the dielectric constant \cite{efros2016origin, cragg2010suppression}. 

According to Eq. (\ref{fermi}), a high dielectric constant screens Coulomb interactions, reducing electron-hole binding strength and suppressing biexciton formation \cite{franceschetti2000addition, franceschetti2000pseudopotential}. As a result, exciton recombination becomes less probable, leading to extended exciton lifetimes ($\tau_{\text{X}}$) and suppression of type-A blinking, which is directly associated with AR.

Specifically, weakly confined QDs are synthesized by increasing the size of the QDs, which reduces the strength of quantum confinement. As QD size increases, the electron-hole pair has greater freedom of movement, weakening Coulomb interactions \cite{franceschetti2000addition, franceschetti2000pseudopotential}. This reduction in Coulomb strength decreases biexciton formation probability, thereby suppressing AR. Fewer nonradiative recombination events lead to less blinking and higher single photon emission purity. Additionally, weak confinement results in a more stable exciton state, where the electron and hole remain loosely bound, preventing rapid recombination or the formation of multiple exciton states that could interfere with single photon emission.

They also have a lower surface-to-volume ratio, reducing the number of charge carriers in contact with the surface. This leads to fewer surface traps where electrons or holes could become localized, thereby mitigating type-B blinking \cite{veamatahau2015origin}. In addition, weak confinement reduces blinking caused by charge carrier dynamics and surface defect density. In $\text{FAPbBr}_3$, Wang \textit{et al.} employed a RT synthesis method that enables the formation of larger QDs. This approach is less aggressive than synthesis techniques aimed at producing smaller, strongly confined QDs, such as hot-injection methods \cite{wang2024weakly, minh2017room}.

It should be noted that dielectric screening does not compromise single photon purity but optimises Coulomb strength. Although weaker Coulomb interactions reduce biexciton binding, biexciton AR remains fast (short \(\tau_{\text{XX}}\)) due to multi-carrier enhancement. Since biexcitons inherently experience stronger interactions, even reduced Coulomb forces sustain rapid AR. By tuning organic cations to increase $\varepsilon_{\text{eff}}$ slightly above that of fully inorganic perovskites, sufficient Coulomb strength is maintained for biexciton AR \cite{chandrasekaran2017nearly}. This ensures that biexciton AR remains efficient, while single-exciton AR is suppressed, reducing blinking and increasing brightness. 

This approach has been demonstrated experimentally by Wang \textit{et al.} in 2024 \cite{wang2001generation}. Their results showed that $\text{FAPbBr}_3$ PQDs with an 11 nm size exhibit an average $g^{(2)}(0)$ of only 0.071, while achieving over 99\% ON time in a 20 ms time bin. This value of \(g^{(2)}(0)\) is lower than that of $\text{CsPbCl/Br}_3$ at low temperatures \cite{fu2018unraveling}. Although the PQD with the best purity is still $\text{CsPbI}_3$, combined with their blinking-free feature, $\text{FAPbBr}_3$ PQDs exhibit truly promising performance \cite{zhu2022room, kaplan2023hong}. Their experiments also confirmed that increasing the x values in mixed $\text{FA}x \text{Cs}_{1-x}\text{PbBr}_3$ PQDs enhances single photon purity and ON-state fraction (cf. Fig. \ref{fig:schemes}(d)-(e)) \cite{wang2024weakly}. These findings suggest that HOIP QDs are a promising alternative for stable, high-purity, and blinking-free SPEs.

\subparagraph{Second Scheme.}
The second scheme prevents AR entirely by leveraging negative biexciton binding energy, thereby suppressing type-A blinking. In 2024, Chuang \textit{et al.} demonstrated a breakthrough in achieving non-blinking, room-temperature single photon emission from a 11 nm $\text{MAPbI}_3$ PQDs using a novel synthesis and structural engineering approach \cite{chuang2024highly}. Unlike two previous methods, this approach focuses on incorporating Cs into organic cations rather than adding organic cations to Cs-based perovskites. Since $\text{Cs}^+$ is smaller than $\text{MA}^+$, it induces compressive strain in the perovskite lattice \cite{liu2022stable, mari2022stability}. This strain alters crystal symmetry and modifies the spatial overlap of electron and hole wavefunctions, weakening Coulomb attraction between excitons.

Strain-induced lattice distortion creates a staggered band structure similar to a type-II heterojunction, where electrons and holes are spatially separated within the QD \cite{liu2022stable}. This separation enhances repulsive interactions between like-charged carriers, such as electron-electron or hole-hole repulsion, destabilizing biexciton formation \cite{kim2020impact}. Similar to the first scheme, Cs doping stabilises large polarons, quasiparticles formed by charge carriers coupled to lattice vibrations, which further screen Coulomb interactions and amplify excitonic repulsion \cite{neukirch2016polaron, sun2021ultrafast}. The combined effect leads to a negative biexciton binding energy, making biexciton formation energetically unfavourable. Without biexcitons, there is no need for AR to quench them, eliminating the root cause of type-A blinking \cite{chouhan2020synthesis, chuang2024highly}.

Furthermore, the reduced multiexciton pathways, high crystallinity, and effective surface passivation help suppress multiple photon emission. The absorption cross-section for biexcitons is orders of magnitude smaller than that for single excitons, measured at $1.80\times 10^{-12}\text{cm}^2$ for excitons and $0.8\times 10^{-15}\text{cm}^2$ for biexcitons \cite{chuang2024highly}. This ensures that even at high pump fluence, single-exciton states dominate, minimising multi-photon emission. Ball-milling synthesis produces QDs with fewer surface defects, which suppresses trap-assisted recombination \cite{cai2023mechanosynthesis, leupold2019high}. Cs doping stabilizes the lattice, further reducing non-radiative decay channels. Compared to the first scheme, which relies on balancing fast AR for purity with dielectric screening to reduce blinking, $\text{MAPbI}_3$ QDs achieve both by eliminating the need for biexciton annihilation and leveraging intrinsic material properties such as strain and polaron effects to suppress multi-exciton states.

Experimentally, Chuang \textit{et al.} demonstrated that this scheme avoids trade-offs between purity and stability, enabling non-blinking emission with very high single photon purity at room temperature (cf. Fig. \ref{fig:schemes}(f)-(j)). The advantages of these $\text{MAPbI}_3$-based PQDs are significant. First, they exhibit near-unity PLQY at the ensemble level, ensuring bright and efficient emission. Second, they achieve a record-high absorption cross-section, enhancing their utility in low-power optoelectronic applications. Third, the HOIP PQDs also demonstrate exceptional structural stability, enduring over an hour of high-resolution TEM electron irradiation without degradation, a notable improvement over conventional $\text{MAPbI}_3$ materials. Most importantly, they exhibit non-blinking single photon emission with 95.3\% purity at RT, a rare feat among quantum emitters \cite{chuang2024highly}.
\subsubsection{Other Approaches}
Apart from QD structure, two possible mechanisms for HOIP SPEs are defect-localized excitons and dimensional confinement. In larger NCs exceeding the Bohr radius, single photon emission can still occur despite the absence of quantum confinement. Instead, emission relies on defects such as colour centres we explained before, or surface states to trap excitons, a phenomenon observed in larger NCs and bulk materials \cite{walke2023unusual, kurtsiefer2000stable, bradac2019quantum, smith2019colour}.

Dimensional confinement, on the other hand, can be utilized in 2D layered structures or nanoplatelets. In particular, 2D perovskites, such as Ruddlesden-Popper phases, form quantum wells that confine excitons within ultrathin layers without requiring QDs \cite{cho2021simulations, su2022tuning, shaik2021optical, peyskens2019integration}. However, no theoretical or experimental research has been conducted on these two approaches. HOIP QDs remain the most promising and extensively studied mechanism, already demonstrating high performance.

Undoubtedly, both weakly confined schemes represent significant advancements, yet they face material-specific limitations. These mechanisms have been verified using only one or two types of materials. While theoretically sound, practical challenges remain largely unexplored. For instance, in the first scheme, balancing AR rates and dielectric screening is a delicate process, slight deviations could reintroduce blinking. Its reproducibility must be further studied. Additionally, the incorporation of organic cations ($\text{FA}^+$) makes the QDs susceptible to environmental degradation (moisture, heat), limiting long-term stability \cite{zhang2012aggregation}.

\begin{figure}[htbp]
    \centering
    \includegraphics[width=0.7\linewidth]{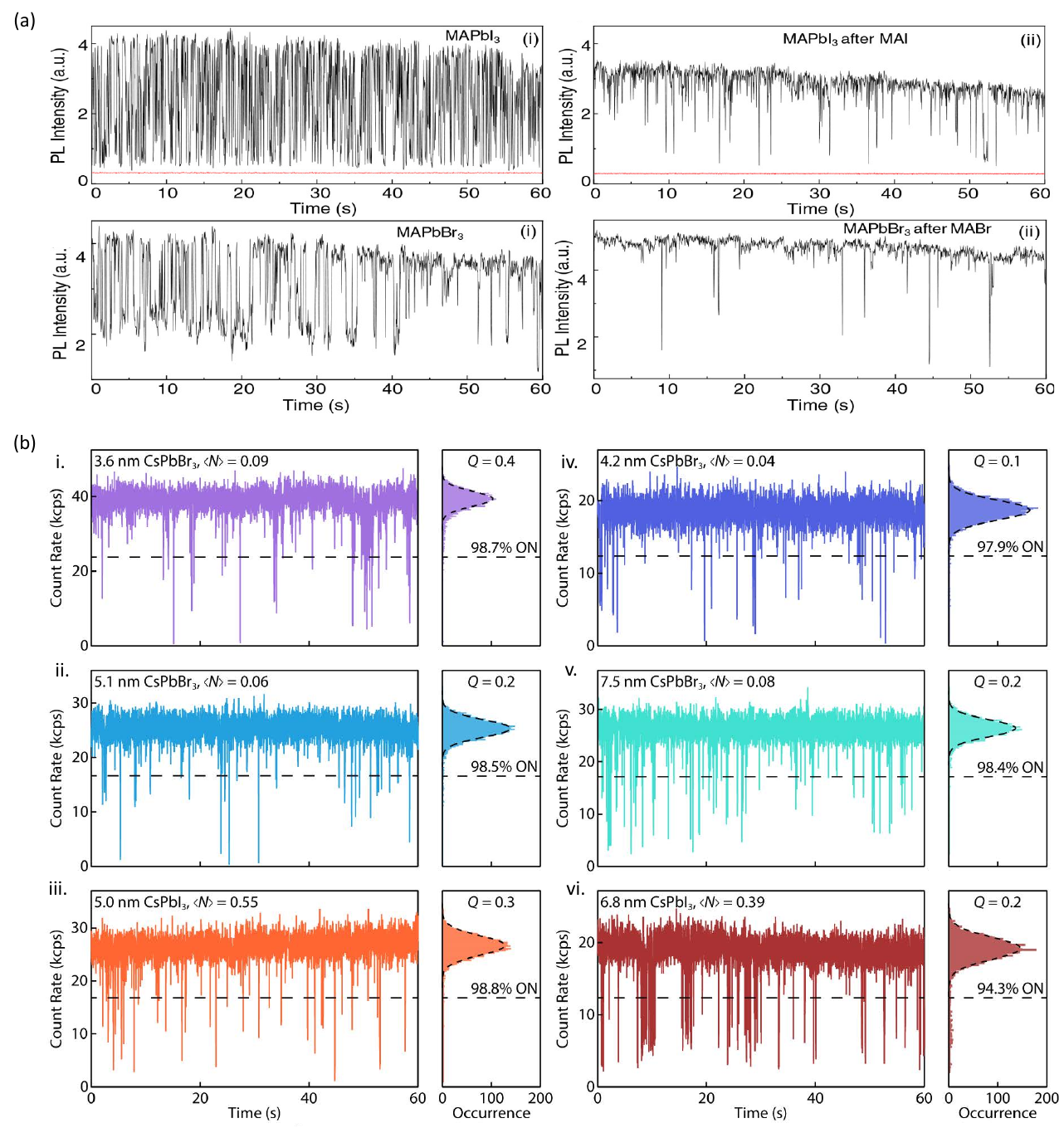}
    \caption{Other blinking-suppressed PQDs approaches. (a) PL intensity trajectories demonstrating blinking suppression in halide vacancy filled PQDs. Upper panel: Trajectories of \(\text{MAPbI}_3\) before and after treatment with a MAI solution. Lower panel: Trajectories of \(\text{MAPbBr}_3\) before and after treatment with a MABr solution \cite{chouhan2021real}. Reproduced with permission from ACS Nano \textbf{15}, 2831 (2021). Copyright 2021 American Chemical Society. (b) Nearly non-blinking PEA-covered \(\text{CsPbBr}_3\) and \(\text{CsPbI}_3\) QDs with varying sizes. \textbf{i-vi} Single QD PL blinking traces and corresponding intensity distribution histograms of four PEA-covered \(\text{CsPbBr}_3\) QDs and two PEA-covered \(\text{CsPbI}_3\) QDs \cite{mi2025towards}. Reproduced with permission from Nat. Commun. \textbf{16}, 204 (2025). Copyright 2025 The Author(s).}
    \label{fig:blinking further research}
\end{figure}

Similarly, $\text{MAPbI}_3$ in the second scheme is notoriously unstable under ambient conditions (moisture, light, heat) \cite{han2015degradation, nagabhushana2016direct, senocrate2019thermochemical}. While Cs doping improves robustness, its long-term stability remains unproven, particularly in device environments. Furthermore, negative biexciton binding energy, crucial for AR suppression, depends heavily on strain engineering and Cs doping. This effect may not translate directly to other cations like Br or Cl analogues, and different cations like $\text{FA}^+$. To explore different compositions, Chouhan \textit{et al.} demonstrated  \(\text{MAPbX}_3\) (X=Br, Cl) with suppressed blinking by filling the halide vacancies \cite{chouhan2021real}. This is achieved by post-synthesis treatment with bromide or iodide donors. Similar further studies should be conducted (cf. Fig. \ref{fig:blinking further research}(a)). To investigate other HOIP material, researchers may be able to incorporate other excellent properties to a single SPE. For example, the bright PL spectrum of $\text{FAPbI}_3$ extends to 800 nm in the near-infrared region (See Table \ref{tab:performance metrics}) \cite{fu2018unraveling}. 

More importantly, rigorous theoretical studies on the effects of incorporating organic and inorganic cations should be conducted, with particular focus on factors such as size and composition. Also, it is important to achieve a better understanding of the excitonic emission dynamics and splitting mechanisms, which may provide alternatives to suppress blinking more fundamentally without complicated dielectric engineering \cite{hou2021revealing}. The size-dependent characteristic of the above approaches are a major drawback in further theoretical research. In other words, the sizes of QDs are strictly tuned to a particular number, not adjustable. Once the size of QD changes, the approach will fail fundamentally, particularly for approach one and three. This is not a desired feature for expanding the research to QDs of other different sizes and integration. In 2025, Mi \textit{et al.} have explored the use of \(\pi -\pi\) stacked PEA ligands that have achieved extraordinary photostability (12 hours of continuous operation) and high purity (98\%) in $\text{CsPbBr}_3$ QDs under both weak and strong confinement \cite{mi2025towards}. The \(\pi -\pi\) stacking of PEA ligands forms a nearly epitaxial layer, lowering surface energy and preventing ligand detachment (cf. Fig. \ref{fig:blinking further research}(b)). This eliminates defect-induced charging and trion formation, which are key drivers of Auger recombination and PL intermittency. The stable ligand coverage ensures excitons recombine radiatively rather than non-radiatively. 

Apart from material composition and ligand strategies, other physical approaches to modifying QD emission have also been explored. In particular, plasmonic coupling strategies enable radiative rate enhancement, directional single-photon output, and coherent light–matter interactions by integrating QDs with metallic nanostructures \cite{karna2016competition, mahat2018plasmonically, neogi2005coupling}. While such methods focus on photonic performance rather than blinking suppression per se, they provide a distinct framework for emission modulation that could potentially be combined with electronic or chemical strategies in future device platforms.

Moreover, a broader challenge for QDs remains: heavy metal toxicity. While this method does not entirely eliminate lead, future engineering efforts should focus on reducing its proportion in A cations or finding alternatives.

Beyond independent research, combining the two weakly confined schemes is particularly compelling. Scheme two achieves complete blinking suppression but has lower purity than scheme one. Therefore, it is natural to consider integrating dielectric screening with strain engineering, which could yield QDs with ultralow $g^{(2)}(0)$ and zero blinking. Moreover, the AR-driven approach results in smaller QDs which are advantageous for integration. Investigating this synergy is certainly worthwhile. Furthermore, comparing their performances with those of \(\text{CsPbX}_3\) is a huge benchmark for their scalable application.  
\section{Future Outlook: BSV SPEs?}
\begin{figure}
    \centering
    \includegraphics[width=0.6\linewidth]{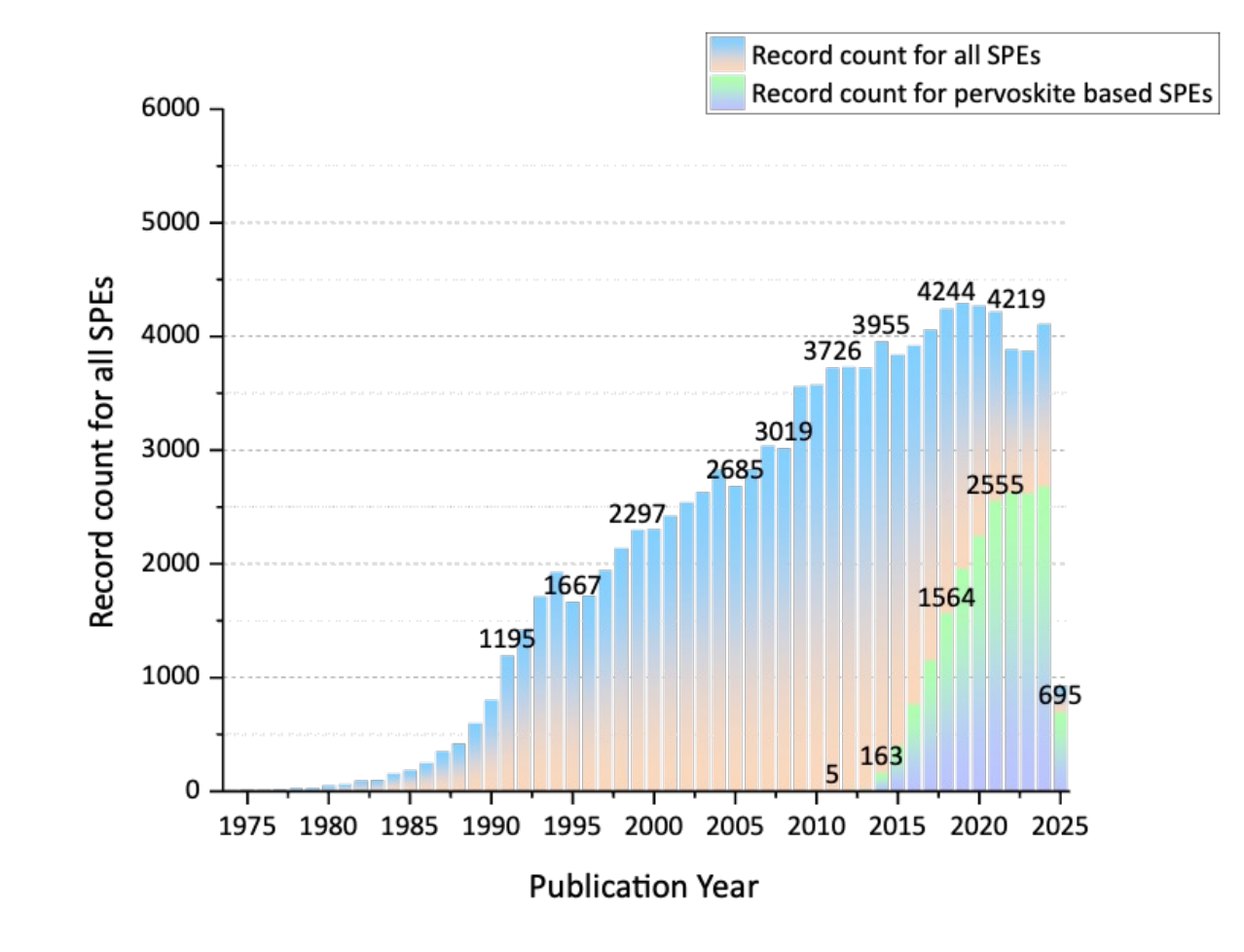}
    \caption{Progress and benchmarks of SPEs: Annual publications on SPEs and perovskite-based SPEs from 1971 to 2025. Data source: Web of Science.}
    \label{fig:progress}
\end{figure}

SPEs have remained a central focus in quantum optics research for over two decades. As early as 2001, annual publication rates on single-photon sources approached 2,297. In recent years, this figure has risen to over 6,500 publications annually (see Fig. \ref{fig:progress}). This growth reflects real technological progress: today’s state-of-the-art SPEs offer significantly improved purity, indistinguishability, and efficiency.

However, despite these advances, most SPEs are still confined to laboratory settings. Achieving commercial viability and scalable deployment remains a long-term objective. While this review has focused primarily on tunability and purity, broader performance metrics also play critical roles in determining whether an SPE platform is scalable. These metrics can be grouped into five fundamental requirements, collectively referred to as the \textit{Robustness, Efficiency, Control, Integrability, and Quality} (RECIQ) framework in this review:
\begin{itemize}
    \item Robustness: stable operation under ambient conditions, including RT and atmospheric moisture.
    \item Efficiency: high QY, photon extraction efficiency, and overall generation rate.
    \item Control: controllable/tunable optical properties.
    \item Integrability: compatibility with existing photonic platforms and supporting technologies
    \item Quality: high single photon purity, indistinguishability, and long coherence time.
\end{itemize}

\begin{figure*}[htbp]
    \centering
    \includegraphics[width=0.8\linewidth]{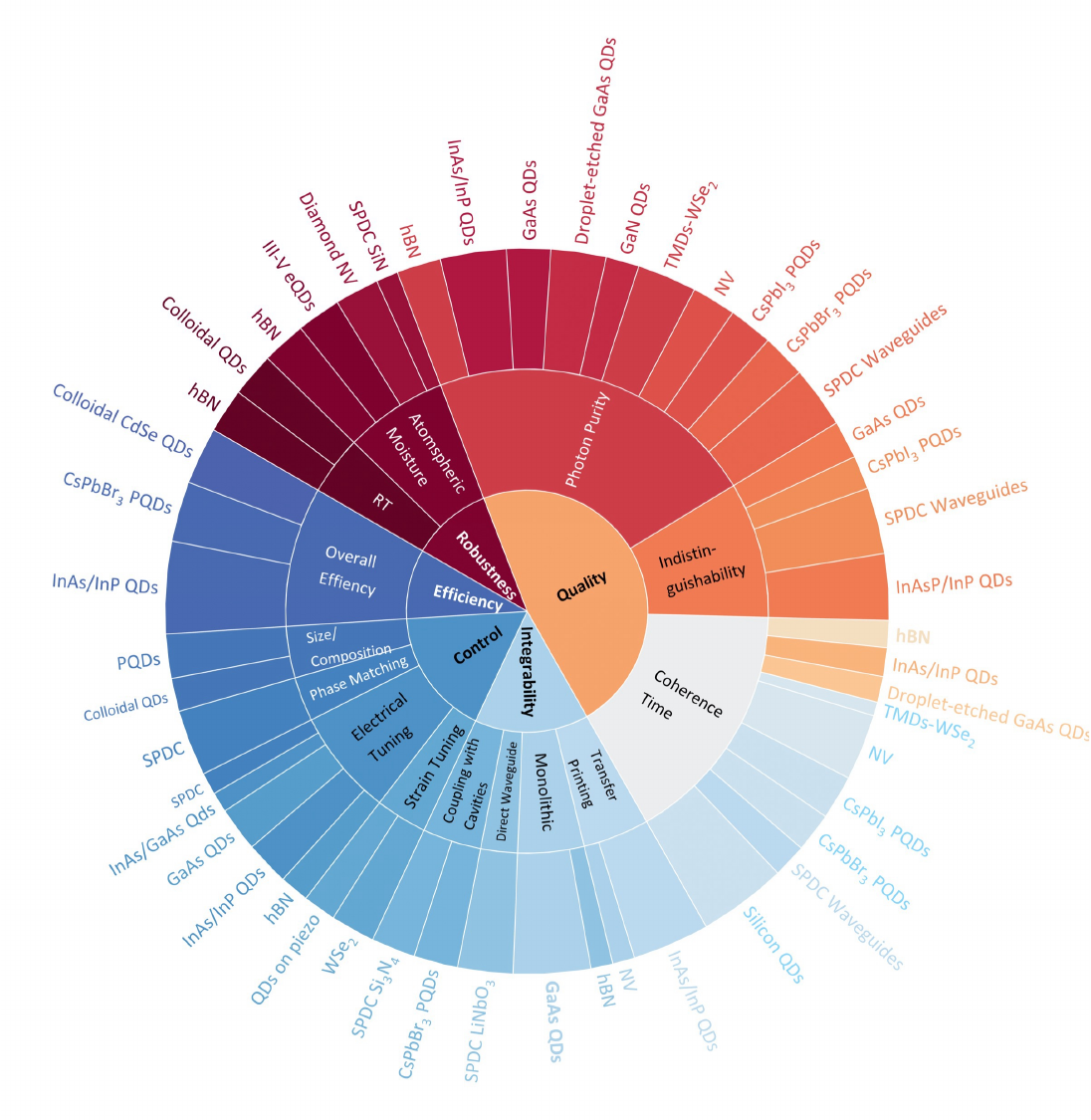}
    \caption{Sunburst plot of the RECIQ framework for SPE performance. Sub-metrics and representative state-of-the-art sources are shown radially in each dimension. Slice sizes scale with relative performance and are indicative.}
    \label{fig:outlook}
\end{figure*}

These criteria provide a generalised structure for evaluating SPEs across diverse platforms and applications, as illustrated in Fig. \ref{fig:outlook}. Depending on the specific use case, the emphasis on individual elements of RECIQ may vary. For example, while high quality is essential for nearly all quantum applications, control, particularly in the form of tunability, is more critical in certain domains than others.

The tunability explored in this review falls under the control category of the RECIQ framework. Specifically, it refers to emission wavelength adjustability and optical property modulation—capabilities that are indispensable for applications such as quantum computing, spectroscopy, and hybrid quantum systems.

In quantum computing, for example, photon interference between independent emitters or optical waveguides is often required. Even small spectral mismatches can destroy interference, compromising entanglement and rendering key quantum operations unreliable. Tunable SPEs allow precise spectral alignment across multiple emitters, which is crucial for protocols such as boson sampling and linear optical quantum computing. In quantum spectroscopy, a tunable photon source can selectively probe different atomic or molecular transitions without the need for separate sources. Similarly, in hybrid systems, where photons interact with other quantum platforms, tuning the photon emission to match the absorption line of the target system is necessary for efficient coupling. 

PQDs currently show some of the most promising tunable performance under RT conditions, alongside high emission quality, explains the reasons behind their high attention \cite{zhu2022room,gorji2025enhanced, kaplan2023hong, farrow2023ultranarrow}. Annual publications on PQDs now account for over 60\% of the total yearly publications on SPEs (cf. Fig. \ref{fig:progress}). However, challenges remain in terms of long-term stability and integration. HOIP QDs, as discussed earlier, offer meaningful improvements in photostability \cite{d2023color, wang2024weakly, chuang2024highly}, and further enhancing their robustness is a key step toward broader system integration. On the integration side, recent demonstrations involving PQDs coupled with nanofibres and nano-antennas \cite{chen2019room, pierini2020highly} show promise, although compatibility with broader photonic infrastructures remains an active area of development.

In fibre-based quantum communication, the notion of control takes on a different meaning. Unlike quantum computing or spectroscopy, wavelength tunability is not required. Telecom systems rely on fixed photon energies, typically around 1550 nm, the centre of the C band, to minimise attenuation in optical fibres. The 1300 nm window is also important due to its zero dispersion and its ability to preserve longer wavelengths for classical data traffic \cite{couteau2023applications1}. In most cases, there’s no need to tune across a range. Rather, what matters is spectral precision and stability at those telecom bands.

Here, instead, \textit{control} refers to the ability to retain quantum properties, most importantly, entanglement. Entanglement is the basis for entanglement-based QKD and other multi-party quantum protocols. Faleo \textit{et al.} recently demonstrated four-photon entanglement at 1550 nm using two high-brightness, high-purity SPDC sources with apodized crystals, achieving spectral purities of at least 98\% \cite{faleo2024entanglement}. Although SPDC sources were demonstrated first, QD-based SPEs have also shown the ability to generate entangled photon pairs \cite{zahidy2024quantum, basso2019entanglement, meng2024deterministic, pennacchietti2024oscillating}, offering deterministic emission with high brightness in the telecom range.

Apart from entanglement-based QKD, prepare-and-measure protocols like BB84 remain widely used. These often rely on QDs as sources and don’t require entanglement. Instead, they encode quantum information into individual photons, typically via polarisation states. These systems can be deployed over fibre or free-space links, for example, in satellite-to-ground communication. In such setups, control refers to polarisation stability and bit fidelity during transmission. Currently, prepare-and-measure QKD has achieved distances up to about 150 km \cite{yang2024high, gao2023atomically, morrison2023single}, and further improvements depend on maintaining photon integrity. 

Returning to tunability, the tuning mechanisms of HOIP QDs remain fundamentally static. Adjusting the emission properties requires modifying the material itself, either through composition, synthesis conditions, or nanocrystal dimensions. This limits real-time control and hinders integration into reconfigurable or heterogeneous quantum systems. These constraints motivate exploring platforms in which tunability arises from the optical process itself rather than from material engineering. There are emerging sources that naturally provide dynamic, hardware-free tunability, a case in point being BSV, offering complementary capabilities to QD-based emitters. BSV provides such kind of tunability, alongside high brightness and multimode structure. 

BSV is a quantum state of light belonging to the squeezed states of light. Unlike traditional heralded schemes, where single photon generation is constrained by the probabilistic nature of photon-pair production, BSV-based approaches leverage the enhanced brightness of squeezed states. This not only improves photon extraction efficiency but also enables precise control over spectral and spatial properties. Their principal strength lies in the ability to generate high-brightness photon states across multiple modes simultaneously. This capability has high potential for hybrid quantum systems and frequency-multiplexed architectures, which benefit from parallel processing and interaction. In theory, such architectures could dramatically improve computational efficiency and throughput. In the remaining section, we explore the possibility of adapting BSV as an SPE and how its unique properties facilitate higher tunability and purity. In the following three sections, we review the theoretical background of BSV, discuss two probable conceptual channels for realizing SPEs using BSV, and finally, we assess the theoretical viability of a multiplexed structure to harness these capabilities.

\subsection{Theoretical Background of BSV}
\begin{figure}[htbp]
    \centering
    \includegraphics[width=0.7\linewidth]{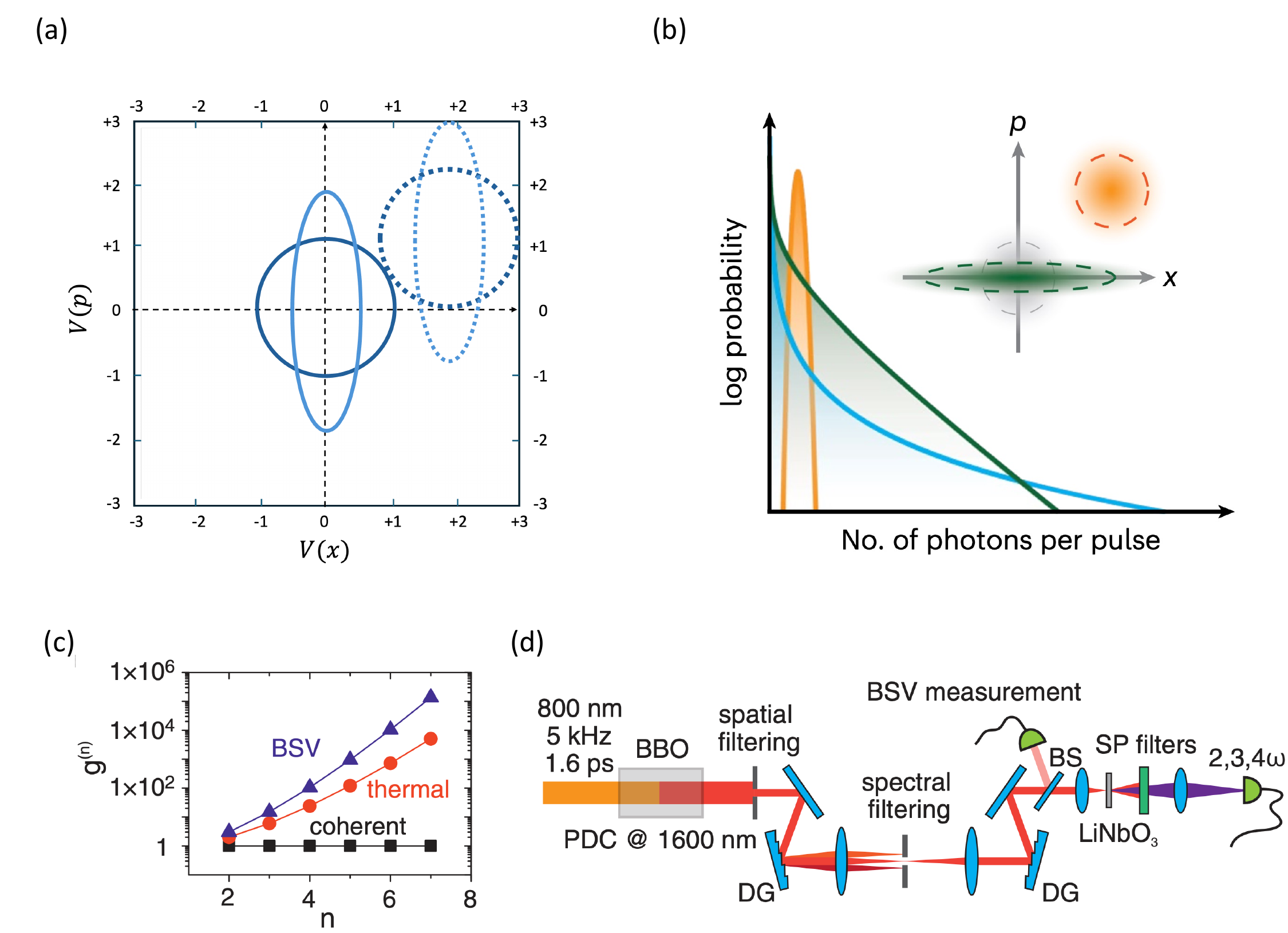}
    \caption{Fundamentals of BSV. (a) Phase-space representations of a coherent state, squeezed state, vacuum state, and squeezed vacuum state. Circles denote non-squeezed states; ellipses represent squeezed states. Solid lines indicate vacuum states, dashed lines indicate non-vacuum states. (b) Photon number distributions for coherent light (orange) and BSV (green), both with equal mean photon numbers per pulse, showing BSV’s broader distribution \cite{heimerl2024multiphoton}. Reproduced with permission from Nat. Phys. \textbf{20}, 945 (2024). Copyright 2024 Springer Nature Limited. (c) Normalized correlation functions illustrating the enhancement of n-photon effects under coherent (black), thermal (red), and BSV (blue) illumination \cite{spasibko2017multiphoton}. Reproduced with permission from Phys. Rev. Lett. \textbf{119}, 223603 (2017). Copyright 2017 American Physical Society. (d) Experimental setup for BSV generation via high-gain PDC \cite{spasibko2017multiphoton}. Reprinted with permission from Phys. Rev. Lett. \textbf{119}, 223603 (2017). Copyright 2017 American Physical Society.}
    \label{fig:phase}
\end{figure}

BSV is a macroscopic quantum state characterised by a large number of photons per mode. Heisenberg's uncertainty principle,
\begin{equation}
    \Delta x\Delta p \geq\frac{\hbar}{2},
\end{equation}
imposes a fundamental limit on the precision of simultaneous measurements of conjugate variables. In quantum optics, the EM field in a given mode is described by two quadrature operators, analogous to position and momentum:
\begin{equation}
    \hat{X} = \frac{1}{\sqrt{2}} (\hat{a} + \hat{a}^\dagger), \quad \hat{Y} = \frac{1}{i\sqrt{2}} (\hat{a} - \hat{a}^\dagger).
\end{equation}
In a coherent state, a quantum analogue of classical light, the variances of these quadratures are equal and normalized to unity, minimizing the uncertainty product \cite{andersen201630, walls1983squeezed, chekhova2015bright}. This results in a circular phase-space distribution (cf. Fig. \ref{fig:phase}(a)).

To distinguish squeezed states from their classical counterparts, we first examine the vacuum state. It refers to the quantum ground state of the EM field, which lacks coherent excitation. Light is the manifestation of excitations (photons) of the EM field, so in the traditional sense, vacuum states do not have any photons on average. However, quantum fluctuations persist due to the Heisenberg uncertainty principle. These fluctuations are like virtual photons popping in and out of existence and cannot be directly detected but are essential to quantum field behaviour \cite{sharapova2020properties}. The vacuum state satisfies
\begin{equation}
    V(x)=V(y)=1,
    \label{vacuum}
\end{equation}
where \(V(x)\) and \(V(y)\) represent the variance for \(x\) and \(y\) respectively.

Researchers naturally sought ways to reduce uncertainty while respecting the Heisenberg principle, leading to the concept of squeezed states of light. A quantum state is considered squeezed when the variance of one quadrature amplitude falls below that of a coherent state, at the cost of increased variance in the conjugate quadrature. This redistribution of uncertainty results in an elliptical phase-space diagram rather than a circular one (cf. Fig. \ref{fig:phase}(a)), reducing quantum noise in a controlled manner. This property makes squeezed states ideal for precision measurements and noise-sensitive applications, such as quantum-enhanced metrology and secure communication \cite{walls1983squeezed, andersen201630}.

A squeezed state is obtained by applying the squeezing operator \( \hat{S}(\zeta)\) to a coherent state. The single-mode squeezing operator is defined as:
\begin{equation}
    \hat{S}(\zeta) = \exp\left[\frac{1}{2} \left(\zeta \hat{a}^2 - \zeta^*\, (\hat{a}^\dagger)^2\right)\right],
    \label{squeezing factor}
\end{equation}
where the complex squeezing parameter \(\zeta=re^{i\phi}\) encapsulates both the squeezing magnitude \(r=\ln{R}\) (with \(R\) being the squeezing factor) and the squeezing phase \(\phi\) \cite{walls1983squeezed, lvovsky2015squeezed}.

Unlike other forms of quantum light, squeezing introduces photon bunching, significantly altering the expectation value of the number operator, $\hat{n} = \hat{a}^\dagger \hat{a}$ \cite{walls1983squeezed, schnabel2017squeezed, spasibko2017multiphoton}. Since the squeezing operator is exponential in the squared creation and annihilation operators, applying it to the vacuum state generates a superposition of even photon-number states, resulting in a squeezed vacuum (SV). BSV is the high-gain regime of SV, characterised by strong squeezing correlations and a macroscopic photon number \cite{chekhova2015bright, rasputnyi2024high, sharapova2020properties}.

For a pure squeezed vacuum state \(|\xi\rangle\), the mean photon number is:
\begin{equation}
    \langle n\rangle =\langle \xi |\hat{a}^\dagger \hat{a}|\xi\rangle=\text{sinh}^2(r),
\end{equation}
where \(r\) is the squeezing operator \cite{ferraro2005gaussian, walls1983squeezed}. Even for small \(r\), \(\text{sinh}^2(r)\approx r^2\), which is nonzero. Therefore, unlike what the term "vacuum" indicates, the squeezed vacuum is not empty and has a non-zero average photon number. BSV, specifically, has very large \(\text{sinh}^2(r)\) and thus exhibits macroscopic photon numbers while retaining squeezing correlations. The "brightness" here emphasises its high intensity. Therefore, most research on BSV has focused on leveraging its multimode nature and brightness for high harmonic generation \cite{heimerl2024multiphoton, rasputnyi2024high, sharapova2020properties, spasibko2017multiphoton}.

\subsection{Utilisation of BSV for SPEs}
\begin{figure*}[htbp]
    \centering
    \includegraphics[width=1\linewidth]{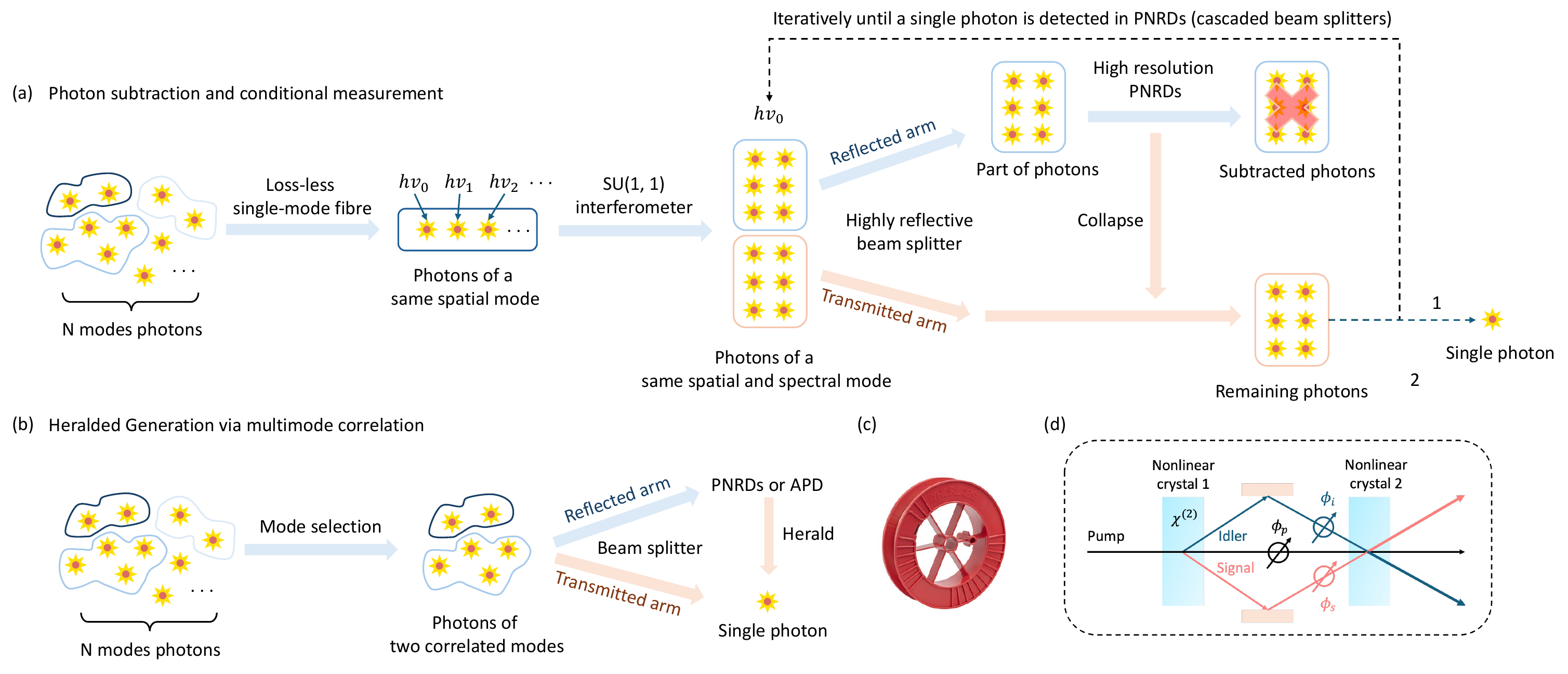}
    \caption{Schematic of two conceptual BSV-SPE channels. (a) Photon subtraction with conditional measurement: a high-reflectivity beam splitter and PNRDs are used to filter single photons from a BSV state. (b) Heralded single photon generation via multimode correlations: correlated photons in distinct BSV modes enable conditional detection. (c) Polarization-maintaining single-mode optical fibre for spatial mode filtering. Image courtesy of Thorlabs Inc. (d) SU(1,1) nonlinear interferometer comprising two high-gain parametric amplifiers, enabling spectral mode control and squeezing enhancement. Adapted from \cite{iskhakov2016nonlinear}. Reproduced with permission from J. Mod. Opt. \textbf{63}, 64 (2016). Copyright 2016 Taylor \& Francis.}
    \label{fig:BSV channels}
\end{figure*}

In this section, we explore two conceptual approaches for single photon emission: isolating single photons within a single mode or maintaining the two correlated modes as heralded sources (See Fig. \ref{fig:BSV channels}). Finally, based on these concepts, we discuss a multiplexed structure for BSV SPEs.

\subsubsection{Two Channels for BSV SPEs}
\subparagraph{Photon Subtraction with Conditional Measurement}
A fundamental prerequisite for high-quality SPEs is indistinguishability, which necessitates that the emitted photon occupies a single optical mode. Since BSV is inherently multimode in the spatial, spectral, and angular domains, adapting it for single-photon emission first requires effective mode isolation.

\begin{figure*}[htbp]
    \centering
    \includegraphics[width=0.8\linewidth]{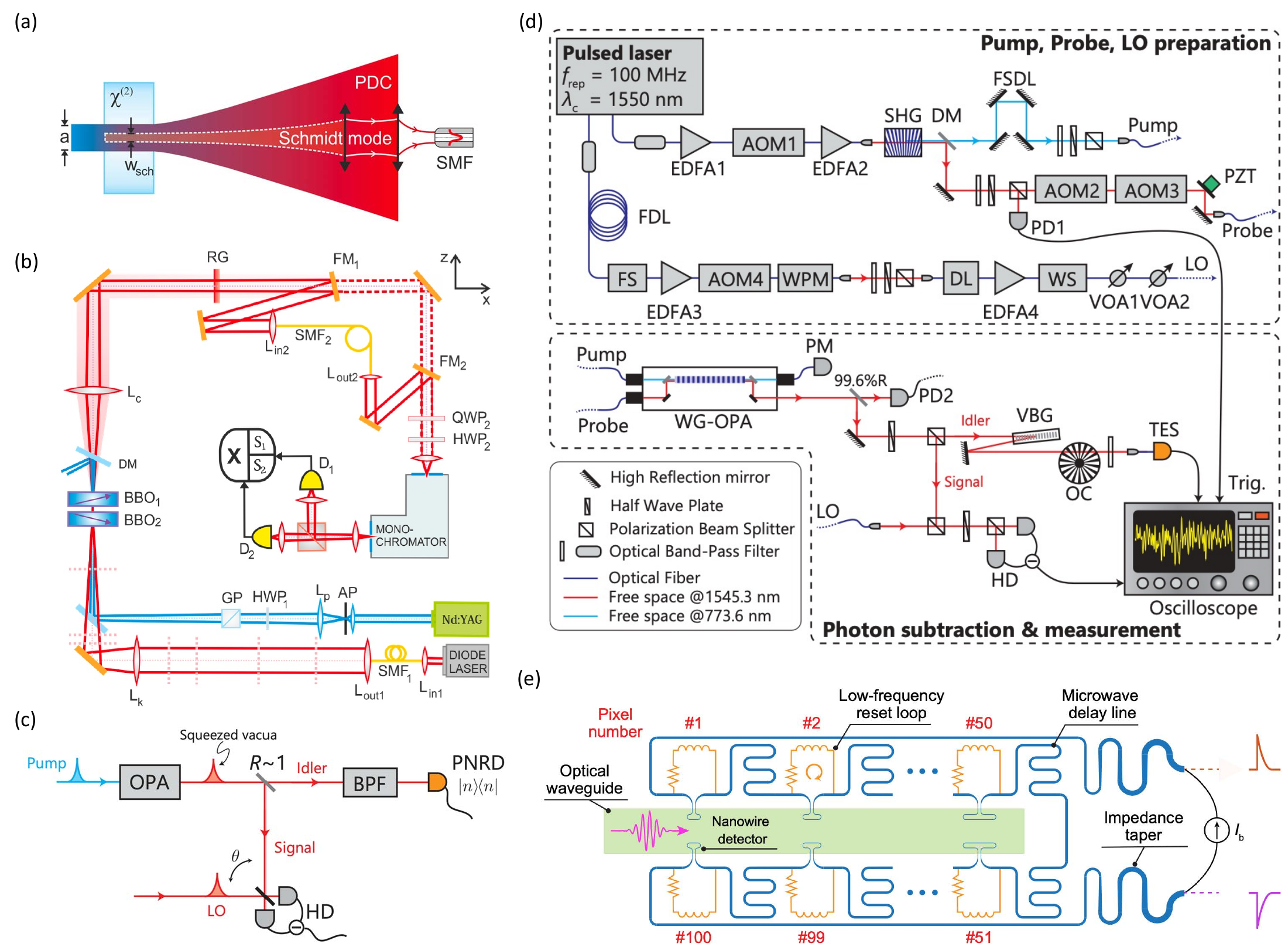}
    \caption{BSV SPEs. (a) Schematic representation of single eigenmode isolation of broadband PDC radiation using a single-mode fibre \cite{perez2015projective}. Reproduced with permission from Phys. Rev. A \textbf{92}, 053861 (2015). Copyright 2015 American Physical Society. (b) Experimental setup for single spatial Schmidt mode isolation: third-harmonic generation from a Nd:YAG laser is directed into BBO crystals to produce PDC. Mode emulation and filtering are performed using a series of lenses and single-mode fibres. The filtered signal is analysed via a HBT interferometer \cite{perez2015projective}. Reproduced with permission from Phys. Rev. A \textbf{92}, 053861 (2015). Copyright 2015 American Physical Society. (c) Conceptual illustration of a photon subtraction experiment: squeezed vacuum light is generated by an optical parametric amplifier (OPA), partially tapped by a beam splitter and detected using a PNRD, while the remaining beam is measured via homodyne detection with variable phase \(\theta\) \cite{endo2023non}. M. Endo et al., Opt. Express \textbf{31}, 12865 (2023); licensed under a Creative Commons Attribution (CC BY) license. (d) Experimental apparatus for multiphoton subtraction featuring erbium-doped fibre amplifiers (EDFAs), acousto-optic modulators (AOMs), second-harmonic generation (SHG), waveguide OPAs, TES, and homodyne detection \cite{endo2023non}. M. Endo et al., Opt. Express \textbf{31}, 12865 (2023); licensed under a Creative Commons Attribution (CC BY) license. (e) On-chip PNRD architecture based on a spatiotemporally multiplexed SNSPD array, capable of resolving up to 100 photons \cite{cheng2023100}. Reproduced with permission from Nat. Photon. \textbf{17}, 112 (2023). Copyright 2023 Springer Nature Limited.}
    \label{fig:BSV exp}
\end{figure*}

Previous studies have established viable pathways for such isolation. For instance, Pérez et al. demonstrated the generation of BSV with an effective spatial mode number of 1.1, approaching a true single-mode state \cite{perez2014spatially}. They further proposed a lossless filtering technique that projects the BSV spatial spectrum onto the eigenmode of a single-mode fibre, demonstrating that maximizing fibre-coupled intensity allows for practical single spatial mode isolation (cf. Fig. \ref{fig:BSV channels}(c), Fig. \ref{fig:BSV exp}(a), (b)) \cite{perez2015projective}. Alongside spatial filtering, spectral control remains critical. The use of $\text{SU}(1,1)$ interferometers, which consist of two nonlinear crystals separated by an air gap, has been shown to mitigate spectral broadening at high pump powers, enabling the tailoring of frequency Schmidt modes (cf. \ref{fig:BSV channels}(d)) \cite{sharapova2018bright, perez2015projective}. When combined with dispersive media, this interferometric approach allows for the manipulation of phase, dispersion, and parametric gain, theoretically ensuring that photons occupy a single spectro-temporal mode to enhance indistinguishability.

However, mode isolation alone is insufficient, as even a single mode of BSV can contain up to 2500 photons \cite{finger2015raman, rasputnyi2024high}. Although this photon number is not macroscopic in the classical sense, it significantly exceeds that of weakly squeezed vacuum states, meaning the vacuum component \(|0\rangle\) and two-photon component \(|2\rangle\) are no longer dominant \cite{perez2015projective,perez2015projective, sharapova2018bright}.. To address this, photon subtraction, a well-established technique in quantum optics, provides a theoretical pathway for conversion \cite{woloncewicz2022improved}.

Standard photon subtraction transforms the even-photon distribution of the squeezed vacuum into an odd-photon superposition, creating non-Gaussian states characterized by negative Wigner function regions (cf. Fig. \ref{fig:BSV exp}(c)) \cite{endo2023non, li2022nonclassicality}. This transformation typically employs a weakly reflective beam splitter ($T \approx 1$, $R \ll 1$) to divert a small fraction of photons to a tapped port. Detecting a photon in the reflected arm infers the subtraction of a photon from the transmitted beam, described mathematically for a single-mode squeezed vacuum as:
\begin{equation}
    |\psi_\text{sub}\rangle\propto\hat a |r\rangle=\sum^\infty_{n=1}\sqrt{n} c_n|n-1\rangle,
    \label{single mode SV}
\end{equation}
where \(c_n\) are coefficients of the original squeezed state. Since BSV only has even photon numbers (\(|0\rangle\), \(|2\rangle\), \(|4\rangle\), ...), subtracting one photon results in odd photon-number states:
\begin{equation}
    a|\psi\rangle\propto |1\rangle+\sqrt{3}\lambda|3\rangle+\sqrt{5}\lambda^2|5\rangle+...
\end{equation}
where \(a\) is the annihilation operator. Only when exactly one photon is detected in the reflected arm can the transmit beam be accepted. 

To adapt this mechanism for high-purity single-photon generation, the subtraction ratio would theoretically need to be inverted. Rather than removing a single photon, a high-reflectivity beam splitter (\(T\ll1, R=1-T\)) could be employed to divert the majority of the BSV flux to a "subtraction" arm, leaving the transmitted arm in a low-photon-number state. In this configuration, high-resolution PNRDs in the reflected arm would measure the photon number $N$, allowing for the conditional acceptance of the transmitted beam only when the remaining photon number is exactly one.

The feasibility of this approach relies heavily on recent advancements in detection technology. For example, Endo \textit{et al.} (2023) demonstrated the subtraction of up to three photons using transition-edge sensors (TES) \cite{endo2023non}, while Cheng \textit{et al.} reported on-chip detectors capable of resolving up to 100 photons via spatiotemporally multiplexed superconducting nanowires, enabling direct measurement of high-order correlation functions \(g^{(N)}\) up to \(N=15\).  \cite{cheng2023100} . Such devices could theoretically support the high-order subtraction required for this scheme (See Fig. \ref{fig:BSV exp}(d), (e)). Technically, in this "inverted" configuration, the number of photons subtracted equals the number detected by the PNRDs. Ideally, a single subtraction event would yield a single photon in the transmission arm, heralded by the PNRD measurement. If the state does not collapse to \(|1\rangle\), the system could theoretically employ iterative loops or, more practically, HBT post-selection to verify single-photon emission, avoiding the optical loss and phase distortion associated with feedback loops. To maintain mode purity, all components would ideally be connected via polarization-maintaining single-mode fibre.

This framework offers distinct advantages regarding tunability and spectral reach. In the high-gain regime, BSV exhibits an exceptionally wide bandwidth, potentially reaching 100–150 THz \cite{katamadze2015broadband}. While this broadband nature challenges single-mode generation, it significantly expands applicable spectral bands. Unlike SPDC and FWM, which rely on specific phase-matching conditions, BSV’s broadband characteristics facilitate access to multiple frequencies, potentially extending to UV ranges if pumped with short-wavelength lasers \cite{spasibko2017multiphoton}. Furthermore, because photon subtraction effectively filters out multiphoton contributions, this method theoretically offers high purity.

Despite these theoretical strengths, this approach remains experimentally untested for SPE purposes. The primary challenge lies in reconciling the inherent multimodality and high photon flux of BSV with the stringent single-mode requirements of SPEs \cite{spasibko2017multiphoton, rasputnyi2024high, sharapova2020properties}. While projective filtering and photon subtraction offer theoretical solutions, practical implementation is hindered by detection limits, optical loss, and the complexity of precise parameter tuning (e.g., squeezing strength and reflectivity) \cite{matsuoka2017generation}. Consequently, extensive experimental validation is required to confirm whether the trade-offs in post-selection efficiency can be overcome to realize a practical source.

\subparagraph{Heralded Generation via Multimode Correlations}
An alternative theoretical framework retains the intrinsic high-dimensional structure of BSV rather than isolating a single operational state. High-gain nonlinear processes, such as PDC or FWM, generate twin beams with strong photon-number correlations distributed across numerous spatial, temporal, or spectral channels \cite{andersen201630, chekhova2015bright, spasibko2017multiphoton}. Conceptually, BSV can be modeled as a macroscopic reservoir of Two-Mode Squeezed Vacuum (TMSV) states. Utilizing this source is therefore analogous to filtering a specific entangled pair from a broader manifold, as illustrated in Fig.  \ref{fig:BSV channels}(b).

The mechanism relies on a probabilistic heralding protocol. In a standard configuration found in the literature, a TMSV component is directed through a 50/50 beam splitter, with one arm monitored by a PNRD \cite{stasi2022enhanced}. Due to the strict correlations characterizing the state, detecting exactly one photon in the monitoring path (Mode A) theoretically confirms the presence of a single photon in the conjugate path (Mode B).

This conditional detection process is fundamentally distinct from the photon subtraction method discussed previously. In photon subtraction, photons are removed to physically collapse the system into a non-Gaussian state. In this heralded scheme, no photons are removed from the target mode; instead, the single photon in Mode B is identified based on the correlated measurement in Mode A. Post-selection logic is then applied to accept only those events where the PNRD registers a single-photon count. While detailed experimental configurations have been investigated for specific low-gain regimes, adapting this logic to the high-gain BSV regime remains an area for further theoretical and experimental development  \cite{stasi2023high, stasi2022enhanced}.

\subsubsection{Multiplexed BSV SPEs Structure}
While the methods of photon subtraction and correlated heralding discussed previously offer pathways to single-photon emission, they inherently necessitate discarding a significant portion of the BSV state to isolate a single mode. This process results in substantial photon loss, mirroring the efficiency challenges faced by unconventional photon blockade systems. However, the defining characteristics of BSV, namely its extreme brightness and multimode structure, suggest that a multiplexed architecture could theoretically mitigate these losses. By drawing parallels to well-established strategies in SPDC sources, where temporal or spectral multiplexing improves the probability of generating a single photon per cycle, a similar framework can be envisioned for BSV to decouple purity from efficiency.

Given that BSV naturally spans multiple spectral modes, effectively forming a broadband frequency comb, a spectral multiplexing strategy appears theoretically viable. Rather than filtering out a single frequency and discarding the rest, this approach would treat the broad BSV spectrum as a series of independent quantum channels. The conceptual architecture would involve utilizing dispersive elements to separate the BSV output into distinct frequency bins, effectively demultiplexing the state. Following this separation, the strong photon-number correlations existing across conjugate spectral modes could be exploited to herald the presence of single photons. Once a specific channel is heralded, active optical switching would then route the photon to a single output, thereby converting the parallel spectral distribution into a high-rate serial stream of single photons.

The feasibility of this approach is grounded in Schmidt mode decomposition, which allows for the precise quantification of spectral entanglement and effective mode number \cite{sharapova2018bright}. Theoretical analysis indicates that the spectral distribution of BSV is dynamic and controllable. Studies by Sharapova \textit{et al.} (2018, 2020) have shown that the $\text{SU}(1,1)$ interferometer allows for precise spectral control of BSV while enabling multi-channel output through cascaded or parallel configurations \cite{sharapova2018bright, sharapova2020properties}. Introducing dispersive media can further optimise spectral separation. By tuning group velocity mismatch, specific frequency bands can be selectively amplified \cite{sharapova2018bright}. For example, an interferometer composed of two nonlinear crystals and dispersive elements can generate multiple independent spectral channels. 

Crucially, the physics of BSV suggests that efficient multiplexing requires operating in a specific gain regime. Counter-intuitively, reducing the pump gain may enhance the utility of this multiplexed structure. At high gain, a "gain narrowing" effect occurs, where the main spectral lobe sharpens and side lobes are suppressed because the fundamental Schmidt mode dominates the interaction \cite{sharapova2018bright}. It can be understood from:
\begin{equation}
    \langle N_s(\omega _s)\rangle=\sum _n |u_n(\omega_s)|^2\Lambda_n,
    \label{narrowing}
\end{equation}
\begin{equation}
        \Lambda_n=\frac{\text{sinh}^2(G\sqrt{\lambda_n})}{\sum_n\text{sinh}^2(G\sqrt{\lambda_n})}.
\end{equation}
where $\Lambda_n$ represents the gain-dependent weight of the $n$-th mode, \(u_n(\omega_s)\) is the frequency of Schmidt modes, $G$ is the parametric gain, and $\lambda_n$ are the fixed Schmidt eigenvalues. As the gain increases, the weight of the primary mode $n=0$ amplifies disproportionately, effectively narrowing the bandwidth and reducing the number of usable channels. Conversely, operating at lower gain preserves a broader bandwidth, thereby increasing the number of accessible spectral modes for multiplexing. Furthermore, lower gain suppresses the multiphoton components within each individual channel, facilitating the isolation of the single photon pairs required for the heralding logic.

It is important to note that this narrowing effect assumes fixed mode shapes in the presence of dispersion and pump delay \cite{sharapova2018bright}. If mode shapes are gain-dependent, a single-crystal setup can exhibit spectral broadening instead \cite{sharapova2020properties}.

From an implementation perspective, while such an architecture is experimentally demanding, the fundamental components required, such as dispersive demultiplexers and fast optical switches, are standard in classical telecommunications. The challenge lies in integrating these technologies into a quantum circuit capable of preserving the specific coherence properties of BSV. If these integration hurdles can be overcome, such a system may theoretically transform the broadband nature of BSV from a liability requiring heavy filtering into an asset enabling parallel generation, significantly enhancing the scalability of the source.

\subsection{Theoretical Viability and Implementation Challenges}
The implementation of a multiplexed architecture theoretically has huge potential to retain BSV’s exceptional tunability and brightness, potentially overcoming the efficiency bottlenecks of single-mode filtering. The theoretical groundwork for such systems is well-supported by the Schmidt mode decomposition framework, which enables the precise quantification of spectral entanglement and effective mode numbers necessary for efficient channel selection \cite{sharapova2018bright}. Furthermore, the utilization of $\text{SU}(1,1)$ nonlinear interferometers provides a robust mechanism for spectral engineering \cite{sharapova2018bright,sharapova2020properties}. By adjusting pump gain and dispersive properties, these interferometric setups can theoretically tailor the BSV output to match specific demultiplexing grids, enhancing the purity of the extracted single photons. Additionally, the strong photon-number correlations inherent to BSV across conjugate modes provide a physical basis for high-fidelity heralded detection, ensuring deterministic output logic once the modes are separated.

However, the theoretical models currently governing these systems face limitations. The assumption that reducing gain consistently preserves or broadens effective bandwidth is highly conditional and dependent on specific dispersion and pump delay configurations. Deviations from these ideal conditions can lead to spectral behaviors that contradict standard predictions, such as unexpected narrowing or splitting \cite{sharapova2020properties}. Consequently, standard Schmidt decomposition may be insufficient to fully describe spectral broadening effects in high-gain regimes, suggesting that more complex integral-differential models are required to accurately predict system performance.

From a practical standpoint, realizing a multiplexed BSV source presents significant engineering hurdles. The generation of BSV typically relies on bulk SPDC or FWM setups, which complicates integration into the compact, scalable photonic circuits required for quantum information processing \cite{andersen201630, chekhova2015bright, spasibko2017multiphoton}. Furthermore, the introduction of demultiplexing components, such as AWGs, and high-speed switching networks inevitably introduces optical losses. In a multi-stage multiplexing setup, these cumulative losses can critically degrade the signal-to-noise ratio. Timing precision poses another challenge; while SNSPDs offer exceptional resolution with timing jitter below 15 ps, synchronization errors across parallel detection channels can still compromise the fidelity of the heralding signal \cite{you2013jitter, abazi2025multi, mccarthy2025high}.

Despite these challenges, the architectural advantages of BSV offer a compelling rationale for continued development. Compared to traditional probabilistic SPDC sources, a multiplexed BSV framework offers a pathway to higher single-photon probabilities without requiring arrays of independent emitters. The frequency-domain encoding intrinsic to BSV aligns naturally with established wavelength division multiplexing technologies in silicon photonics, positioning it as a potentially scalable resource for quantum networks. Moreover, such a structure inherently supports parallel quantum processing, where independent quantum channels operate across different spectral modes, theoretically enabling simultaneous multi-photon operations that could significantly improve system throughput \cite{manceau2017detection}.
\section{Conclusion}
This review has presented a mechanism-based analysis of SPEs, emphasizing the two dominant physical paradigms: quantum emitter transitions and nonlinear optical processes. We critically examined their current progress and performance metrics, with particular focus on tunability and purity—two key benchmarks for scalable quantum applications. Despite notable advances, the intrinsic trade-off between these metrics remains a major obstacle to practical deployment.

To address the limitations of material-based tuning, we highlighted HOIP QDs as a standout platform. Our analysis confirms that HOIP QDs uniquely bridge the gap between high purity and broad tunability at room temperature, leveraging three distinct physical mechanisms to suppress blinking and enhance stability. Additionally, to standardize the evaluation of such diverse systems, we introduced the RECIQ framework. This guideline provides a cohesive metric for assessing robustness, efficiency, control, integrability, and quality, offering a reference point for future research aimed at designing scalable, high-performance sources.

However, while material engineering in HOIPs offers significant near-term gains, fundamental limits in scalability persist for single-mode emitters. Looking beyond material composition, our review identifies optical state engineering as a complementary frontier. Specifically, we discussed BSV as a high-potential candidate for next-generation sources. Rather than relying solely on single-emitter isolation, we discussed how the intrinsic multimode structure and macroscopic brightness of BSV could be harnessed through theoretical multiplexed architectures. By synthesizing concepts from nonlinear optics and quantum networking, our discussion suggests that such frameworks could theoretically decouple efficiency from purity constraints. While currently at a conceptual stage, this perspective offers a compelling roadmap for overcoming the stochastic limitations of traditional heralded sources, potentially expanding the scope of practical quantum technologies.

\section*{Author Declarations}
\subsection*{Conflict of Interest}
The authors have no conflicts to disclose.
\subsection*{Author Contributions} 
\noindent
\textbf{Galy Yang:} Data curation (equal); Formal analysis (equal); Investigation (equal); Methodology (equal); Writing - original draft (equal). \textbf{Eric Ashalley:} Investigation (equal); Methodology (equal); Writing - original draft (equal). \textbf{Zhiming M. Wang:} Supervision (equal). \textbf{Abolfazl Bayat:} Writing - review \& editing (supporting). \textbf{Arup Neogi:} Conceptualization (equal); Project administration (equal); Resources (equal); Supervision (equal); Validation (equal); Writing - review \& editing (equal).
\section*{Data Availability}
Data sharing is not applicable to this article as no new data were created or analyzed in this study.

\bibliographystyle{aipnum4-1}
%

\end{document}